\documentclass[twocolumn]{aastex631}

\usepackage{graphicx}	% Including figure files
\usepackage{amsmath}	% Advanced maths commands
\usepackage{amssymb}	% Extra maths symbols
\usepackage{verbatim}
\usepackage{multirow}
\usepackage{multirow,bigdelim}
\usepackage{soul}
\usepackage{threeparttable}
\usepackage{float}
\usepackage{babel}

\newcommand{\HI}[0]{{}H\,{\sc i}}

\newcommand{\OI}[0]{{}O\,{\sc i}}
\newcommand{\CII}[0]{{}C\,{\sc ii}}
\newcommand{\SiII}[0]{{}Si\,{\sc ii}}
\newcommand{\SiIII}[0]{{}Si\,{\sc iii}}
\newcommand{\SII}[0]{{}S\,{\sc ii}}
\newcommand{\NV}[0]{{}N\,{\sc v}}

\newcommand{\calcos}{{\sc CalCOS}\,}

\newcommand{\oiii}{[O\,\textsc{iii}]}
\newcommand{\nii}{[N\,\textsc{ii}]}
\newcommand{\sii}{[S\,\textsc{ii}]}

\newcommand{\oii}{[O\,\textsc{ii}]}

\newcommand{\ha}{H$\alpha$}
\newcommand{\hb}{H$\beta$}
\newcommand{\Msun}{${\rm{M_\odot}}$}

\newcommand{\usc}{Department of Physics \& Astronomy, University of South Carolina, Columbia, SC 29208, USA}
\newcommand{\unc}{Department of Physics \& Astronomy, University of North Carolina Asheville, Asheville, NC 28804, USA}
\newcommand{\tcu}{Department of Physics \& Astronomy, Texas Christian University, Fort Worth, TX 76109, USA}
\newcommand{\esa}{European Southern Observatory, Karl-Schwarzschildstrasse 2, D-85748 Garching bei M{\"u}nchen, Germany}
\newcommand{\mars}{Aix Marseille Universite, CNRS, LAM (Laboratoire d'Astrophysique de Marseille) UMR 7326, F-13388, Marseille, France}
\newcommand{\uw}{University of Wisconsin—Madison, Department of Astronomy, 475 N. Charter Street, Madison, WI 53706-1582, USA}
\newcommand{\saao}{ South African Astronomical Observatory, PO Box 9, Observatory 7935, Cape Town, South Africa}
\newcommand{\uct}{Department of Astronomy, University of Cape Town, Private Bag X3, Rondebosch 7701, South Africa}

\shorttitle{The Baryonic Content of MaNGA Galaxies}
\shortauthors{Klimenko et al.}

\begin{document}

\title{The Baryonic Content of Galaxies Mapped by MaNGA and the Gas Around Them}

\author[0000-0001-6730-2917]{Viacheslav V. Klimenko}
\affiliation{\usc}

\author{Varsha Kulkarni}
\affiliation{\usc}

\author{David A. Wake}
\affiliation{\unc}

\author{Suraj Poudel}
\altaffiliation{\usc}
\affiliation{\tcu}

\author{Matthew A. Bershady}
\affiliation{\uw}
\affiliation{\saao}
\affiliation{\uct}

\author{Celine P\'eroux}
\affiliation{\esa}
\affiliation{\mars}

\author{Britt Lundgren}
\affiliation{\unc}

\begin{abstract}
We analyze the cool gas in and around 14 nearby galaxies (at $z$$<$0.1) mapped with the SDSS-IV MaNGA survey by measuring absorption lines produced by gas in spectra of background quasars/AGN at impact parameters of 0-25 effective radii from the galaxy center. Using HST/COS, %we detect far-UV absorption lines of species associated with the cool gas. 
we detect absorption at the galaxy redshift and measure or constrain column densities of neutral (\HI, N\,{\sc i}, O\,{\sc i}, Ar\,{\sc i}), low-ionization (Si\,{\sc ii}, S\,{\sc ii}, C\,{\sc ii}, N\,{\sc ii} Fe\,{\sc ii}), and high-ionization (Si\,{\sc iii}, Fe\,{\sc iii}, N\,{\sc v}, O\,{\sc vi}) species for 11 galaxies. We derive the ionization parameter and ionization-corrected metallicity using {\sc cloudy} photo-ionization models. 

The \HI\ column density ranges from $\sim$$10^{13}$ to $\sim$$10^{20}\,{\rm cm^{-2}}$ and decreases with impact parameter for $r\ga R_{e}$. Galaxies with higher stellar mass have weaker \HI\ absorption. 
%We confirm the known trends of decreasing $N(\rm HI)$ with increasing impact parameter and increasing galaxy stellar mass. 
Comparing absorption velocities with MaNGA radial velocity maps of ionized gas line emissions in galactic disks, we find  that the neutral gas seen in absorption co-rotates with the disk out to $\sim$10 $R_{e}$. Sight lines with lower  elevation angles show lower metallicities, consistent with the metallicity gradient in the  disk { derived from} 
%\st{measured by IZI modelling of} 
MaNGA maps. Higher elevation angle sight lines show higher  ionization, lower \HI-column density, super-solar metallicity, and velocities consistent with the direction of galactic  outflow. 
Our data offer the first detailed comparisons of CGM properties (kinematics and metallicity) with extrapolations of detailed galaxy maps from integral field spectroscopy; similar studies for larger samples are needed to more fully understand how galaxies interact with their CGM.

\end{abstract}

%%%%%%%%%%%%%%%%%%%%%%%%%%%%%%%%%%%%%%%%%%%%%%%%%%

%%%%%%%%%%%%%%%%% BODY OF PAPER %%%%%%%%%%%%%%%%%%

\section{Introduction}
\label{sec:intro}

Galaxies interact with their surroundings through gas flows. Inflows of cool gas bring in fresh material for star formation. Outflows of enriched gas  carry the chemical elements produced by star formation back into the  intergalactic medium (IGM). These gas flows pass through the circumgalactic medium (CGM) that acts as an interface between the galaxy and the IGM.  Many aspects of the physical interactions between 
%physical problems related to these interactions between 
galaxies and the IGM are  not well-understood. Examples include %For example, 
how galaxies acquire their gas, what processes affect the chemical abundances of stars and gas, and how processes such as accretion, mergers, and secular evolution affect the growth of galaxy components. 

Constraining these physical processes observationally requires spatially resolved information about the kinematics and chemical composition within and around galaxies and CGM. Integral field spectroscopy (IFS) enables spatially resolved measurements of emission line fluxes and line ratios, allowing construction of maps of important physical properties and their gradients such as gas kinematics, ionization, metallicity, and star formation rate (SFR). Comparisons of these rich datasets with predictions of galaxy structure and evolution models can then shed light on how disks and bulges assemble and how baryonic components of galaxies interact with their dark matter halos. A number of interesting studies using IFS have been carried out at intermediate and high redshifts to investigate the gas flows passing through the CGM %(e.g.,  Bouch\'e et al. 2007; P\'eroux et al. 2011, 2016, 2019, 2022; Schroetter et al. 2016, 2019; Fumagalli et al. 2016; Lofthouse et al. 2020, 2022)
\citep[e.g.,][]{Bouche2007, Peroux2011, Peroux2016, Peroux2019, Peroux2022, Schroetter2016, Schroetter2019, Fumagalli2016, Lofthouse2020, Lofthouse2023}. Many of these studies were based on absorption-selected samples. However, connecting these studies to local galaxies requires a parallel study of low-redshift, galaxy-selected samples. 

The Mapping Nearby Galaxies at Apache Point Observatory (MaNGA; \citealt{Bundy2015}) survey of the Sloan Digital Sky Survey IV (SDSS IV; \citealt{Blanton2017}) is particularly useful in this context. MaNGA has obtained IFS data for 10,000 nearby (0.01$<z<$0.15) galaxies with 19 to 127 fibers, spanning 3600-10300 {\AA} with a resolution of $\sim$2000. This survey has led to a number of interesting results relevant to the CGM. For example, extraplanar ionized gas with a variety of emission lines has been detected in edge-on or highly inclined MaNGA galaxies out to $\sim$4-9 kpc \citep[e.g.,][]{Bizyaev2017,JonesNair2019}. In a significant fraction of MaNGA galaxies, this gas appears to lag in rotation compared to the gas closer to the galactic plane \citep{Bizyaev2017}.

While the MaNGA survey provides information about the structure and kinematics of the disk and bulge components in the inner 1.5$-$2.5 effective radii of the galaxies, it does not offer much insight about the gaseous halos of these galaxies.  The H\,{\sc i}-MaNGA program is obtaining 21-cm observations for a large fraction of MaNGA galaxies \citep[e.g.,][]{Masters2019, Stark2021}. However, this 21-cm emission survey is sensitive to relatively high \HI\ column densities ($\ge10^{19.8}\,{\rm cm^{-2}}$)\,\footnote{The estimate corresponds to a $3\,\sigma$ upper limit of the H\,{\sc i} mass surface density for non-detections, see \cite{Masters2019}.}, making it difficult to access the lower column-density outskirts of galaxies and the CGM. 

A powerful technique to probe the gaseous galaxy halos and the CGM is by means of the absorption signatures from the gas against the light of background sources such as quasars or gamma-ray bursts (GRBs). 
Indeed, halo/CGM gas has been detected extending to $\gtrsim$200 kpc around individual galaxies \citep[e.g.,][]{Tumlinson2013} and in large samples of stacked spectra out to $\sim$10\,Mpc \citep[e.g.,][]{Perez-Rafols2015}. Probing the outskirts of MaNGA galaxies with this absorption-line technique provide a mean to establish a local sample of exquisitely imaged galaxies studied in neutral and ionized gas. Such a local sample is essential for placing IFS observations of high-redshift galaxies and their CGM in perspective, and thereby developing a systematic understanding of the interactions between galaxies and the gas flows around them, and the evolution of these interactions with time.   

With these improvements in mind, we have started to investigate the 
ISM and CGM of MaNGA galaxies using the Hubble Space Telescope (HST) Cosmic Origins Spectrograph (COS). Here we describe the results from our COS observations of 14 MaNGA galaxies, and compare the gas properties deduced from the COS data to the properties of the ionized gas measured from the MaNGA data. 

This paper is organized as follows. Section\,\ref{sec:data} describes the sample selection, observations, and data reduction. Section\,\ref{sec:results} describes results from our COS spectroscopy. Section\,\ref{sec:discussion} presents a discussion of our results, including comparisons with the MaNGA data and other studies from the literature. Section\,\ref{sec:conclusions} presents our conclusions. Throughout this paper, we adopt the ``concordance'' cosmological parameters ($\Omega_{m}$=0.3, $\Omega_{\Lambda}$=0.7, and $H_{0}$=70 km s$^{-1}$ Mpc$^{-1}$). 

%\section{Method}
%\label{sec:method}

\section{Observations and data reduction}
\label{sec:data}
\subsection{Sample selection}
\label{sec:sample}

Our sample consists of 14 nearby galaxies (at $z<0.1$) mapped in the SDSS/MaNGA survey with UV-bright quasars/AGNs at impact parameters between 0 to 140 kpc from the galaxy centers. The targeted quasars have GALEX FUV mag $<$19.5, and their impact parameters range from 0 to 25 times the effective radii of the corresponding  MaNGA galaxies. For 1-635629, a bonus galaxy covered
in the same setting as 1-180522, the impact parameter is 38.7 $R_e$. %Another motivation behind selecting these sight lines was that an examination of the SDSS spectra of the sample quasars showed possible weak Na\,{\sc i} and/or Ca\,{\sc ii} absorption lines at the galaxy redshift in most cases, suggesting the existence of cool gas along these sight lines.

\begin{figure*}
\includegraphics[width=\linewidth]{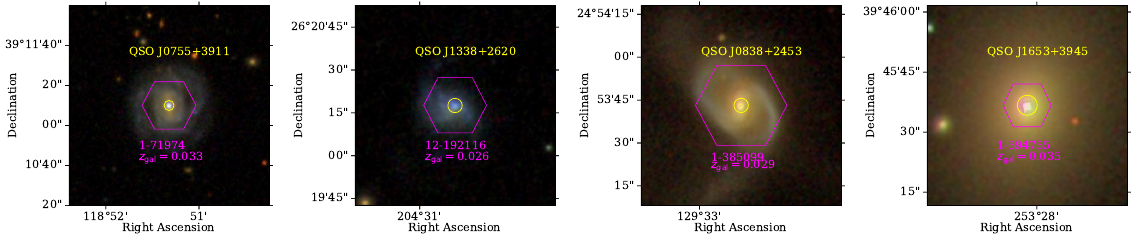}\hfill
%\begin{subfigure}[b]{0.23\textwidth}
%  \centering
%  \includegraphics[width=\linewidth]{image+1-71974.pdf}\hfill  
%\end{subfigure}
%\begin{subfigure}[b]{0.23\textwidth}
%  \centering
%  \includegraphics[width=\linewidth]{image+12-192116.pdf}\hfill  
%\end{subfigure}
%\begin{subfigure}[b]{0.23\textwidth}
%  \centering
%  \includegraphics[width=\linewidth]{image+1-385099.pdf}\hfill
%\end{subfigure}
%\begin{subfigure}[b]{0.23\textwidth}
%  \centering
%  \includegraphics[width=\linewidth]{image+1-594755.pdf}\hfill
%\end{subfigure}
\caption{The images of four galaxies with zero-impact parameter. Each panel shows the SDSS three-color image of the area near the MaNGA galaxy. The image is centered at the position of the galaxy. The pink hexagons show the sky coverage of the MaNGA IFU. The yellow circles show the position of the HST/COS aperture centered on the galaxy nucleus.}
\label{fig:quasar-gal-pairs2}
\end{figure*}

\begin{figure*}
\includegraphics[width=\linewidth]{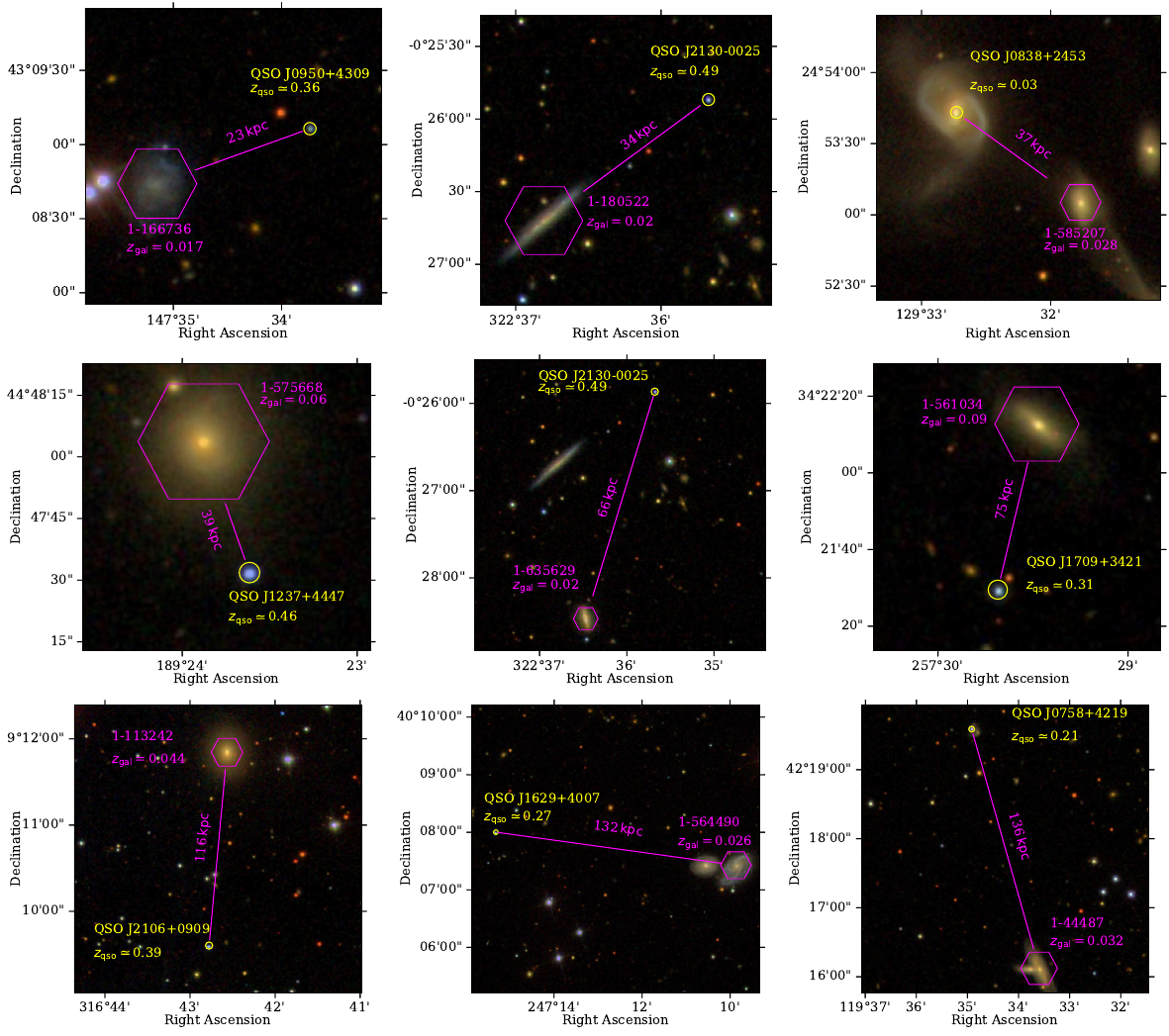}\hfill
%\begin{subfigure}[b]{0.33\textwidth}
   %\centering
   %\includegraphics[width=\linewidth]{image+1-166736.pdf}\hfill
    %\end{subfigure}
    %\begin{subfigure}[b]{0.33\textwidth}
     %\centering
     %\includegraphics[width=\linewidth]{image+1-180522.pdf}\hfill
     %\end{subfigure}
     %\begin{subfigure}[b]{0.33\textwidth}
      %\centering
      %\includegraphics[width=\linewidth]{image+1-585207.pdf}\hfill  
    %\end{subfigure}
    %\begin{subfigure}[b]{0.33\textwidth}
      %\centering
      %\includegraphics[width=\linewidth]{image+1-575668.pdf}\hfill 
    %\end{subfigure}
    %\begin{subfigure}[b]{0.33\textwidth}
     % \centering
      %\includegraphics[width=\linewidth]{image+1-635629.pdf}\hfill  
    %\end{subfigure}
    %\begin{subfigure}[b]{0.33\textwidth}
     % \centering
     % \includegraphics[width=\linewidth]{image+1-561034.pdf}\hfill  
    %\end{subfigure}
    %\begin{subfigure}[b]{0.33\textwidth}
     % \centering
     % \includegraphics[width=\linewidth]{image+1-113242.pdf}\hfill  
    %\end{subfigure}
    %\begin{subfigure}[b]{0.33\textwidth}
     % \centering
     % \includegraphics[width=\linewidth]{image+1-564490.pdf}\hfill  
    %\end{subfigure}
    %\begin{subfigure}[b]{0.33\textwidth}
     % \centering
     % \includegraphics[width=\linewidth]{image+1-44487.pdf}\hfill
    %\end{subfigure}
    \caption{The images of galaxy-quasar pairs with non-zero impact parameter. The panels are arranged in order of increasing impact parameter. Each panel shows the SDSS three-color image of the area near the MaNGA galaxy. The image is centered at the position between the quasar and the galaxy. The pink hexagon shows the sky coverage of the MaNGA IFU. The yellow circle represents the position of the quasar and has the size of the HST/COS aperture. The distance between the quasar and the center of the MaNGA galaxy at the galaxy redshift (the impact parameter) is shown by the pink link.}
    \label{fig:quasar-gal-pairs}
\end{figure*}

We performed HST COS spectroscopy for these quasars/AGNs, as described in section 2.2. The targets are listed in Table\,\ref{tab:sample}. {We divide the targets into two groups by the value of impact parameter. The first group contains four objects with zero impact parameter, because in these cases we observed the  central source, hereafter referred to as AGNs (see Fig.\,\ref{fig:quasar-gal-pairs2}). In these cases, the absorbing gas can be at any distance from the central source along the line of sight, and could potentially be associated with the central engine. Physical conditions in the absorbing gas in such cases can be very different from those in the CGM and ISM. The objects in the second group introduce %\st{have} 
non-zero impact parameter (more than 20\,kpc) and likely probe gas related to the CGM of galaxies (see Fig.\,\ref{fig:quasar-gal-pairs}).} %, {so-called ``down-the-barrel"{}}. 
%For galaxy 1-90242, we used archival HST/COS observations. 
%{ The galaxy-quasar pairs and AGNs from our sample are shown in Fig.\,\ref{fig:quasar-gal-pairs} and Fig.\,\ref{}, respectively}. 
In two cases ({J2130$-$0025 and J0838$+$2453}), the quasar sight lines cross two galaxies at different impact parameters and redshifts. 
%For these galaxies, we clearly detect strong absorption lines of H\,{\sc i}, Si\,{\sc ii}, Si\,{\sc iii} at redshifts close to the galaxy redshifts. Since there are no other galaxies at these redshifts around the quasar sight lines, our HST COS spectra probe the CGM of the selected MaNGA galaxies.  The profile fits to the HST COS absorption line data and the measurements of column densities are presented in detail in Appendix\,\ref{app:A}. 

\setlength{\tabcolsep}{2pt}
\begin{table*}
\begin{center}
\caption{The physical properties of the sample of MaNGA galaxies and background quasars}
\label{tab:sample}
\begin{tabular}{llllllllcll}
\hline
MaNGA & $z_{\rm gal}$   & $\log M_{\star}$ & SFR                        &  log sSFR            & $D_n(4000)$$^{\rm a}$ &  quasar & $z_{\rm quasar}$& Imp Par.   & $R_e$ &$b/R_e$\\
ID     &                 & $[M_{\odot}]$    & $M_{\odot}\,{\rm yr}^{-1}$ &  $[{\rm yr}^{-1}]$ & &    &               &  $(b)$ kpc & kpc &\\
\hline
1-71974             & 0.03316 & 10.30 & 1.72 & $-10.06$ & 1.30 & J0755$+$3911 & 0.0332 & 0   & 4.9 & 0\\
1-385099$^{\rm b}$  & 0.02866 & 10.69 & 0.20 & $-11.38$ & 1.54 & J0838$+$2453 & 0.0287 & 0   & 5.4 & 0 \\
12-192116           & 0.02615 & 8.80  & 0.59 & $-9.03$  & 1.22 &J1338$+$2620 & 0.0261 & 0   & 3.3 & 0 \\
1-594755            & 0.03493 & 10.78 & N/A  & N/A      & 1.23 & J1653$+$3945 & 0.0349 & 0   & 1.3 & 0 \\
%1-90242            & 0.03023 &       &       &         & J\,1535$+$5754 & 0.0302 & 0   & 0\\ 
1-575668            & 0.06018 & 11.02 & N/A  & N/A      & 1.87 & J1237$+$4447 & 0.4612 & 39  & 10.6& 3.8 \\
1-166736            & 0.01708 & 8.98  & 0.08 & $-10.07$ & 1.31 & J0950$+$4309 & 0.3622 & 23  & 3.4 & 6.9 \\
1-180522$^{\rm  c}$ & 0.02014 & 9.31  & 0.55 & $-9.56$  & 1.26 & J2130$-$0025 & 0.4901 & 34  & 4.1 & 8.5\\
1-561034            & 0.09008 & 10.86 & N/A  & N/A      & 1.86 & J1709$+$3421 & 0.3143 & 75  & 6.0 & 12.8 \\
1-585207$^{\rm b}$  & 0.02825 & 10.09 & N/A  & N/A      & 1.93 & J0838$+$2453 & 0.0287 & 37  & 2.4 & 15.5\\
1-113242            & 0.04372 & 10.88 & N/A  & N/A      & 2.14 & J2106$+$0909 & 0.3896 & 116 & 5.5 & 22.9 \\
1-44487$^{\rm d,e}$ & 0.03157 & 10.47 & 1.34 & $-10.34$ & 1.74 & J0758$+$4219 & 0.2111 & 136 & 6.2 & 22.6\\
{ 1-44487$^{\rm d,e}$} & 0.03174 & N/A   & N/A  & N/A      & N/A      & J0758$+$4219 & 0.2111 & 137 &   N/A  & N/A \\
1-564490            & 0.02588 & 10.15 & 2.60 & $-9.73$  & 1.27 & J1629$+$4007 & 0.2725 & 132 & 5.7 & 23.8\\
1-635629$^{\rm c}$  & 0.01989 & 9.56  & 1.26 & $-9.45$  & 1.34 & J2130$-$0025 & 0.4901 & 66  & 1.7 & 38.7\\
\hline
 \\
\end{tabular}
 \begin{tablenotes}
      \small
      \item $^{\rm a}$ The 4000 \AA\ break ($D_n(4000)$) was measured between $1<R/R_e<1.5$ to avoid the AGN contamination.
      \item $^{\rm b,  c,d},$ Quasar-galaxy pairs have the same quasar sight lines.
      \item $^{\rm e}$ The case of merged galaxies.
      \item N/A - not applicable, or not available. 
    \end{tablenotes}
\end{center}
\end{table*}

\subsection{HST/COS Data}
\label{sec:hstdata}

Our sample of 11 quasars was observed with HST COS under program ID\,16242 (PI V. Kulkarni) during September-December 2020. These observations are summarized in Table\,\ref{tab:hst_log}. The FUV channel of COS was used in 
TIME-TAG mode. The G130M FUV grating and the 2.5'' Primary Science Aperture (PSA) were used. The grating was  centered at 1222\,\AA\, and 1291\,\AA\ to cover the absorption lines of interest. %($1067-1363$\,\AA\, and $1134-1431$\,\AA, respectively).
This leads to a resolving power across the dispersion axis of $R\sim10,000-18,000$.  
%depending on the grating offset position (FP-POS = 1,2,3,4). 
The grating settings were optimized so that the key lines do not fall in the gaps in the wavelength coverage or in geocoronal emission lines. 
Target acquisitions were performed using the ACQ/IMAGE modes, after  which 3 to 11 exposures ranging from 515 s to 1339 s each were obtained for each target. 
%Each visit started from the target acquisition procedure (ACQ/SEARCH and ACQ/IMAGE), that guarantee that the telescope aperture is centered on the object and maximum incoming flux is received. %In three visits observations were failed and then these targets were re-observed in additional visits.  

\setlength{\tabcolsep}{2pt}
\begin{table*}
\begin{center}
\caption{HST COS observing log}
\label{tab:hst_log}
\begin{tabular}{llccllll}
\hline
Quasar & $z_{\rm quasar}$ & RA  & DEC  & Date & COS Setting & $T_{\rm exp}^{\rm a}$ & SNR$^{\rm b}$ \\
 &  & (J2000.0) & (J2000.0) &  &  & (s) &  \\
\hline
J0755$+$0311 & 0.0332  & 118.85 & 39.18   & 2020 Sep 11 & G130M/1291 & 2192 & 13.5 \\
J0758$+$4219 & 0.2111  & 119.58 & 42.32   & 2020 Sep 06 & G130M/1222 & 2147 & 5.3 \\
J0838$+$2453 & 0.0287  & 129.55 & 24.89   & 2020 Oct 29 & G130M/1222 & 7436 & 7.4 \\
J0950$+$4309 & 0.3622  & 147.56 & 43.15   & 2020 Nov 27 & G130M/1222 & 15254& 9.4 \\
J1237$+$4447 & 0.4612  & 189.39 & 44.79   & 2020 Dec 16 & G130M/1222 & 4986 & 9.9 \\
J1338$+$2620 & 0.0261  & 204.51 & 26.34   & 2020 Dec 17 & G130M/1222 & 2062 & 4.5\\
J1629$+$4007 & 0.2725  & 247.26 & 40.13    & 2020 Sep 02 & G130M/1222 & 7631 & 10.4 \\
J1653$+$3945 & 0.0349  & 253.47 & 39.76   & 2020 Oct 08 & G130M/1222 & 2086 & 13.7 \\
J1709$+$3421 & 0.3143  & 257.49 & 34.36   & 2020 Sep 03 & G130M/1291 & 9592 & 12.0\\
J2106$+$0909 & 0.3896  & 316.71 & 9.16    & 2020 Oct 10 & G130M/1222 & 7376 & 4.5 \\
J2130$-$0025 & 0.4901  & 322.59 & $-0.43$ & 2020 Sep 12 & G130M/1222 & 19310& 7.8  \\
\hline
\end{tabular}
 \begin{tablenotes}
      \small
      \item  $^{\rm a}$ The total integration time (summed over all exposures).
      \item $^{\rm b}$ The signal to noise ratio (SNR) was calculated at 6-pixel binning at $\simeq1250$\AA.
      %\item Additional observations were performed for J\,0950$+$4309 on 2020 Dec 22, J\,1709$+$3421 on 2020 Sep 07,  J\,2130$-$0025 on 2020 Sep 13 and 2020 Oct 11.
    \end{tablenotes}

\end{center}
\end{table*} 

For a majority of the targeted galaxies, we clearly detect strong absorption lines of H\,{\sc i}, Si\,{\sc ii} and Si\,{\sc iii} at redshifts close to the galaxy redshifts (within $\pm200\,{\rm km s^{-1}}$). Since there are no other galaxies at these redshifts around the quasar sight lines (within $|z_{\rm photom}-z_{\rm gal}|<0.05$ and  $\sim100$\,kpc and down to SDSS magnitude $r\simeq22$), our HST COS spectra probe the CGM of the selected MaNGA galaxies.\footnote{In the case of the quasar-galaxy pair  J1629$+$4007-1-564490, there is another galaxy (SDSS J162842.25+400726.1)  which is closer to the quasar sightline than the targeted MaNGA galaxy, but it has a higher redshift $z=0.03357$ versus $z=0.02588$. For this quasar sightline we detect a weak \HI\ absorption  at the redshift $z=0.033$, associated with the SDSS J162842.25+400726.1  galaxy, and do not detect any absorption within $\pm$1000\,km s$^{-1}$ at the redshift of MaNGA galaxy 1-564490.}  The profile fits to the HST COS absorption line data and the measurements of column densities are presented in detail in the Appendix\,\ref{app:B}.

\subsubsection{Data Reduction and Spectral Extraction}

The original \calcos pipeline v3.4.0 was first used to reduce the HST COS exposures and extract the one-dimensional spectra. However, a reanalysis of the data was found to be necessary, because some of our exposures had low counts ($N_{\rm counts}\simeq1-10/{\rm pix}$). For these exposures, the flux uncertainties in individual exposures were found to be overestimated using the original pipeline. The flux variance was $\sim$2 times lower than the flux uncertainties estimated by the standard pipeline, and the difference was found to be correlated with the flux value. The procedure for estimating the flux uncertainty in the original pipeline was therefore modified in our reanalysis of the data.

This problem was described in the ISR COS 2021-03 \citep{COS-2021-03-v1}, where it is shown that the number of received counts is described by a binomial distribution with an asymmetric shape at low count levels ($N\le10$). The \calcos pipeline uses the method developed by \cite{Gehrels1986} to estimate flux uncertainties in this case. However, we found that the 1-$\sigma$ uncertainties derived by \cite{Gehrels1986} corresponds to 63.8\% quantile interval, which is shifted to positive values relative to the uncertainties derived with the maximum likelihood estimate. This shift slightly reduces the negative uncertainty and increases the positive uncertainty. In spectra corrected for this shift, the 1-$\sigma$ uncertainties correspond well to the flux dispersion values. Therefore, flux errors were reevaluated based on the modified estimates. Further details are provided in Appendix\,\ref{app:A}.  

The procedure for the subtraction of the noise background also does not work well in a low-counts regime, and was therefore also modified. Originally, for each exposure the average background flux was calculated from the detector area free from the science target and wavelength calibration lamp signals. This average background flux was subtracted from the science spectrum in each exposure. However, in cases of low flux levels, %exposure time
the number of noise counts is also %catastrophically 
very low (e.g., 1-2 counts per 10 pixels), therefore the average background flux corresponded to a fractional number of counts (about $\sim0.2$ counts/pixel) and its subtraction shifted the zero-flux level to negative values, which was also observed in the final spectrum of the coadded exposures in the cores of saturated absorption lines. Therefore, instead of using this method for background subtraction, we used the following approach: we derived the background flux from the final spectrum of coadded ``background{"} exposures, which were extracted by the same method as the method used for the extraction of the science exposures, but the method was applied to a shifted extraction box in the detector area free from the science target and wavelength calibration lamp signals (and with a minimum content of bad quality pixels, including the gain sag hole and poorly calibrated regions). The ``background{"} exposures were next coadded, rebinned, smoothed by 10 pixels and then subtracted from the science exposure. This approach allows for a more accurate estimate of the average background flux level (since it gives 
a number of noise counts per bin strongly exceeding 1) and enables the calculation of the wavelength-dependence of the background.  
%and make it dependent on the wavelength.    

%The number of background counts is mostly 0 with the average about  $\sim0.2$ counts/pixel, that makes difficult to derive which counts correspond to noise or object signal inside the extraction box. The standard pipeline subtracts the average background flux for each exposure. However the background level is also quite uncertain and the subtraction of fraction number of counts shifts the zero flux level to a negative value, that is observed in the final spectrum ({\sc x1d.fits}) in cores of saturated absorption lines. 

%To estimate the background flux for each exposure, we repeated this procedure for an extraction box of the same size, shifted to a detector area free from both the object  wave-calibration lamp signals with a minimum content of bad quality pixels (including gain sag hole and poorly calibrated regions). After coaddition and binning of individual exposures background flux were then smoothed by 10 pixels and  subtracted from science object 1D spectrum. This method allows a more accurately estimate of the background flux, which fills the wavelength range more evenly than with individual exposure.

\begin{figure*}
\begin{center}
        \includegraphics[width=1\textwidth]{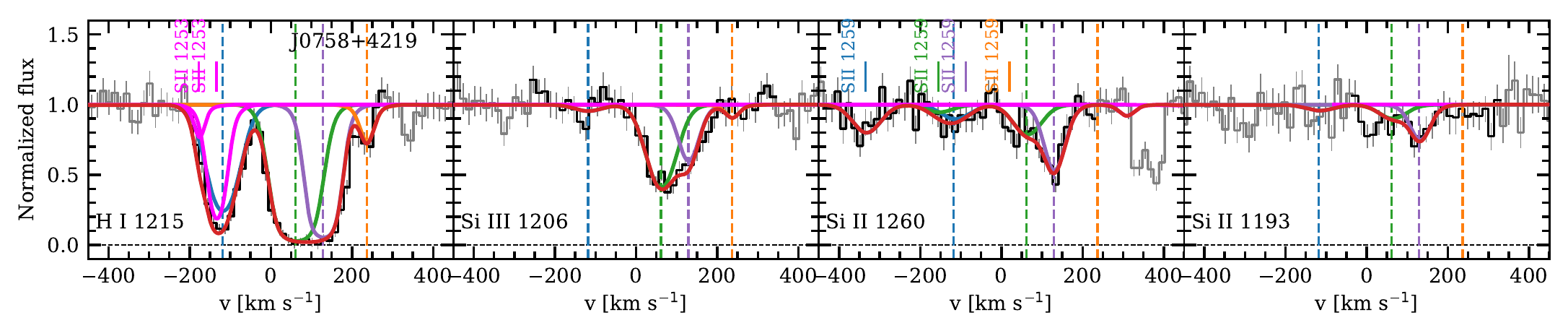}
        \includegraphics[width=1\textwidth]{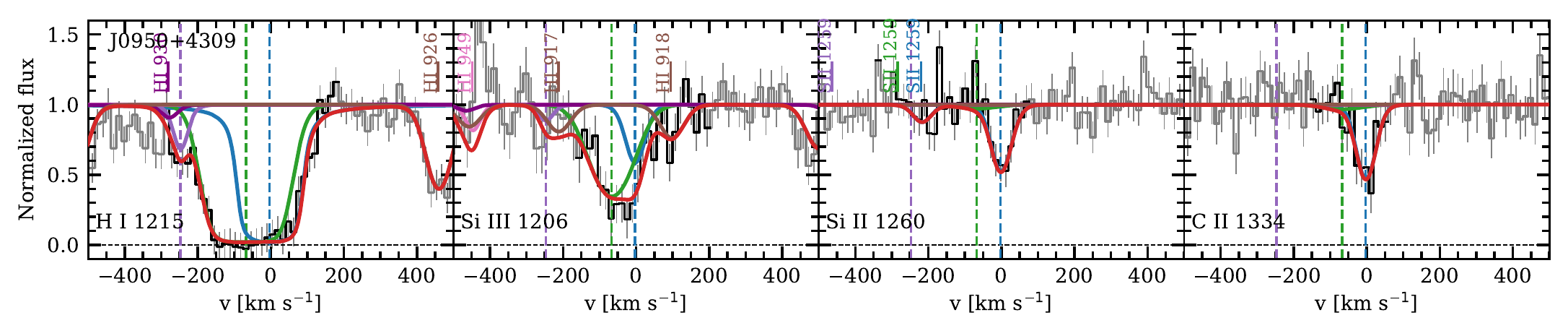}
        \includegraphics[width=1.0\textwidth]{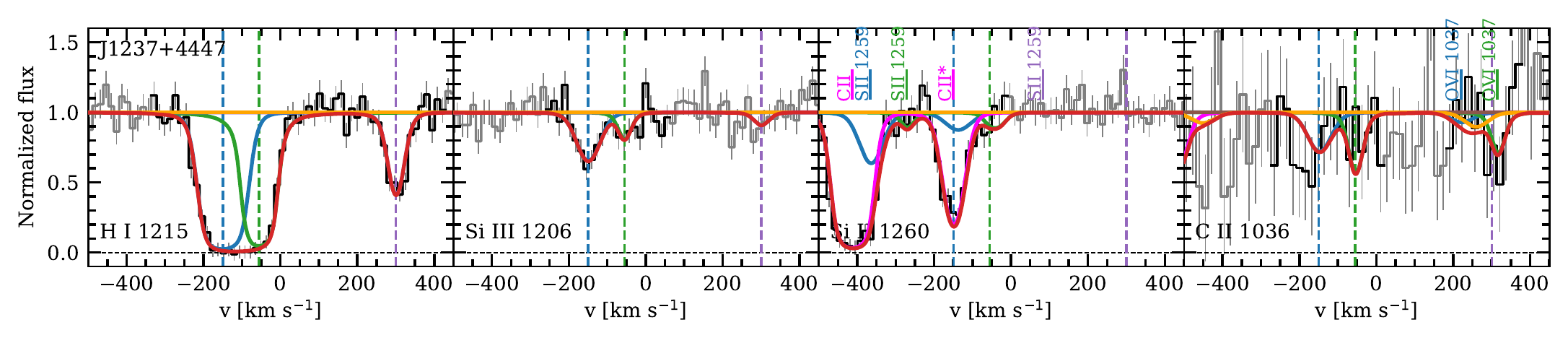}
        \includegraphics[width=1\textwidth]{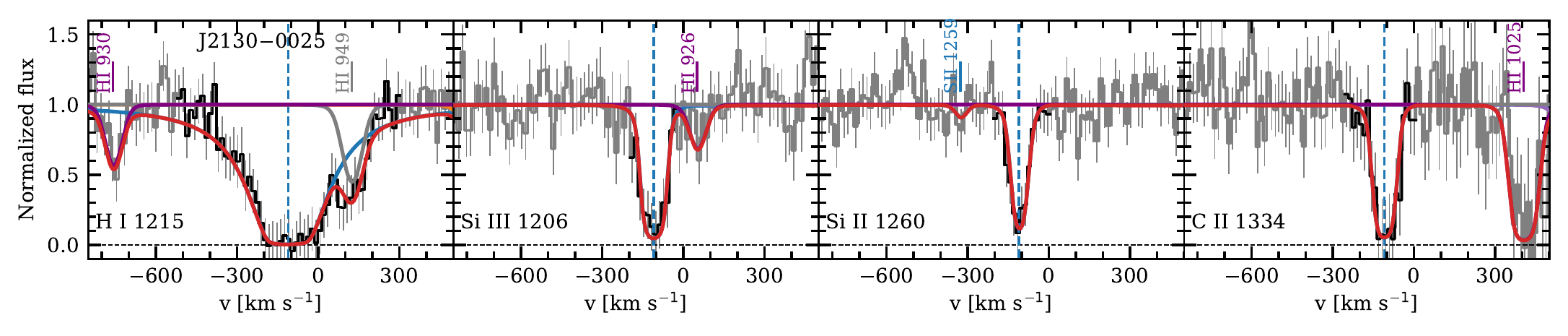}
        \caption{The absorption lines of H\,{\sc i}, Si\,{\sc iii}, Si\,{\sc ii}, C\,{\sc ii} in the HST COS spectra of J0758$+$4219, J0950$+$4309, J1237$+$4447 and J2130$-$0025 at the redshifts of the corresponding galaxies (1-44487, 1-166736, 1-575668 and 1-180522, respectively). The synthetic profile is shown in red and the contribution from each component, associated with the studied galaxy, is shown in green, blue, purple and orange. Dashed vertical lines represent the position of each component. Vertical ticks indicate the position of absorption lines, associated with the Milky Way (MW, magenta sticks) and remote galaxies.
        }
        \label{example:hst-data}
\end{center}
\end{figure*}

\subsubsection{Spectral Analysis}
\label{sec:spec_analysis}
For each quasar/AGN, we analyzed the absorption systems at the redshifts of associated MaNGA galaxies. Fig. \,\ref{example:hst-data} shows examples of our analysis for the four galaxies shown in Fig. \,\ref{fig:quasar-gal-pairs}. Fits for the remaining sight lines are shown in Appendix\,\ref{app:B}. To perform this analysis, we used a custom modification of the Voigt profile fitting code\footnote{https://github.com/balashev/spectro} to derive the redshifts, column densities and Doppler parameters of velocity components for H\,{\sc i}, N\,{\sc i}, O\,{\sc i}, Ar\,{\sc i} and various low-ionization  (Si\,{\sc ii}, S\,{\sc ii},  C\,{\sc ii},  N\,{\sc ii}, Fe\,{\sc ii}) and high-ionization (Si\,{\sc iii}, Fe\,{\sc iii}, N\,{\sc v}, O\,{\sc vi}) metal absorption lines. The wavelengths and oscillator strengths for these transition were taken from \cite{Morton2003}. Further details about this code and examples of its usage can be found in \cite{Balashev2017,Balashev2019}. 

For each absorption line (except {a case of damped} H\,{\sc i} Ly$\alpha$ described below) we derived the local continuum using a B-spline interpolation matched on the adjacent unabsorbed spectral regions. The spectral pixels used to constrain the fit were selected visually. The number of velocity components was also defined visually and increased in case of remaining structure in the residuals. {Since our spectra have low signal-to-noise ratio ($\sim$1-10) and given medium spectral resolution ($\sim20\,{\rm km\,s^{-1}}$), we can not resolve the velocity structure in detail, therefore the redshifts and Doppler parameters were tied for all species for each velocity component.} The fit to each absorption lines was calculated by the convolution of the synthetic spectrum with the COS line spread function (LSF) chosen for the appropriate COS setting.\footnote{The COS LSF has broad non-Gaussian wings and the shape varies with the wavelength. We used the approximation by the piece-wise function taken from the COS documentation, see e.g. https://www.stsci.edu/hst/instrumentation/cos/performance/spectral-resolution}  
%Parameter values and their uncertainties were estimated using $\chi^2$ likelihood function, which assumes the probability distribution function of uncertainties in the spectrum to be Gaussian. The shape of the likelihood function is defined by the Monte Carlo Markov chain approach with implementation of the affine-invariant ensemble sampler \citep{Goodman2010}. Such a technique ensures that we confidently find the global maximum in the many parametric space and provides reliable estimates of statistical errors on the parameters. Furthermore, the correlation between different parameters can be easily traced by this method. All the errors quoted here are given using 63.8\% quantile interval, which formally corresponds to $1\sigma$ interval for normal distribution. In each figure presenting a fit to absorption lines (except Figs ..), we grey out pixels of spectrum not used to constrain the fit and show the residuals (data minus model, divided by uncertainty).
 %The velocity structure was derived mainly by fitting strong absorption lines (e.g. H\,{\sc i},  Si\,{\sc iii}, Si\,{\sc ii}, C\,{\sc ii}). 
For weak lines, we present measurements of column densities where possible,  and 3-$\sigma$ upper limits in cases of non-detections.

The H\,{\sc i} column density was measured from the Voigt profile fit to the Ly$\alpha$ line\footnote{The Ly$\beta$ \HI\ line is covered by HST spectra for only one galaxy, 1-575668, which has a higher redshift $z_{\rm gal}\simeq0.06$ than others.}. {For most of our spectra the \HI\ line is not damped (with $N({\rm HI})\le10^{18}\,{\rm cm^{-2}}$) and corresponds to the linear or flat parts of the curve of growth. In these cases, the number of components and $b$-values should be accurately constrained. Therefore the number of components was defined visually from fitting to the profiles of associated metal lines (\SiII, \SiIII, \CII), and increased in case of remaining structure in the residuals. 
The range of $b$-parameters was constrained to $15-100\,{\rm km\,s^{-1}}$ (the values of $b$-parameters measured for \HI\ absorbers at $z\le1$ in the COS CGM Compendium  \citep{Lehner2018}, where \HI\ lines were fitted using several transitions in the Lyman series). 
Then the redshift, $b$-parameter and \HI\ and metal column densities for each component were varied together using the AffineInvariantMCMC sampler by MADS\footnote{http://madsjulia.github.io/Mads.jl/} to obtain the posterior probability density function (PDF) of the fitting parameters. The column density of \HI\ and metals and the $b$-parameters can be rather uncertain for individual components that are blended, however the total column densities summed over the components are usually well constrained (with uncertainty $\sim0.3$ dex). We demonstrate the posterior PDF of fitting parameters for systems shown in Fig.\,\ref{example:hst-data} in Appendix\,\ref{app:B} (see Fig.\,\ref{fig:PDF}), along with comments to fits for individual systems.}

%\st{For most of our spectra the H\,{\sc i} line is usually weak and  not damped (with $N({\rm HI})\le10^{17}\,{\rm cm^{-2}}$), therefore we usually fitted only the Ly$\alpha$ line profile, while the local continuum was derived from a B-spline interpolation.} 

For one case, {the galaxy} J1338$+$2620, the {absorption} Ly$\alpha$ line %\st{is saturated} 
{shows damping wings} and is located close to the galaxy Ly$\alpha$ emission line. {An interesting  feature of this spectrum is that both the emission Ly$\alpha$ line and absorption Ly$\alpha$ line are shifted relative to their expected positions. The emission Ly$\alpha$ line is redshifted by $150\,{\rm km\,s^{-1}}$ with respect to the the galaxy redshift derived by positions of other emission lines (H$\alpha$, H$\beta$, S\,{\sc ii}, N\,{\sc ii}, O\,{\sc iii}) seen in the SDSS spectrum. The emission lines are very narrow $\sim300\,{\rm km\,s^{-1}}$ (similar to those for type II Seyfert galaxies), which allows the redshift to be well-constrained. The Ly$\alpha$ absorption line has a broad core ($\sim 300\,{\rm km\,s^{-1}}$ wide) and its center is blue-shifted by $-150\,{\rm km\,s^{-1}}$ with respect to the strongest component of the metal absorption lines (\SiII, \SII, \OI). Additionally we detect the decrease of the local continuum near the Ly$\alpha$ absorption and Ly$\alpha$ emission lines, which is consistent with the presence of the damped Ly$\alpha$ (DLA) absorption line with broad damping wings and high \HI\ column density ($10^{20.3}\,{\rm cm^{-2}}$).  However, such a \HI\ Ly$\alpha$ line is expected to have a broad bottom $\sim600\,{\rm km\,s^{-1}}$, twice the observed value. We believe this situation is similar to studies of proximate DLA absorption systems in quasar spectra, which work as a natural coronagraph for the Ly$\alpha$ emission from the accretion disk, while the leaking Ly$\alpha$ emission remains partially blended in the wings of the DLA system \citep[see e.g.][]{Noterdaeme2021}.} In this case we simultaneously fitted both the absorption profile and the unabsorbed quasar continuum. 

\begin{figure*}
\begin{center}
        \includegraphics[width=1\textwidth]{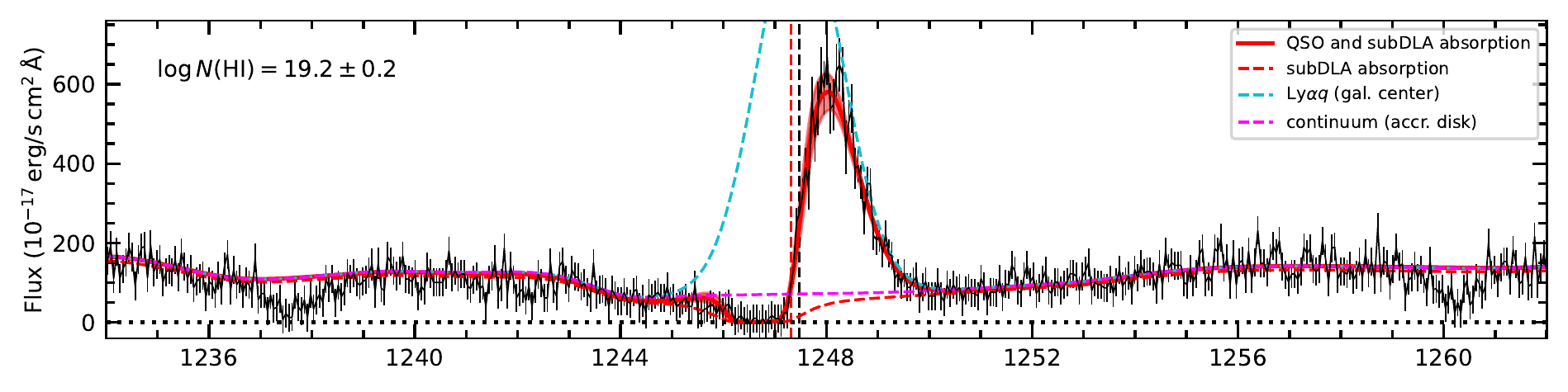}
        \includegraphics[width=1\textwidth]{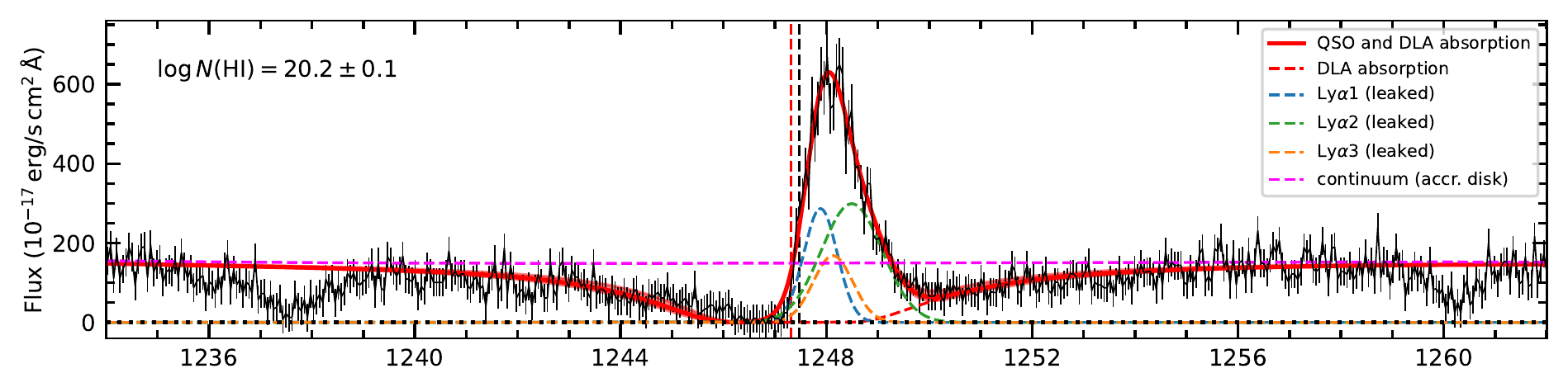}
        \includegraphics[width=1\textwidth]{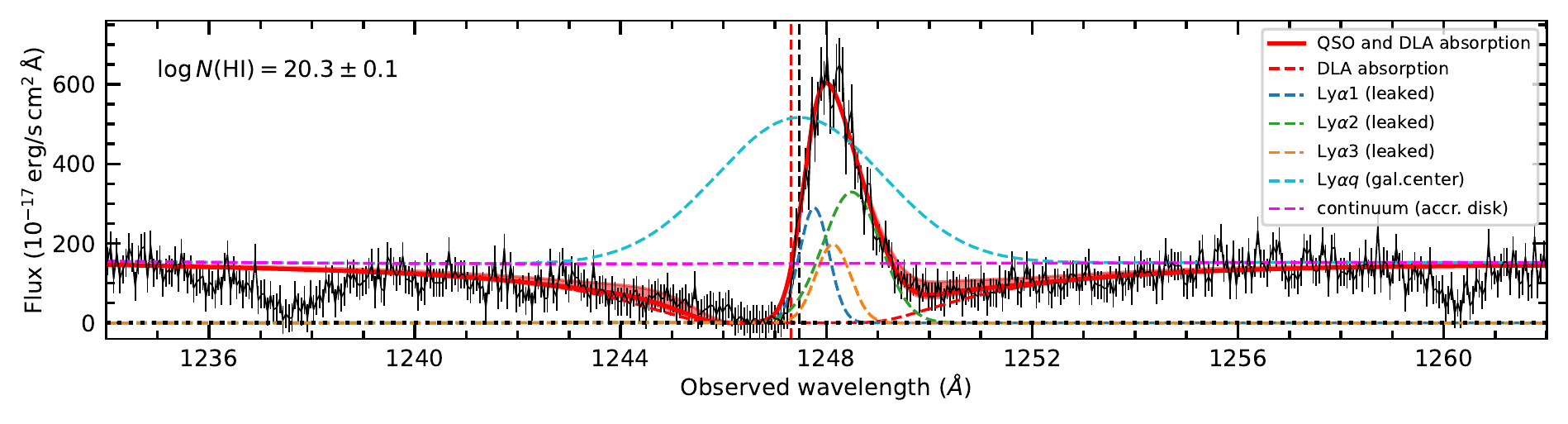}
        \caption{\rm  Fit to the H\,{\sc i} Ly$\alpha$ absorption line in the spectrum of galaxy J1338$+$2620. Top, middle and bottom panels shows different solutions: sub-DLA + galactic Ly$\alpha$ line, DLA + leaked Ly$\alpha$ emission, DLA + leaked Ly$\alpha$ emission + galactic Ly$\alpha$ line, respectively (see details in the text). The black line represents the observed HST/COS spectrum, the red line shows the best fit. The red shaded area represents a variation of the synthetic fit due to the variation of H\,{\sc i} column density within the derived uncertainty. The profile of the absorption Ly$\alpha$ line is shown by the red dashed curve. The smooth part of the reconstructed continuum is shown by the dashed pink curve. The cyan dashed curves in the top and bottom panels represent the reconstructed emission Ly$\alpha$ lines from the galactic center. The blue, orange and green dashed lines in middle and bottom panels show the components used to fit the leaked Ly$\alpha$ emission. The red and black vertical lines denote the redshift of the strongest metal absorption component and the redshift of quasar, respectively. The derived total H\,{\sc i}  column density is given in the top left corner of each panel.}
        \label{fig:HILyaline}
\end{center}
\end{figure*}

{ We consider two potential ways to fit the Ly$\alpha$ line in this spectrum: (i) it could be a sub-DLA system which covers the Ly$\alpha$ emission line only partially. In this case} the { local} continuum was modeled as the sum of a smooth component and a Gaussian emission line. The smooth component represents the flat part of the quasar continuum and was reconstructed locally by fitting with a B-spline interpolation. %{ In this case we  decrease the continuum near Ly$\alpha$ line manually at $\Delta v>1000\,{\rm km \,s^{-1}}$.}  
The Ly$\alpha$ emission line was fitted by a Gaussian function centered on the redshift of the quasar. 
%\st{We also added a linear component to take into account a possible curvature of the continuum near the Ly$\alpha$ emission line. The \HI\ Ly$\alpha$ profile together with the estimated quasar continuum for J\,1338$+$2620 are shown in Fig.} 
{ The \HI\ Ly$\alpha$ absorption line was fitted by the sum of four velocity components, whose  redshifts were tied to the redshifts of the velocity components in metal absorption lines (\OI, \SiII, \SII, Fe\,{\sc ii}, \SiIII). The detailed fit to metal lines is provided in Appendix\,\ref{app:B}. In this case, we derive the total \HI\ column density ($10^{19.2\pm0.1}\,{\rm cm^{-2}}$), and the best fit is shown in the top panel of Fig.\,\ref{fig:HILyaline}. We note, however, that (i) we needed to decrease the continuum level manually in the vicinity of the sub-DLA system and (ii) the strongest \HI\ component ($10^{19.2\pm0.1}$) is shifted to $-150{\,\rm km\,s^{-1}}$ relative to the strongest component in the metal absorption lines.}

{The second possibility is a combination of a broader, more damped Ly$\alpha$ absorption line and leaked Ly$\alpha$ emission. The damped Ly$\alpha$ line was centered at the redshift of the strongest metal component at $z=0.026043$. The profile of the leaked Ly$\alpha$ emission is non-Gaussian, therefore it was fitted by a sum of three Gaussian lines. In this case we derive the \HI\ column density $10^{20.2\pm0.1}\,{\rm cm^{-2}}$. The fit is shown in the middle panel of Fig.\,\ref{fig:HILyaline}. The absorption Ly$\alpha$ line is broad ($\sim600\,{\rm km\,s^{-1}}$) and likely cover the emission Ly$\alpha$ line from the galactic center completely. In the bottom panel we also show the fit by a model, where the galaxy Ly$\alpha$ emission line at the quasar redshift is added to the fit. However the difference in the fit profile and the derived \HI\ column density is small, compared to for the fit (middle panel) without including the galaxy Ly$\alpha$ emission  ($\sim0.1$ dex). }

{ The advantage of the fitting approach including the leaked Ly$\alpha$ emission is that (i) it can describe the decrease of the local continuum near the Ly$\alpha$ absorption  without manual  modification of the smooth B-spline fit and (ii) the redshift of \HI\ component matches the redshift of the strongest metal components well. Therefore we adopt the column density of \HI\ for this system to be $10^{20.2\pm0.1}\,{\rm cm^{-2}}$.}
%\st{The number of velocity components and their redshifts were determined from the fits to the associated metal lines (\SiIII, \SiII, \SII, Fe\,{\sc ii}).} 

\subsection{SDSS-IV/MaNGA Data}
\label{sec:mangadata} 

The Mapping Nearby Galaxies at APO (MaNGA; \citealt{Bundy2015}) survey is one of the three main components making up the Sloan Digital Sky Survey IV (SDSS-IV; \citealt{Blanton2017}). Completed in June 2020, MaNGA made integral field unit (IFU) spectroscopic observation of just over 10,000 galaxies using the 2.5m Sloan Telescope at Apache Point Observatory \citep{Gunn2006}. These galaxies were selected from the extended version of the NASA-Sloan Atlas  \citep[NSA;][]{Blanton2017,Wake2017} 
to be in the redshift range of $0.01 < z < 0.15$ and have an approximately flat number density distribution as a function of stellar mass between $10^9$ and $10^{12}$ \Msun. The targets were further chosen so that they could be covered by the MaNGA IFU bundles out to either a radius of 1.5 or 2.5 times the effective radius ($R_e$). Full details of the MaNGA sample selection are given in \cite{Wake2017}. 

The 17 MaNGA IFU bundles are hexagonal in shape with sizes ranging from 12{"} to 32{"} matched to the typical angular size distribution of the target sample of galaxies. In addition, there are 12 seven-fiber mini-bundles which are placed on flux calibration stars, and 92 single fibers for sky subtraction \citep{Drory2015}. All the fibers feed the dual-channel Baryon Oscillation Spectroscopic Survey (BOSS) spectrographs \citep{Smee2013}, which cover a wavelength range of 3,622\AA\ to 10,354\AA\ with a median spectral resolution of $\sim$2,000.

In this paper we use the reduced MaNGA data produced by the MaNGA Data Reduction Pipeline \citep[DRP;][]{Law2016}  as well as derived data products produced by the MaNGA Data Analysis Pipeline \citep[DAP;][]{DAP2019}. These derived products include maps of various emission lines (\oii, \hb, \oii, \nii, \ha, \sii), and emission line and stellar velocities. We access and interact with MaNGA data using the Marvin \citep{Cherinka2019} Python package.

We also make use of integrated galaxy properties included in the MaNGA dataset that are derived from the extended version of the NSA. These include redshift, total stellar mass ($M_{*}$), elliptical effective radius ($R_e$), and inclination, all derived from elliptical Petrosian aperture photometry \citep[see][for details]{Wake2017}.

\subsubsection{Nebular Metallicity and Ionization Parameter}
\label{sec:mangamet} 

In order to connect the properties of the CGM absorption systems detected in our COS spectra with the gas within the MaNGA galaxies, we derive maps of the metallicity and ionization parameter of emission lines originating in the nebulae photoionized by massive stars.   

To make these measurements, we use of the Bayesian strong emission line (SEL) fitting software IZI initially presented by \cite{Blanc2015} and extended to utilize MCMC, additionally fitting for extinction by \cite{Mingozzi2020}. IZI compares a grid of photoionization models with a set of SELs and their uncertainties, to derive the marginalized posterior probability density functions (PDFs) for the metallicity ($12 + \log({\rm O/H})$), the ionization parameter ($\log(q)$), and the line-of-sight extinction ($E(B-V)$). Such an approach takes into account the covariance between these parameters, which is not insignificant.

\begin{figure*}
\begin{center}
        \includegraphics[width=1.0\textwidth]{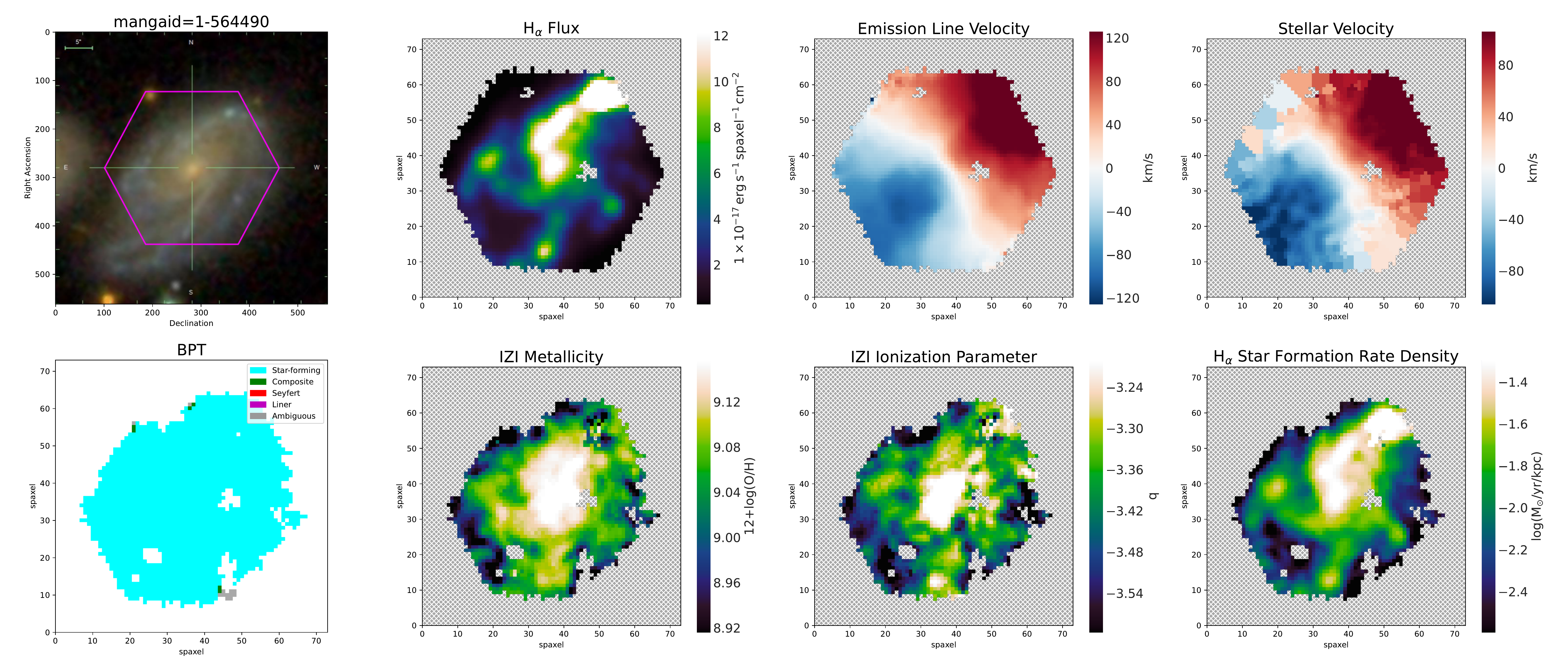}
        \caption{\rm The physical properties of the SDSS MaNGA galaxy 1-564490. Top panels show from left to right the SDSS three-color image, the H$_{\alpha}$ line emission map, the H$_{\alpha}$ line  velocity map, and the stellar velocity map. Bottom panels shows the BPT diagram and maps of IZI metallicity, IZI ionization parameter and the density of star formation rate.}
        \label{fig:manga_ex}
\end{center}
\end{figure*}

In this work we follow the approach of \cite{Mingozzi2020}, who ran IZI on a subset of the MaNGA sample. We use the photoionization model grids presented in \cite{Dopita2013} fitting for \oii$\lambda3726,3729$, \hb, \oii$\lambda4959,5007$, \nii$\lambda6548,6584$, \ha,
\sii$\lambda6717$, and \sii$\lambda6731$. We only fit spaxels that are classified as star-forming according to either the \nii- or \sii-Baldwin, Phillips\&Terlevich (BPT)
diagrams (using the regions defined by \cite{Kauffmann2003} and by \cite{Kewley2001}. We further restrict to spaxels with a \ha\, $S/N>15$, which ensures sufficient $S/N$ in the remaining SELs we use.
Fig.\,\ref{fig:manga_ex} presents the example of MaNGA observations of the galaxy 1-544490. It shows maps of H$_{alpha}$ flux, gas kinematics, SFR and the physical conditions derived from IZI modelling.

\subsubsection{Galaxy Rotational Velocities}
\label{sec:mangavel}

Beyond the ionization properties described above we are also interested to see if there is any association between the velocity of the CGM absorption systems and galaxy rotational velocity. One might imagine the absorption systems tracing the gas dynamics at large radii.

In order to make such a connection we fit disk rotation models to the stellar and gas velocity fields using models similar to those described in \cite{Bekiaris2016}. We assume a flat thin disc in all cases linking the observed coordinates ($x,y$) to the projected major and minor axes coordinates of the disc ($x_{\mathrm{e}}, y_{\mathrm{e}}$) using:

\begin{equation}
x_{\mathrm{e}} = - (x-x_{\mathrm{o}})\sin{PA} + (y-y_{\mathrm{o}})\cos{PA},
\end{equation}
\begin{equation}
y_{\mathrm{e}} = - (x-x_{\mathrm{o}})\cos{PA} - (y-y_{\mathrm{o}})\sin{PA},
\end{equation}

where PA is position angle and ($x_{\mathrm{0}}, y_{\mathrm{0}}$) are the coordinates of the center of the disk.

We define the radius of the disk r in the disk plane at the observed coordinates ($x,y$) as:
\begin{equation}
r = \sqrt{x_{\mathrm{e}}^2 + \left(\frac{y_{\mathrm{e}}}{\cos{i}}\right)^2 },
\end{equation}

where is the inclination of the disc to the line of sight.

The position angle relative to the major axis of the disc, $\theta$, at ($x,y$) is given by

\begin{equation}
\cos{\theta} = \frac{x_{\mathrm{e}}}{r}.
\end{equation}

To model the rotation curve we make use of a two-parameter $arctan$ profile \citep{Courteau1997}:

\begin{equation}
V_{\mathrm{rot}}(r) = \frac{2}{\pi}V_{\mathrm{t}}\arctan{\frac{r}{r_{\mathrm{t}}}},
\end{equation}

where $V_{\mathrm{rot}}(r)$ gives the rotation velocity at radius $r$, $r_{\mathrm{t}}$ is the turnover radius and $V_{\mathrm{t}}$ is the asymptotic circular velocity. At large radii beyond $r_{\mathrm{t}}$ this model represents a very slowly rising rotation curve.

Our final model for the velocity in the plane of the sky is given by

\begin{equation}
v_{\mathrm{model}}(x,y) = V_{\mathrm{sys}} + V_{\mathrm{rot}} (r)\sin{i}\cos{\theta}.
\end{equation}

where $V_{\mathrm{sys}}$ represents any velocity offset from the systemic redshift used to generate the MaNGA velocity field. This model contains seven free parameters that we must fit for.

For all galaxies we attempt to fit both the emission line and stellar velocity maps provided by the MaNGA DAP. We make use of the default MILESHC-MASTARSSP hybrid maps, which use a Voronoi binning scheme for the stellar velocities and individual spaxels for the emission line velocities \citep[see][for details]{DAP2019}. For the stellar velocity maps we fit to all Voronoi bins that have not been masked by the DAP and have a S/N $>$ 10. For the emission line maps we again exclude all masked spaxels and fit to those spaxels where any of the  \ha, \oii, or \oiii~ lines have a S/N $>$ 5.  We also mask any regions of the maps not associated with the target galaxy, for instance the very close satellite galaxy of 1-44487. 

We fit our model using the MCMC code emcee \citep{Foreman-Mackey2013}. We make an initial simpler fit to estimate the position angle and use that as our initial guess for that fit parameter. For the center, inclination, and $r_{\mathrm{t}}$ we make initial estimates based on the NSA photometry. For $V_{\mathrm{sys}}$ our initial estimate is the median velocity within 0.5 $R_e$. Finally, we set $V_{\mathrm{t}}$ to 200 km s$^{-1}$ as our initial guess for all galaxies. For each fit, use 64 walkers each with 20,000 steps, discarding the first 10,000. We fit both the emission and stellar velocity maps simultaneously and each independently, potentially giving three fits for each galaxy.

\section{HST/COS Fitting Results}
\label{sec:results}

We detect associated absorption for 11 out of the 14 MaNGA galaxies. H\,{\sc i} Ly$\alpha$ absorption is detected in all 11 of these cases, while \SiII\ and \SiIII\ are detected in 7 of the 11 cases. For two sight lines, each of which has two galaxies with closely spaced redshifts, we detect absorption in \HI, \SiII, \SiIII\ (and \CII\ in one case), but we can not reliably determine which galaxy corresponds to which velocity component in the detected absorption. In three cases, we do not detect any absorption (in H I or any of the metal ions) within the range of $\pm800\,{\rm km s^{-1}}$ relative to the galaxy  redshifts; in these cases, we set upper limits on $N({\rm HI})\sim10^{13}\,{\rm cm^{-2}}$. For two of these sight lines, J1709$+$3421 and J2106$+$0909, the absence of any absorption may be because of high values of { the impact parameters $75$ and $116\,{\rm kpc}$, respectively} (with $b/R_{e}$ of 12.8 and 22.9). The absence of any lines is more surprising in the third case J1653$+$3945, a  galaxy sight line with zero impact parameter, and may be a result of high ionization of the gas. We discuss ionization corrections in Section 3.1 below. 

Table\,\ref{tab:hst_results} summarizes the results of our fits. We present the absorption redshifts, total column densities of \HI\ and associated strongest metal ions (\SiII, \SiIII\ and \CII) and results of the photo-ionization code simulations. We refer to the sight lines with zero impact parameters as the ``galactic'' sight lines and list them in the first five lines of Table\,\ref{tab:hst_results} before the remaining  sight lines that we refer to as ``quasar sight lines''.

\setlength{\tabcolsep}{1pt}
\begin{table*}
\begin{center}
\caption{Neutral gas and metallicity measurements from HST observations.}
\label{tab:hst_results}
\begin{tabular}{lllccccccccccc}
\hline
& quasar & $z_{\rm abs}$  &  $\log N({\rm H I})$ & $\log N({\rm Si II})$& $\log N({\rm Si III})$& $\log N({\rm C II})$ & Si\,{\sc iii}/Si\,{\sc ii} & [X/H] & $\log q$   &  $\log N(H_{\rm tot})$ &  $\log f({\rm HI})$ & $F_{\star}^a$ \\
&    &                &  $[{\rm cm}^{-2}]$   & $[{\rm cm}^{-2}]$    & $[{\rm cm}^{-2}]$     & $[{\rm cm}^{-2}]$    &                            &        &            &  $[{\rm cm}^{-2}]$ & & \\
\hline
\parbox[t]{3mm}{\multirow{4}{*}{\rotatebox[origin=c]{90}{AGNs}}} & J0755$+$0311  & 0.0330 & $13.6^{+0.1}_{-0.1}$ & $12.7^{+0.3}_{-0.5}$ & $12.4^{+0.2}_{-0.5}$ & $13.0^{+0.2}_{-0.3}$ & $-0.3^{+0.5}_{-0.6}$  & $1.2^{+0.2}_{-0.2}$ & $-1.5^{+0.3}_{-0.3}$ & $16.5^{+0.3}_{-0.3}$ & $-2.8^{+0.3}_{-0.3}$ & $0.0^{+0.4}_{-0.0}$\\
& J0838$+$2453A & 0.0280 & $13.2^{+0.1}_{-0.1}$ & $13.4^{+0.2}_{-0.2}$ & $13.8^{+0.6}_{-0.5}$ & N/C                  & $0.4^{+0.6}_{-0.5}$ & $1.9^{+0.3}_{-0.3}$ & $-0.8^{+0.2}_{-0.5}$ & $16.7^{+0.3}_{-0.3}$ & $-3.5^{+0.3}_{-0.3}$ & $0.0^{+0.1}_{-0.0}$\\
& J0838$+$2453B & 0.0256 & $14.0^{+0.1}_{-0.1}$  & $<13.6$              & $13.6^{+0.5}_{-1.0}$              & N/C                  & N/A          & $0.5^{+1.5}_{-0.5}$     & $-1.1^{+0.5}_{-0.7}$ & $17.1^{+1.5}_{-0.3}$ &  $-3.1^{+0.3}_{-1.5}$ & $<1$\\
& J1338$+$2620  & 0.0260 & $20.2^{+0.1}_{-0.1}$ & $13.9^{+0.2}_{-0.2}$ & $13.9^{+0.4}_{-0.2}$ & N/C                  & $0.0^{+0.5}_{-0.3}$ & $-0.4^{+0.4}_{-0.1}$ & $-2.8^{+0.3}_{-0.2}$ & $20.5^{+0.2}_{-0.2}$ &  $-0.4^{+0.4}_{-0.3}$ & $0.8^{+0.2}_{-0.2}$ \\
& J1653$+$3945  & 0.0341 & $<12.8$ & $<12.7$ & $<12.0$ &  N/C &  N/A & N/A & N/A & N/A & N/A & N/A \\
\hline
\parbox[t]{3mm}{\multirow{8}{*}{\rotatebox[origin=c]{90}{Quasars}}} &
{ J1237$+$4447A}  & 0.0597 & $17.2^{+0.3}_{-0.4}$ & $<13$ & $12.9^{+0.1}_{-0.1}$ & $<14.7$ & N/A & $-1.8^{+0.8}_{-0.8}$  & $-2.9^{+0.9}_{-0.6}$ & $19.4^{+0.8}_{-0.5}$ & $-2.2^{+0.6}_{-1.1}$ & $0.0^{+0.2}_{-0.0}$\\
& { J1237$+$4447B}  & 0.0597 & $15.2^{+0.5}_{-0.3}$ & $<13$ & $12.9^{+0.1}_{-0.1}$ & $14.0^{+0.3}_{-1.2}$ & N/A   & $-0.2^{+0.5}_{-1.0}$ & $-2.5^{+0.6}_{-0.4}$ & $18.1^{+0.9}_{-0.5}$ & $-2.9^{+0.7}_{-0.9}$ & $0.0^{+0.2}_{-0.0}$\\
&J0950$+$4309  & 0.0170 & $17.6^{+0.3}_{-0.7}$ & $13.0^{+0.1}_{-0.1}$ & $13.6^{+0.1}_{-0.1}$ & $14.1^{+0.2}_{-0.2}$ & $0.6^{+0.2}_{-0.2}$ & $-0.6^{+0.2}_{-0.7}$ &  $-3.2^{+0.2}_{-0.2}$ & $19.0^{+0.6}_{-0.2}$ & $-1.4^{+0.6}_{-0.6}$ & $0.0^{+0.2}_{-0.0}$\\
& J2130$-$0025  & 0.0195 & $18.8^{+0.1}_{-0.1}$  & $13.8^{+0.1}_{-0.2}$ & $14.6^{+0.9}_{-0.5}$ & $15.3^{+0.9}_{-0.5}$ & $0.8^{+0.5}_{-0.9}$  & $-1.1^{+0.2}_{-0.2}$  & $-2.1^{+0.4}_{-0.5}$ & $21.1^{+0.4}_{-0.6}$ & $-2.3^{+0.6}_{-0.4}$ & $0.1^{+0.3}_{-0.1}$\\
& J1709$+$3421  & 0.0880 & $<13.5$              &  $<12.5$             & $<13.0$              &  N/C                 & N/A            & N/A                  & N/A                 & N/A   & N/A  & N/A\\
& J2106$+$0909  & 0.0442 & $13.7^{+0.2}_{-0.2}$ & $<13.0$              & $<13.2$              & N/C                  & N/A             & N/A                 & N/A                 & N/A   & N/A & N/A\\
& J0758$+$4219  & 0.0320 & $15.3^{+0.3}_{-0.2}$ & $13.2^{+0.1}_{-0.1}$ & $13.4^{+0.1}_{-0.1}$ & N/C                   & $0.2^{+0.1}_{-0.1}$         & $0.8^{+0.3}_{-0.3}$   & $-2.0^{+0.3}_{-0.4}$ & $18.1^{+0.3}_{-0.4}$ & $-2.8^{+0.2}_{-0.3}$ & $0.1^{+0.5}_{-0.1}$\\
& J1629$+$4007  & 0.0240 & $<13.2$              &  $<12.5$             & $<12.6$              &  N/C                 & N/A            & N/A                 & N/A                  & N/A                  & N/A & N/A \\
\hline
\end{tabular}
\begin{tablenotes}
      {{ Note.} The first five rows (above the horizontal line) correspond to { AGN} %\st{galaxies} 
      sight lines with zero impact parameter, and the rest (below the horizontal line) correspond to quasar sight lines with non-zero impact parameter.  \\
      { Indices A,B in the first column indicate two distinct absorption systems with a high velocity separation ($\sim800$ ${\rm km\,s^{-1}}$) in the spectrum of the AGN J0838$+$2453 and two different solutions for the fit to the \HI\ profile in the spectrum of J1237$+$4447 (see detail in Appendix\,\ref{app:B}).}
      \small
        \item $^a$ $F_*$ - The parameter of the model of dust depletion  by \cite{Jenkins2009}.
        \item N/C - The lines are not covered by the HST/COS spectrum.
        \item N/A - In this case, physical parameters can not be constrained by the photo-ionization  model.
        }
\end{tablenotes}
\end{center}
\end{table*}

The \HI\ column density ranges from $\sim10^{13}\,{\rm cm^{-2}}$ to $\sim10^{20.2}\,{\rm cm^{-2}}$ for the { AGN} %\st{galactic} 
sight lines with \HI\ detections and from $\sim10^{14}\,{\rm cm^{-2}}$ to $\sim10^{19}\,{\rm cm^{-2}}$ for the quasar sight lines with \HI\ detections (i.e. with $N({\rm HI})>10^{13}\,{\rm cm^{-2}}$). 
%\st{The range probed by most of the quasar sight lines corresponds to the Lyman limit systems (LLSs) and sub-damped Lyman-$\alpha$ systems (sub-DLAs).}
For most of the systems  we detect associated absorptions of low (\SiII, \CII, N\,{\sc ii})  and high (\SiIII) ionization ions and also set upper limits on weak  absorption by N\,{\sc i}, N\,{\sc v}, O\,{\sc i}, and Fe\,{\sc ii}.  The detailed fit for each system is shown in Appendix\,\ref{app:B} and the fit results to individual velocity components are presented in Table\,\ref{tab:detailedfit}.

\begin{figure*}
\begin{center}
        \includegraphics[width=0.8\textwidth]{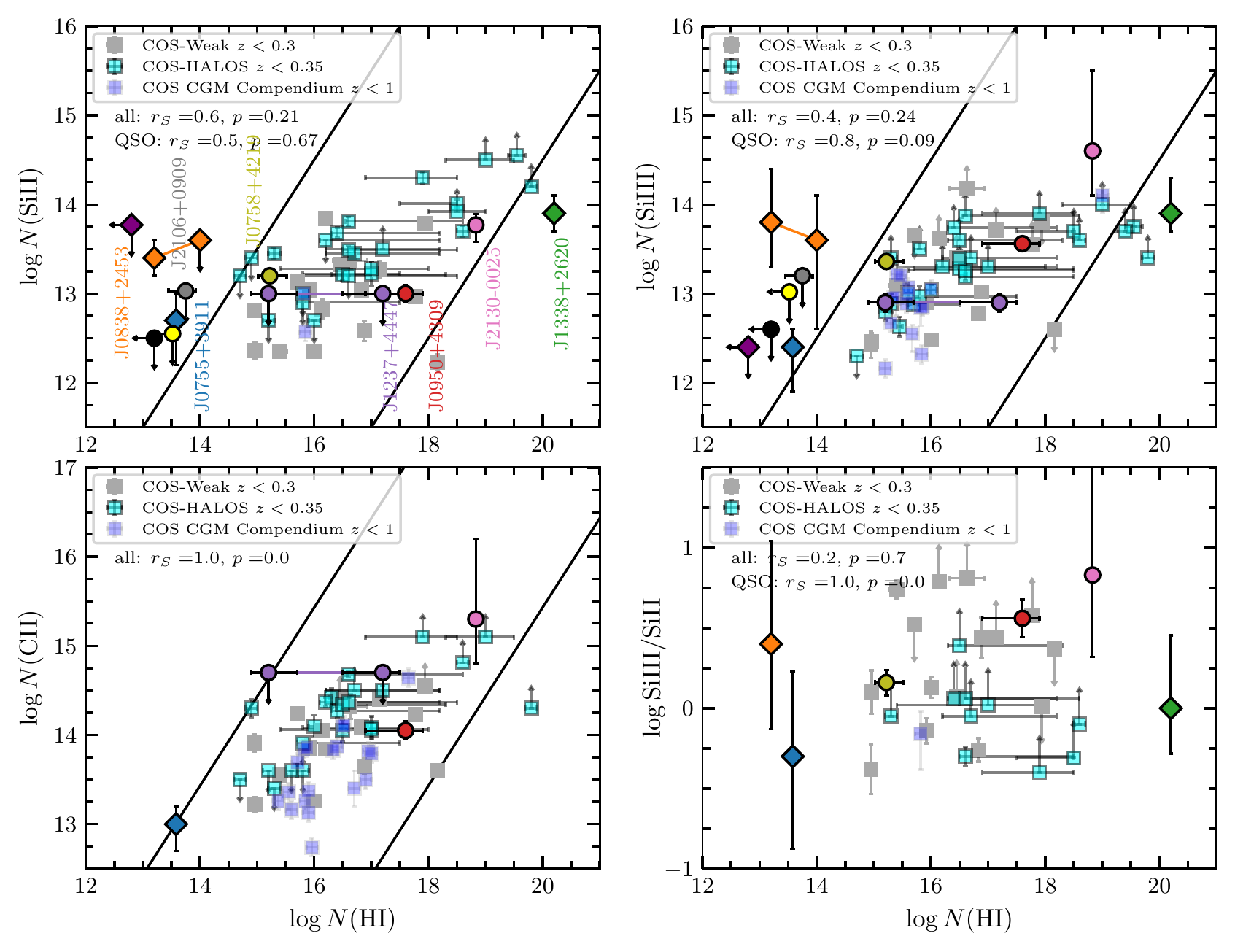}
        \caption{\rm The comparison of {\rm total} column densities of Si\,{\sc ii}, Si\,{\sc iii},  C\,{\sc ii} and the  Si\,{\sc iii}/ Si\,{\sc ii} ratio with $N({\rm HI})$. { Our systems are shown by diamonds (AGN  sight lines) and circles (quasar sight lines).} The color of points encodes the name of the systems in our sample. We also show upper limits for three non-detections: J1709$+$3421 (yellow circle), J1629$+$4007 (black circle), J1653$+$3945 (purple diamond).  Grey, blue and cyan squares represent data from different COS surveys: COS-Weak \citep{Muzahid2018}, COS CGM Compendium \citep{Lehner2018} and COS-Halos  \citep{Tumlinson2013, Werk2013} respectively. Spearman rank order correlation coefﬁcient $r_s$ and the $p$-value for our sample { (all systems and only QSO absorptions)} are given at the left top corner of each panel. { Black lines indicate the range of theoretical constraints on the column densities of melal ions and H\,{\sc i} for the parameters ($Z$, $f_{\rm HI}$ $f_{\rm X}$) typical for the ISM/CGM (see text)}. }
        \label{fig:sp_vs_hi}
\end{center}
\end{figure*}

In Fig.\,\ref{fig:sp_vs_hi}, we examine the dependence of the column densities of the strongest metal ions detected in absorption (\SiII, \SiIII\ and \CII) on the \HI\ column density. 
There is an { overall} increase of  \SiII, \SiIII\ and \CII\ column densities with $N({\rm HI})$. A similar trend was also reported previously by \cite{Lehner2018}, \cite{Muzahid2018} and \cite{Werk2013} in the HST COS surveys of { \HI\ absorption systems in quasar spectra: COS CGM Compendium (at $z_{\rm abs}<1$ ),  COS-Weak ($z_{\rm abs}<0.3$) and COS-Halos ($z_{\rm abs}<0.35$), respectively}. The results { for our quasar sightlines} are consistent with { these trends}. %\st{those of  Lehner2018  and Muzahid2018.} %{ i.e. they cover the same area and therefore probe similar physical conditions}. { The sightlines from the COS-Halos survey with measured \SiII\ column densities show slightly higher values for $N({\rm SiII})$ with respect to our sight lines that may be a selection effect, since }
A difference is observed for { the AGN} 
%\st{galaxy} 
sight lines: { for J0838$+$2453 and J0755$+$0311} we detect higher metal column densities than those predicted by the trend for quasar absorption systems, { for J1338$+$2620, the metal column densities are slightly lower}. { To demonstrate this in detail, we also show in Fig.\,\ref{fig:sp_vs_hi} theoretical constraints on the metal and \HI\ column densities calculated under simple assumptions: $N({\rm X})/N({\rm HI})=({\rm X/H})_{\odot} Z f_{\rm X}/f_{\rm HI}$, where $({\rm  X/H})_{\odot}$ is the solar abundance of element $X$, $Z$ is the metallicity relative to the solar level from \cite{Asplund2009}, $f_{\rm HI}=N({\rm HI})/N({\rm H_{\rm tot}})$ is the \HI\ fraction, and $f_X=N_X/N(X_{\rm tot})$ is the  fraction of element X in the particular 
ionization stage considered. For physical conditions expected in the ISM and CGM, we assume the ranges $Z=0.1-1$, $f_{\rm HI}=10^{-3}-10^{-1}$ and $f_{\rm X} = 0.1-1$, and vary the factor $Z f_{\rm X}/f_{\rm HI}$ between $10^{-1}$ and $10^3$. These constraints are fulfilled for all quasar absorption systems from our sample and from other COS surveys, while the detections and upper limits for our AGN sight lines  are beyond these constraints.} 

%{ The absorption system in the AGN J\,1338$+$2620 spectrum has completely different properties than absorptions in other  AGNs (low metallicity and the high $N({\rm HI})=10^{20.2}\,{\rm cm^{-2}}$). This system is likely related to gas absorption in the host galaxy rather than gas absorption in high ionization outflow or IGM.}

%\st{The ratio of $N({\rm SiIII})/N({\rm SiII})$ does not show a strong correlation with $N({\rm HI})$ and is generally consistent with $1$ with a scatter of $\sim0.5$\,dex.} 
We also present the Spearman rank-order correlation coefficient ($r_S$) and the probability that the observed value of $r_S$ could arise purely by chance ($p$-value) for all our systems and quasars only in the left top corner of each panel in Fig.\,\ref{fig:sp_vs_hi}. 
%\st{It is seen that a statistically significant correlation is observed only between SiIII and HI column densities.} 
{ The ratio $N({\rm SiIII})/N({\rm SiII})$ shows  a statistically significant correlation ($r_{\rm S}=1.0$ and $p=0.0$) with the \HI\  column density for our quasar sight lines (although we caution that our sample consists of only three measurements). This correlation is consistent with the correlation seen in the COS-Weak survey, which, however, had  low statistical significance ($r_{\rm S}=0.13$ and $p=0.63$). For sightlines in the COS-Halos survey, there are mainly lower limits. } 
%(XXXX May want to give correlation test results--e.g., Spearman rank order correlation-- and state the median value with error bar.XXXXX)

We also note that the samples of { quasar absorption systems from the COS-Weak and COS CGM Compendium surveys} were selected by a blind method { (or based on availability in the HST archives)}, while { the quasar sight lines in our sample and from the COS-Halos survey were} selected to have  relatively small impact parameters. The consistency of our results with those of these other studies suggests that, on average, 
%\st{LLSs  at $z<1$} 
{ \HI\ absorption with $N({\rm HI})>10^{15}\,{\rm cm^{-2}}$ and associated metal features} can correspond to the CGM of  galaxies with impact parameters $\le\sim140$\,kpc.

\subsection{Ionization corrections}
\label{sec:met}
Since our systems are not self-shielded from ionizing UV radiation, we need to calculate the ionization corrections to estimate the physical conditions and metallicity. 

We used the photo-ionization code {\sc cloudy} to infer the ionization structure of systems and estimate the metallicity, ionization parameter and total hydrogen column density. We assumed a constant density model in a plane parallel geometry illuminated by the radiation field and cosmic rays (CRs).  The radiation field was modeled as consisting of two parts: the extra-galactic UV background (UVB) radiation at $z=0.1$ as computed by \cite{Khaire2019}\footnote{ We used the UVB model Q18 recommended by \cite{Khaire2019} as a fiducial.}, and the %\st{interstellar radiation field} 
{ galaxy light component modelled by the interstellar radiation field} as per the {\sc cloudy} template, which is consistent with the Draine model in the UV range \citep{Draine1978}. The interstellar radiation field was scaled by the factor $I_{\rm UV}$ to characterize the strength of the UV radiation from the nearby galaxy. This factor is especially important for our { AGN} %\st{galactic} 
sight lines (i.e., those with zero impact parameter), { for which the distance of the absorbing region from the galaxy center  is unknown.}

{ The UV and X-ray radiation produced by the galaxy (by stars and AGN) is generally ignored for the CGM absorption systems because the \HI\ ionizing photons produced within the galaxy are assumed to be absorbed by the neutral hydrogen and dust within the galaxy. Indeed the average escape fraction of the \HI\ ionizing photons from galaxies is assumed to be very low ($<1\%$) at $z<1$, see e.g. \citep{Khaire2019}. However, the UV spectral observations of nearby galaxies (including our AGN sightlines) do not show the strong damped Ly$\alpha$ absorption line associated with the neutral hydrogen in those galaxies. Moreover, the spectra usually have strong Ly$\alpha$ emission lines. This indicates that the \HI\ ionizing radiation can leak out of the galaxies along these sightlines and increase the UV background around the galaxies. The intrinsic spectral energy distribution (SED) of the galactic radiation is unknown for our galaxies, therefore we chose one of the {\sc cloudy} templates to model this radiation. The SED of this interstellar radiation model is similar to that for the UVB model, and about 100 times more intense in the range of 1-100 eV at $I_{\rm UV}=1$. Therefore we varied this parameter in the range of  $-3 \le \log I_{\rm UV}\le 1.0$ to allow for a wide range of values of the escape fraction and the galactic star formation rate/AGN activity.}
%{ The  galaxy light component can also be modelled by the AGN radiation from {\sc cloudy} templates, however we have found that the ism and AGN templates have similar spectral energy distribution in UV and X-ray ranges.}
%{ Also we did not worry about the escape fraction of the \HI\ ionising radiation, because the intensity of the galactic component is a free parameter, but not fixed to some value at the position of the absorption system.} %Moreover this approach allow us to estimate the distance to the absorption systems in AGN sight lines by comparing their UV magnitude with the modelled value.}
%We will estimate it for AGNs later from the comparison of the observed and modelled fluxes in UV range.

%absorption systems observed towards AGNs. 
We also took into account the ionization of the CGM by cosmic rays (CR), given that simulations predict a strong effect of CR on the evolution of the CGM up to the distance about several hundreds of kpc from the galaxy \citep{Salem2016}. The intensity of CR ionization rate was consistently scaled with the same factor $I_{\rm UV}$. The initial value of CR ionization rate was set to to the average value in the MW ($2\times10^{-16}\,{\rm s^{-1}}$). 
The number density in the models is characterized by the parameter ($n_{\rm H}$) and the chemical composition by the parameter of gas metallicity [X/H]. The element abundance pattern was chosen according to the model by \cite{Jenkins2009}, where the parameter $F_{\star}$ regulates the value of dust depletion. $F_{\star}$ is varied from 0 to 1, where $F_{\star}=0$ and $F_{\star}=1$ denote the minimum and maximum level of depletion, respectively. The depletion pattern in these cases roughly corresponds to typical values seen in the MW halo and MW ISM \citep{Welty1999}. 

The size of the model cloud is calculated by  {\sc cloudy} in such a way that the modeled \HI\ column density was equal to the observed value. { Since the observed values of the \HI\ column density (total and for individual components) are not very well constrained for the quasar sightlines (within $\sim$ $0.1-0.7$ dex), we set the \HI\ column density as an additional fitting parameter.} Then we calculated a grid of models that uniformly covers the parameter space in the ranges of $-3.5 \le \log n_{\rm H}/{\rm cm}^{-3}\le 1.0$ (with a 0.5\,dex step), $-3 \le \log I_{\rm UV} \le 1.0$ (with a 0.5\,dex step), $-3 \le \log [{\rm X/H}] \le 2.0$ (with a 0.5\,dex step), $0<F_{\star}<1$ (with a 0.25 step), { and $13<\log N({\rm HI})<20.5$ (with a 0.5 dex step)}. 
For each node of the grid, we saved the column densities of metals (\SiII, \SiIII, \SII, \CII, N\,{\sc i}, N\,{\sc ii}, N\,{\sc v}, Fe\,{\sc ii}, O\,{\sc i}) and the ionization parameter $q=Q/4\pi R^2 n_{\rm H}c$, and calculated interpolations of metal column densities and $q$ on the grid.

Then we calculated the likelihood function for the fitting parameters ($n_{\rm H}$, $I_{\rm UV}$, $[{\rm X/H}]$, $F_{\star}$, $N({\rm HI})$) based on a least-squares  comparison of the observed and modeled column densities for the various ionic species. For this, we used the Monte Carlo Markov Chain approach with implementation of the affine-invariant ensemble sampler. The parameters were varied simultaneously to derive maximum probability values and their uncertainties corresponded to 63.8\% interval. The results are presented in Table\,\ref{tab:hst_results} for the total column densities and Table\,\ref{tab:detailedfit} for the individual velocity components. The comparison of metal column densities predicted by {\sc  cloudy} with the observed one in the absorption systems is shown in Fig.\,\ref{fig:_hist} in Appendix\,\ref{app:B}.

The {\sc cloudy} models allow us to describe the observed column densities  relatively well. As can be seen from Fig.\,\ref{fig:_hist}, the observed column density values and their uncertainties show good consistency with the predicted ranges of values for the ions of \SiII, \SiIII, and \CII\ (which show strong absorption lines and are detectable in our data even at relatively low S/N ratio). For other ions whose lines are relatively weak (e.g.,  N\,{\sc i}, N\,{\sc v}, S\,{\sc ii}), the {\sc cloudy} models predict lower column densities than the observed values. This difference may be caused by an overestimate of the measured column densities from the noisy spectra. 
In Table\,\ref{tab:cloudyfitparam} we also present estimates of the metallicity and the ionization parameter in the individual velocity components.  In some systems (e.g. J1237$+$4447 or J0950$+$4309), large differences are observed  
between the different components which may indicate substantial differences in physical conditions between the components (e.g., in the ionization parameters, which can cause different \SiIII/\SiII\ ratios). { In such cases, the fit to the total column densities is not reliable, and we need to analyze the parameters in individual components.} 
In some cases, we are not able to resolve individual absorption components in the \HI\ Ly$\alpha$ line and therefore cannot accurately measure their \HI\ column densities, although the total column density is well-constrained (e.g. J1338$+$2620). { In the case of J1338$+$2620, we analyze the physical conditions assuming equal metallicity for the components.} 
We also note that, in some cases (e.g., J0758$+$4219), there is good consistency between the physical conditions inferred for the different velocity components.

%\begin{figure}
%\begin{center}
%        \includegraphics[width=0.5\textwidth]{hst_cloudy_4panels.pdf}
%        \caption{\rm The parameters of ionization model ([X/H], $\log q$, H$_{\rm tot}$) and the total \HI\ column density of absorption systems. The circles represent our data. The color scheme is the same as in Fig.\,\ref{fig:sp_vs_hi}. The grey squares represent data from \cite{Muzahid2018} .}
%        \label{fig:_corner}
%\end{center}
%\end{figure}

\begin{figure*}
\begin{center}
        \includegraphics[width=0.75\textwidth]{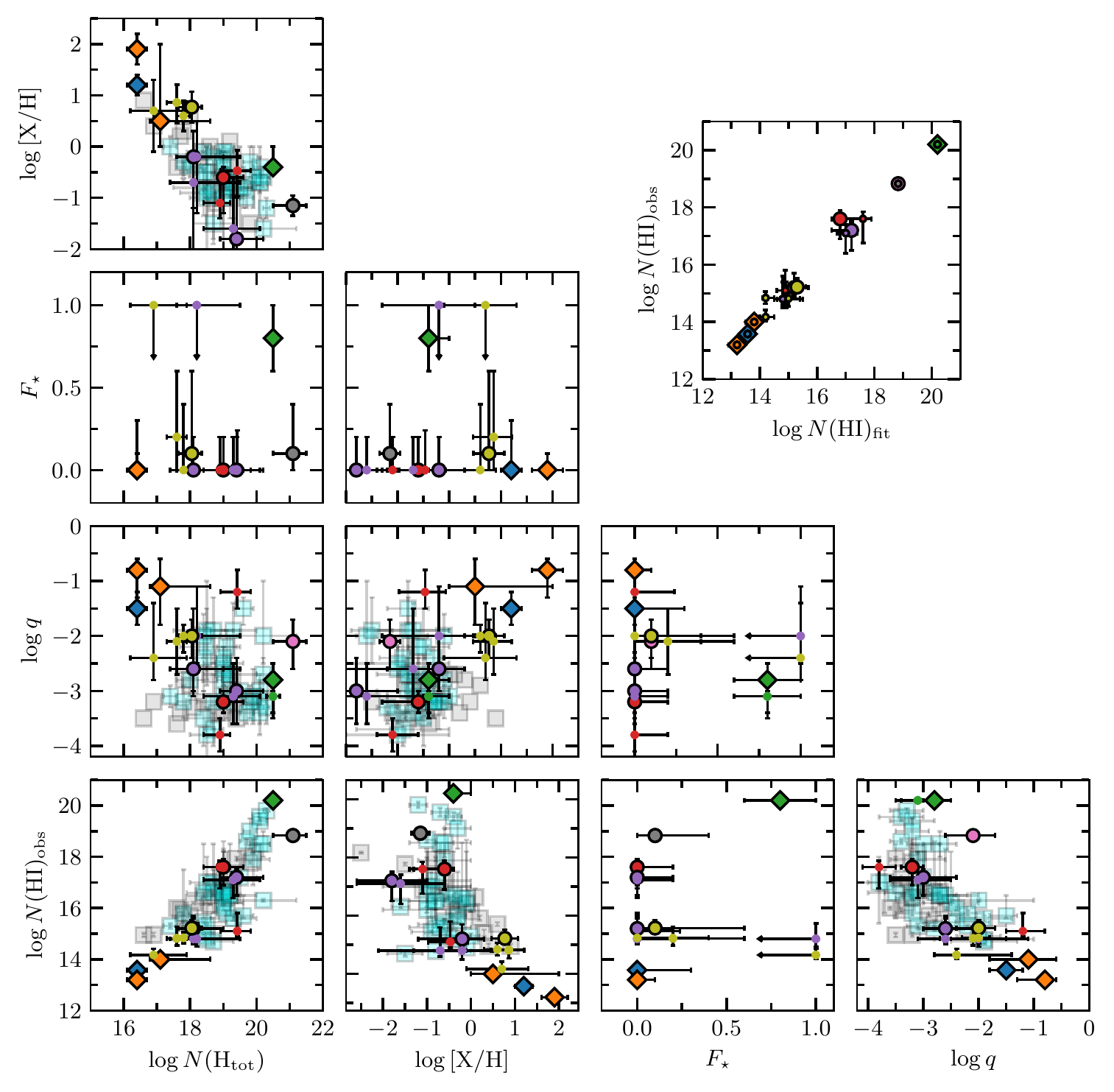}
        \caption{\rm The parameters of ionization model ([X/H], $F_{\star}$, $\log q$, H$_{\rm tot}$) and the observed \HI\ column density of absorption systems. The circles and diamonds represent results for our sample. { The large symbols show the average values (from fitting to the total column densities), the small symbols show the values for  individual components.} The color scheme is the same as in Fig.\,\ref{fig:sp_vs_hi}. The grey and cyan squares represent data from \cite{Muzahid2018} and \cite{Werk2014}. {The additional panel in the top right corner shows the observed \HI\ column density versus the fitted value from our {\sc cloudy} simulation. The model values of $N({\rm HI})$ correspond  well with the observed values.}}
        \label{fig:_corner}
\end{center}
\end{figure*}

Fig.\,\ref{fig:_corner} shows the comparison of the parameters derived with {\sc cloudy} (metallicity, depletion level, ionization parameter, total hydrogen column density) and column density of \HI. 
%For one system, J\,0950$+$4309, we can not accurately measure \HI\ column density from fitting to only H\,{\sc i}\,Ly$\alpha$ line, and therefore present results obtained for two limit cases of low and high $N({\rm HI})$. 
The metallicity spans over three orders of magnitude from -1.5 dex to 2 dex and is anti-correlated with the hydrogen column density. Low H\,{\sc i} column density systems tend to have higher metallicities and higher ionization. Similar results were found by \cite{Muzahid2018} and \cite{Werk2014}. %We show the comparison of our data and their trends for $N({\rm HI})$ versus  $N({\rm H_{\rm tot}})$ and  $N({\rm HI})$ versus [X/H]. 
There is a good agreement, but it is probably caused by fitting with the same photo-ionization code. Also we should remind that {\sc cloudy} indeed contains many assumptions: perhaps most importantly, it is a 1-dimensional calculation. 

{ We note an interesting difference in the physical conditions between the absorbing regions in our quasar sight lines and AGN sight lines. The AGN absorbing regions are located at different ends of the distributions of the physical conditions. For J0755$+$0311 and J0838$+$2453, we detect high values of the metallicity and the  $q$ parameter, whereas J1338$+$2620 has a low metallicity and a low $q$ parameter.}
The depletion level for J1338$+$2620 is also unusually high $F_{\star}=0.8\pm0.2$, while it is low for other systems ($F_{\star}<0.3$). A high value of $F_{*}$ is typical for the cold neutral phase of the ISM of the MW, while lower depletion level ($\sim0.2$) corresponds to gas in the warm phase and galaxy halo \citep{Welty1999}. { We speculate that the absorption in 
the J0755+0311 and J0838+2453 may be associated with outflowing gas driven by those AGN, while the absorption in J1338+2620 may 
be associated with inflowing cold gas falling into the AGN.}
%\st{This quasar sight line has the lowest ratio $b/R_e=3.8$ between sight lines with non-zero impact parameter, therefore in this case we are likely to detect gas corresponding to the CGM or ISM, and not the halo of the galaxy.} 
%{ The system J\,1338$+$2620 has unique properties, which differ from the properties of other AGN absorptions in our sample (see Table\,\ref{tab:hst_results}). It makes this system is likely related to the ISM of the host galaxy.%, while absorptions in other AGNs (J\,0755$+$0311 and J\,0838$+$2453) can represent high ionization gas in galactic halo or central outflow. 

Also, we note that there is no detection of low metallicity and low $N({\rm HI})$ gas, which could correspond to infalling metal free gas { in the outskirts of the galaxies}. This may be caused by a selection effect due to the difficulty of detecting weak metal lines: for low $N({\rm HI})$ and low metallicity, we can set only upper limits on the metal column densities, that do not allow us to constrain physical conditions with {\sc cloudy}, so the estimates for such absorbers 
are very uncertain. 

%\subsection{ The distance to AGN absorptions}

\section{Discussion}
\label{sec:discussion}
The combination of the HST COS spectroscopic data for the targeted sight lines and the MaNGA maps of the galaxies provides a powerful way to directly compare the CGM properties of the sample galaxies with their stellar properties. We now consider the relations between the various galaxy and CGM properties derived from the available data and discuss our results.  To put our work in broader perspective, we compare our results along with those for 
other galaxies from the literature \citep{Kulkarni2022, Muzahid2018, Tumlinson2013} and references therein.

%Galaxy sight lines more likely probe gas in close vicinity to the galaxies. Weak \HI\ absorptions  and in the case of J\,1653$+$3945 we could not find associated absorptions within $\pm800$\,km s$^{-1}$ relative to the galaxy redshift.  Only in one case, the galaxy J\,1338$+$2620, we detected strong \HI\ absorptions with  $N({\sc HI})=10^{19.3}\,{\rm cm^{-2}}$ and associated metal absorption lines. %which can be related to gas located near the galactic disk. 
%The second group of sight lines towards background quasars probes gas at higher impact parameters. We detected strong \HI\ absorptions and associated metal lines in 5 out of 8 spectra. In other two cases we set upper limit $N({\rm HI})<10^{13}\,{\rm cm^{-2}}$ and in one spectrum of the quasar J2106$+$0909 we 

\subsection{H I column density and impact parameter}
\label{sec:impactpar}
\begin{figure*}
\begin{center}
        \includegraphics[width=1\textwidth]{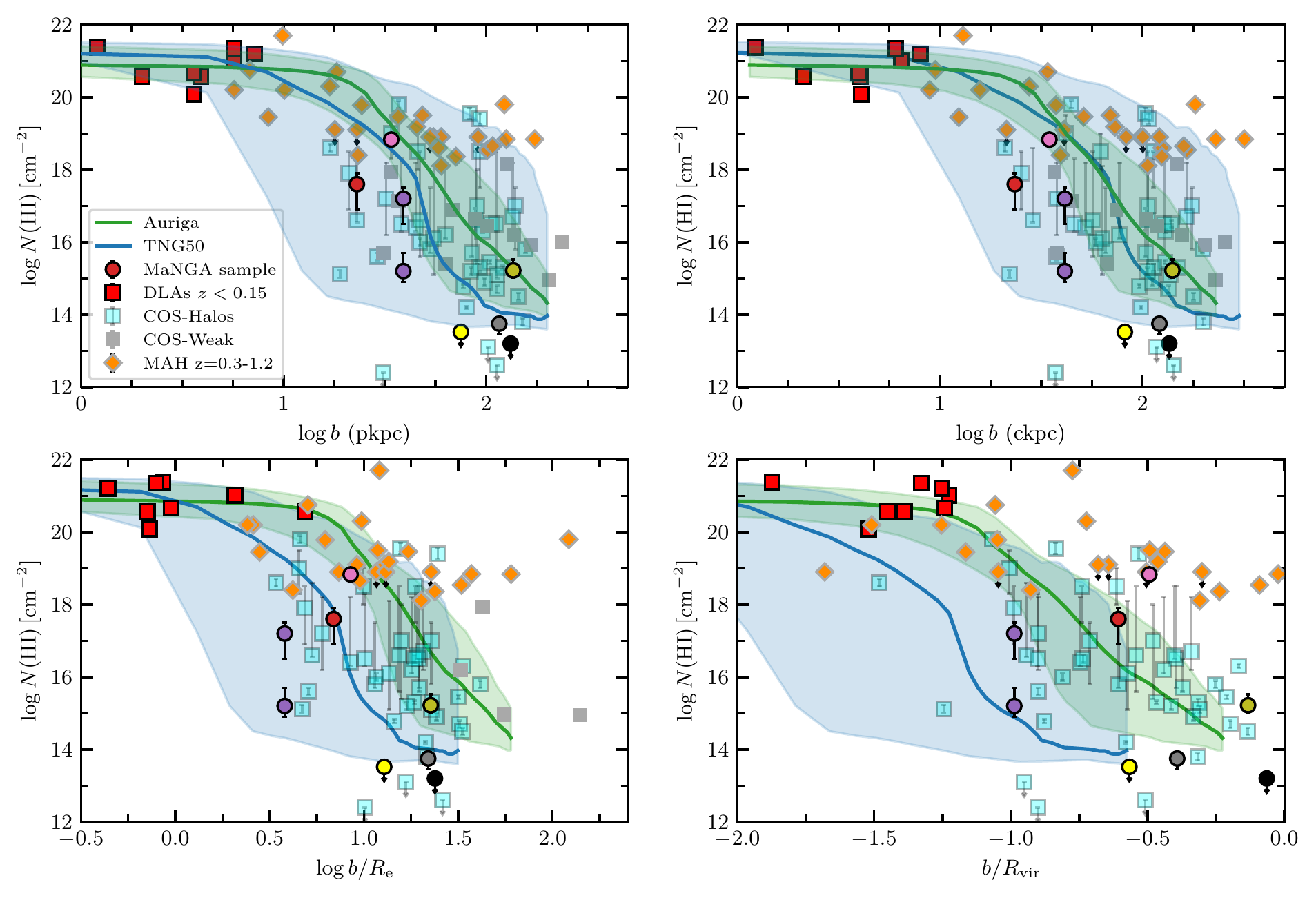}
        \caption{\rm The comparison of \HI\ column density against the impact parameter measured in { physical kpc (top left panel), in comoving kpc (top right panel), effective radii (bottom left panel) and virial radii (bottom right panel).} Our systems are shown by circles (quasar sightlines). The color scheme is the same as in Fig.\,\ref{fig:sp_vs_hi}. 
        Red  squares represent ``galaxies on top of quasars'' from \citealt{Kulkarni2022} ($z<0.15$), cyan squares are from the COS-Halos survey \citep{Tumlinson2013, Werk2013} ($z=0.14-0.35$), grey squares are from the COS-Weak survey \citep{Muzahid2018} ($z<0.32$), and the orange diamonds are from the MUSE-ALMA Halos survey \citep{Karki2023, Weng2023} ($z=0.3-1.2$). The effective radii of the COS-Halos and MUSE-ALMA Halos galaxies were estimated based on the relation with stellar mass from \cite{Mowla2019}. The green and blue curves with the shaded areas show the median radial profiles of the \HI\ column density and the 1 $\sigma$ scatter around those from high resolution galaxy simulations: Auriga project \citep{VandeVoort2019} and TNG50 \citep{Nelson2020}.}
        \label{fig:HIvsb}
\end{center}
\end{figure*}

First, we check the correlation of the total \HI\ column density with the impact parameter\footnote{ The impact parameter denotes the lower limit to the distance between the galactic center and the absorption system along the quasar sightline. The real distance can be higher, however it is believed that the distribution of gas around galaxies strongly decreases with the distance, and therefore the impact parameter has the highest probability of gas detection.}. We find that the quasar and galaxy sight lines in our sample show different behaviors. The quasar sight lines  probe gas around galaxies with impact parameters ranging from 20 to 130\,kpc. For them we find a decrease in the \HI\ column density with increasing impact parameter.
{ This result is in line with other studies of quasar-galaxy pairs at low redshift, such as the COS-Halos, COS-Weak, and the Galaxies on Top of Quasars  (within impact parameters $\sim1-7$ kpc) surveys \citep[e.g.,][]{Tumlinson2013, Muzahid2018, Kulkarni2022}, and at higher redshift  \citep[$z=0.3-1.2$, e.g., the MUSE-ALMA Halos (MAH) survey,][]{Weng2023,Karki2023}. A comparison of our results with these other studies is shown in Fig.\,\ref{fig:HIvsb}.} 
% We note that studies of galaxies detected in the fields of strong \HI\ absorbers at high redshift (the literature sample presented in \citealt{Kulkarni2022}) show many cases with high impact parameters.} %{ The fact that we do not detect such strong \HI\ absorbers with high impact parameter at low redshift may be explained by the selection of high redshift absorbers, and small statistic    }

{ The AGN sight lines have, technically, zero impact parameter, but the  absorbing gas can be separated from the galactic center at any distance along the sight line. In one case, J1338$+$2620, we found a high \HI\ column density ($\simeq10^{20.2}\,{\rm cm^{-2}}$), consistent with what is seen in quasar sightlines at very low impact parameters \citep{Kulkarni2022}. In other AGNs, the \HI\ absorption lines are weak ($\simeq10^{13}\,{\rm cm^{-2}}$) or not detected. A natural explanation in these latter cases may be a high ionization of the gas in the central outflow or (less likely) highly ionized gas in the IGM. %Therefore, AGN absorptions probe gas in different physical condition than quasar absorptions, and we should consider them separately. 
To avoid confusion, we do not show the cases of the AGN sight lines in Fig.\,\ref{fig:HIvsb}.}

%\st{On the contrary, our galaxy sight lines (which have zero impact parameter) probe gas in close vicinity to the inner parts of the galaxies. In one case, J\,1338$+$2620, we found high  HI column density absorption. In other cases, HI absorption is weak ($\simeq10^{13}\,{\rm cm^{-3}}$) or not detected. This difference between the properties of quasar and galaxy sight lines can be explained by assuming that the galaxy sight lines probe the gas outflow at a high elevation angle { We define the elevation angle as $(90^{\circ}-{\rm polar\,angle})$, i.e. as the angle of absorption system with respect to the disk plane. See detail in Section\,\ref{sec:azimodel}}, while the quasar sight lines probe gas in the galactic disk or halo with smaller elevation angles.}

%\st{A comparison of our results with other studies at low redshift (Kulkarni2022, Muzahid2018, Tumlinson2013) and high redshift (references in {Kulkarni2022}) is shown in Fig.. The trend of decreasing $N({\rm HI})$ with increasing impact parameter in our sample is in the line with measurements at low redshift, while galaxies at high redshift show stronger \HI\ absorption at high impact parameters.}

%It can be explained if the neutral galactic disk continues up to $\sim10$\,kpc, and then the gas becomes more diffuse and ionized, and the \HI\ column density sharply decreases with an impact parameter with a slope of about $\sim-7$. The measurements in galaxies at higher redshifts (orange squres in Fig.\,\ref{fig:HIvsb}) show that neutral gas in galactic disks continues up to higher distances $\sim50-100$\,kpc.  
{
The top left and right panels of Fig.\,\ref{fig:HIvsb} show the \HI\ column density plotted versus the impact parameter in physical (proper) and comoving\footnote{ The impact parameter can be presented in proper (physical) kpc by multiplying the  angular separation (in arcsec) by the angular diameter distance ($D_A$), and in comoving kpc (ckpc) by multiplying the angular separation by the comoving distance ($D_M$), which is larger than $D_A$ by a factor of $(1+z)$.} units, respectively. The physical units correspond  to the distance in the rest frame of each galaxy and can be meaningfully compared to simulations in physical units. The comoving units factor out the cosmological expansion, allowing a comparison of the properties of galaxies at different redshifts. The difference between the physical and comoving impact parameters is not significant for our MaNGA galaxies due to their low redshift ($z=0.01-0.10$), but is larger for higher-redshift 
galaxies in the literature ($z=0.15-0.35$ for the COS-Halos galaxies, $z=0-0.3$ for the COS-Weak galaxies, and $z=0.3-1.2$ for the MAH galaxies), and for the simulations of the higher-redshift CGM.  
We also plot in Fig.\,\ref{fig:HIvsb} the median radial profile of the \HI\ column density and the 1 $\sigma$ scatter around that from  magnetohydrodynamic simulations of an isolated Milky Way-mass galaxy at $z=0-0.3$  by \cite{VandeVoort2019} (based on the Auriga project, \citealt{Grand2017}), and from the study of the distribution of cold gas in the CGM around galaxy groups { ($\log M_{\rm halo}\sim13.2-13.8$ and $\log M_{\rm *}\sim11.3-11.8$)} { at $z=0.5$} by post-processing TNG50 simulations by \cite{Nelson2020}. 

Combining the observational data from the different studies mentioned above, we cover relatively well a wide range of $b$ parameters from 1 to 150 kpc. There is agreement between most of the observations and the radial profile from the simulations within the uncertainties, although we %Most of the observations are in 1\,$\sigma$ uncertainty of simulations. 
note higher \HI\ column densities compared to the simulations for a few MAH galaxies at high impact parameters. These outliers have $z>0.7$, that is higher than for the rest of the observed galaxies. A similar increase in the $N({\rm HI})$ profile at high impact parameters for high-redshift galaxies was  reported earlier by \cite{Kulkarni2022}.  In the comoving coordinates, the median $N({\rm HI})$ radial profiles from the Auriga and TNG50 simulations are in better agreement with each other, suggesting that the  difference between them  is probably due to the difference in redshift. Most of our MaNGA\ galaxies show  weaker \HI\ absorption than the simulated Auriga galaxy. The agreement with the simulated TNG50 galaxies is better than with the Auriga galaxy. We note, however, that  our  
MaNGA galaxies are lower in redshift than the TNG50 galaxies, and are also lower in stellar mass than the Auriga and TNG50 galaxies. 
}

The { bottom left and right} panels of Fig.\,\ref{fig:HIvsb} show the \HI\ column density plotted versus the impact parameter normalized to the effective radius ($R_{\rm e}$) and virial radius ($R_{\rm vir}$). The virial radius was estimated as $(3M_{\rm halo}/200\rho_{\rm cr}4\pi)^{1/3}$, where $M_{\rm halo}$ was estimated from the $M_{\star}-M_{\rm halo}$ relation by \cite{Girelli2020} and $\rho_{\rm cr}$ is the critical density at the redshift of the galaxy. For galaxies at low redshift, { all absorbers classified as DLAs and several classified as sub-DLAs are associated with the region within $\sim3$ effective radii. Most LLSs appear to correspond to the region from $\sim3$ to $\sim30$ effective radii.} The trend of $N({\rm HI})$ with $b/R_e$ is similar to { the trend with $b$}. Comparison of  $N({\rm HI})$ with $b/R_{\rm vir}$ shows that all the detected \HI\ { absorbers} are within the virial radius.

\subsection{H I column density versus stellar mass, sSFR and $D_n(4000)$}
\label{sec:hisfr}
\begin{figure*}
\begin{center}
        \includegraphics[width=1\textwidth]{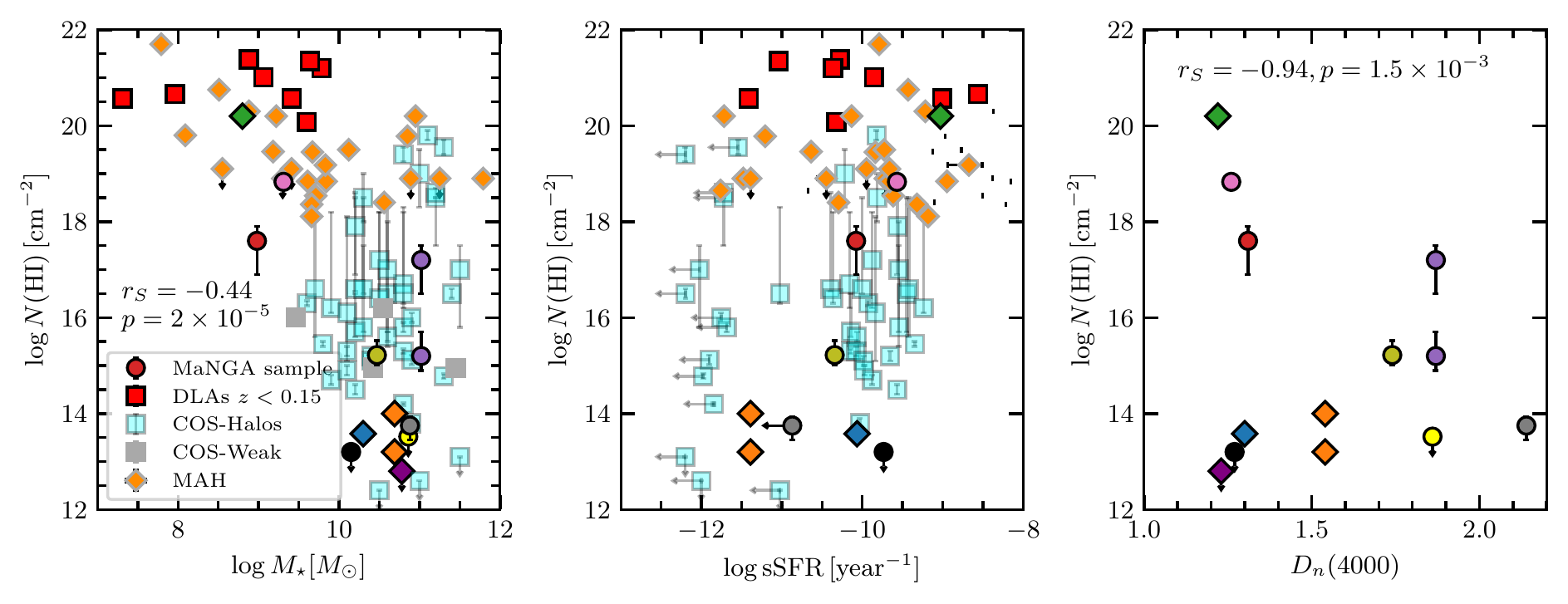}
        \caption{\rm The comparison of \HI\ column density against the stellar mass (left panel), the specific star formation rate (middle panel) and the $D_n(4000)$ index (right panel). Symbols are the same as in Fig.\,\ref{fig:HIvsb}. { In addition, the large diamonds represent our AGN sight lines with ``zero impact parameter''. The \HI\ column density is anti-correlated with the galaxy stellar mass, but not with sSFR. The quasar sightlines suggest $N({\rm HI})$ decreasing sharply with increasing $D_n(4000)$ (excluding the non-detection of 1-549490 due to the large impact parameter). This indicates the connection of the gas content at very large radii to the star-formation history in the center of the galaxy.}}
        \label{fig:HIvsMs}
\end{center}
\end{figure*}

Fig.\,\ref{fig:HIvsMs} shows the relations between the \HI\ column density of the associated absorbers, and the stellar mass, the specific star formation rate (sSFR = SFR/$M_{*}$) and the $D_n(4000)$ index of the host galaxies, based on our sample and the literature. The stellar mass and sSFR in our sample range from $10^7\,M_{\odot}$ to $10^{12}\,M_{\odot}$ and from $10^{-12}\,{\rm yr^{-1}}$ to $10^{-9}\,{\rm yr^{-1}}$, respectively. Most of our galaxies are star-forming (${\rm sSFR>10^{-11}\, yr^{-1}}$). The value of $D_n(4000)$ index ranges from 1.27 to 2.14 and characterizes the star formation history in the center of the galaxy.

Absorption systems with a high \HI\ column density are more likely related to low stellar mass galaxies \citep{Kulkarni2022}, while systems with a lower \HI\ column density are associated with the halos of more massive galaxies \citep[e.g.,][]{Kulkarni2010, Augustin2018, Tumlinson2013}. { } 
%{ NOT SURE WHAT THE POINT OF THESE SENTENCES IS--WHAT BIMODALITY AND WHAT IS BEING EXPLAINED? Indeed we see the bimodality in the $N({\rm HI})$ distribution in previous studies \citep{Tumlinson2013, Kulkarni2022}. It can be explained by the difference in the strategy by \cite{Tumlinson2013} and \cite{Kulkarni2022} in the sample selection. \cite{Kulkarni2022} targeted galaxies with small impact parameters and found them to be associated with DLAs, and combined this sample with the literature sample of galaxies found in the fields of known DLAs. \cite{Tumlinson2013} select a sample in mass and impact parameter and then measure $N({\rm HI})$. As it happens, the \cite{Tumlinson2013} sample doesn't probe below $b=10-15$\,kpc or below $9.5$\,dex in solar mass. The \cite{Kulkarni2022} sample is agnostic to stellar mass, but their literature sample focuses on  high \HI\ columns (DLAs). NOT SURE WHAT YOU MEAN BY TRYING TO RECTIFY---I DON'T THINK WE ARE TRYING TO RECTIFY ANYTHING WITH THE MANGA SAMPLE--THEY ARE TOTALLY DIFFERENT APPROACHES. The preponderance of this work (quasar-galaxy pairs) from previous studies is that we are trying to rectify that with the MaNGA\ sample, and the results appear to be tracking the simulations (Fig.\,\ref{fig:HIvsb}). Do we know what the simulations predict for $N({\rm HI})$ versus stellar mass, sSFR, $D_n(4000)$?}

The MaNGA galaxies from our sample also follow this trend. Three systems  with the highest \HI\ column density ($N({\rm HI})\ge10^{18}\,{\rm cm^{-2}}$) are observed near galaxies with $M_{*}\le10^9\,M_{\odot}$, while the other systems correspond to high stellar mass galaxies with  $M_{*}\simeq10^{10}-10^{11}\,M_{\odot}$. For the sample of low-redshift $z<0.35$ systems (our data, \citealt{Tumlinson2013, Kulkarni2022}) we obtain a correlation coefficient $r_S=-0.34$ and $p$-value of $5\times10^{-3}$, for the entire sample, including MUSE-ALMA observations, $r_S=-0.44$ and $p=2\times10^{-5}$. We do not see a difference in the \HI\ content for galaxies with low or high  specific star formation rates. However, a strong negative correlation is observed between the sSFR and stellar mass for { AGNs} 
%\st{galaxies} 
in all the samples examined here, including our own galaxies and those from the literature ($r_S=-0.66$, $p=2\times10^{-9}$). %PLEASE GIVE THE r_S AND p VALUES. 

We also report a strong dependence of $N({\rm HI})$ decreasing with increasing $D_n(4000)$ for quasar sight line - discounting the non-detection of 1-564490 due to large impact parameter. The Spearman rank order correlation coefficient is $r_S=-0.94$ and $p$-value is 0.015. 
It is interesting that this correlation connects the gas content at very large radii to the star-formation history in the center of the galaxy. We suspect we would see the same thing with sSFR if we had measures of sSFR for the three quasar sight line galaxies with $D_n(4000)>1.7$. For galaxy sight lines (when we get spectrum from the central AGN) we do no see a such dependence probably due to the contamination of the observed spectra by AGN.

{ Comparing the correlations between $N({\rm HI})$ and $b$, $M_{*}$, $D_n(4000)$ we believe that the primary correlation is likely with stellar mass. Higher mass galaxies have higher $D_n(4000)$ index and their  higher past star formation activity is expected to affect cool gas around them, both because cool gas is consumed in star formation and because AGN radiation and stellar winds blow out gas around these galaxies. 
%, while it is still present in the case of low mass galaxies. % have consumed more \st{of the neutral} gas. Furthermore, the higher past activity in more massive galaxies affects cool \st{neutral} gas around them, since AGN radiation and stellar winds blow out cool \st{neutral} gas from the close vicinity of these galaxies. 
%, while it is still present in the case of low mass galaxies. 
We do not have many observations of cool gas around massive galaxies at low impact parameters to check this statement in further detail. %BUT HIGH-MASS GALAXIES NEED NOT BE FORMING STARS MORE ACTIVELY NOW OR RECENTLY. THEY PROBABLY HAD MORE STAR FORMATION IN THE PAST. 
} 

\subsection{Galaxy geometry and kinematics}
\label{sec:geometry}

At the spectral resolution of COS G130M, we can obtain fairly reliable
velocity profiles for the absorbing gas along the sight line through the galaxy. This can
be compared with kinematics of the ionized gas from MaNGA data. With this in mind, we used the radial velocity maps of the stellar disk and H$_{\alpha}$ gas for our galaxies to reconstruct the position of the quasar sight lines relative to the gaseous disks of the galaxies and examined the correspondence between the radial velocities of the absorbers and the rotation of the galactic gaseous disks. 
%, and (ii) correspondence of the measured metallicity of the absorbers to the metallicity gradient of the galaxies.

First, we fitted both the gas velocity map (H$_{\alpha}$ line emission) and the stellar velocity map with symmetric models of thin disk rotation. The formalism was described in Section\,\ref{sec:mangavel}. 
%We followed the formalism presented by \cite{Bekiaris2016} and used the model of "arctan"{} profile rotation curve \citep{Courteau1997}, which has 7 fitting parameters: PA (position angle of the disk), $i$ (inclination angle of the disk), $x_0$ and  $y_0$ (coordinates of the center of the disk), $V_t$ (asymptotic circular velocity), $r_t$ (turn-over radius), and $V_{\rm sys}$ (the systemic correction for velocity offset relative to the galaxy redshift). Then we  calculated the likelihood function using the MCMC approach and estimated the parameter values. 
We choose the best fit giving priority to the joint fit (stellar+H$_{\alpha}$ emission) first, followed by the H$_{\alpha}$ emission line fit, using only the stellar fit if others didn't fit. The position angle ($PA$), inclination angle ($i$), and maximal rotation velocities ($V_{\rm max}$) are presented in Table\,\ref{tab:mangafit}. 
The fitted velocity map and rotation curve for each galaxy are shown in Figs.\,\ref{Final-1-166736}-\ref{Final-12-192116}.

%The third row of panels shows 3d model of the thin rotation disk and quasar sight line, and $Y-Z$ and $X-Y$ projections. It was calculated by rotating the thin disk model and quasar sight line by the PA angle around $Z$ axis (right to left) and by the inclination angle around $Y$ axis. Color of disk points corresponds to the initial colormap of radial velocities along the quasar sight line. In this model the observer is located at the top, and the quasar is located at the bottom. 

\setlength{\tabcolsep}{2pt}
\begin{table*}
\begin{center}
\caption{The properties of MaNGA galaxies}
\label{tab:mangafit}
\begin{tabular}{llllllllcccccccccc}
\hline
MaNGA &  $z_{\rm gal}$ & $R_e$ & [O/H] & $\nabla_{\rm R}[{\rm O/H}]$ & $\log q_{\rm ion}$ &  $\nabla_{\rm R}[{\rm q}]$ & $V_{\rm max}^a$     & PA  & Incl.& $\phi_{\rm stand}^b$ &  $\phi_{\rm model}^c$ \\
ID     &                & kpc   &       & $10^{-3}$kpc$^{-1}$         &                    &  $10^{-3}$kpc$^{-1}$       & ${\rm km\,s^{-1}}$ & deg & deg  & deg  & deg  \\
\hline
1-71974   & 0.03316 & 4.9  & $0.31\pm0.04$  & $-7\pm2$  & $7.16\pm0.07$ & $-13\pm2$ &    151   &  147  & 33 & 57$^{+4}_{-4}$ & $57^{+4}_{-4}$ \\
1-385099  & 0.02866 & 5.4  & N/A            & N/A       & N/A           & N/A       &    200   &  21  & 32  & 58$^{+2}_{-2}$ & $58^{+2}_{-2}$ \\
1-585207  & 0.02825 & 2.4  & N/A            & N/A       & N/A           & N/A       &    155   &  13  & 47  &  34$^{+2}_{-2}$ & $18^{+20}_{-20}$  \\
12-192116 & 0.02615 & 3.3  & $-0.25\pm0.08$ & $-18\pm2$ & $7.05\pm0.20$ & $-40\pm2$ &     64   &  141  & 36  & 54$^{+1}_{-1}$ & $54^{+2}_{-2}$ \\
1-594755  & 0.03493 & 1.3  & N/A            & N/A       & N/A           & N/A       &    144   &  162  & 22  & 68$^{+4}_{-4}$ & $68^{+4}_{-4}$  \\
%1-90242   & 0.03023   &  & &    &       &       &    &    &   && J\,1535$+$5754 & 0.0302 & 0  & 0\\ 
1-575668  & 0.06018 & 10.6 & N/A            & N/A       & N/A           & N/A       &    500   &  172  & 8  & 10$^{+4}_{-4}$ & $0^{+24}_{-24}$  \\
1-166736  & 0.01708 & 3.4  & $-0.16\pm0.10$ & $-18\pm8$ & $6.96\pm0.18$ & $-8\pm1$  &    58    &  156  & 53  & 50$^{+8}_{-8}$ & $28^{+25}_{-23}$  \\
1-180522  & 0.02014 & 4.1  & $0.06\pm0.07$  & $-26\pm2$ & $7.02\pm0.12$ & $3\pm1$   &    124   &  122  & 74  & 4$^{+12}_{-12}$ & $4^{+15}_{-15}$ \\
1-635629  & 0.01989 & 1.7  & $0.35\pm0.05$  & $-14\pm4$ & $7.04\pm0.10$ & $-50\pm5$ &    124   &  16  & 65  & 28$^{+2}_{-2}$ & $27^{+15}_{-15}$ \\
1-561034  & 0.09008 & 6.0  & N/A            & N/A       & N/A           & N/A       &    236   &  61  & 50  &  44$^{+3}_{-3}$ & $28^{+27}_{-18}$  \\
1-113242  & 0.04372 & 5.5  & N/A            & N/A       & N/A           & N/A       &    550   &  12  & 26  & 17$^{+2}_{-2}$ & $5^{+22}_{-23}$ \\
1-44487   & 0.03157 & 6.2  & $0.33\pm0.05$  & $-14\pm1$ & $7.06\pm0.10$ & $-11\pm1$ &    225   &  25  & 78  & 6$^{+2}_{-2}$ & $9^{+6}_{-8}$ \\
%1-44487$^{\rm c,d}$ & 0.03174   & &  10.22 & & NA  & &  &     &    && J\,0758$+$4219 & 0.2111& 137 & \\
1-564490  & 0.02588 & 5.7  & $0.36\pm0.05$  &   $-20\pm2$       & $7.10\pm0.13$ &        $-60\pm3$   &    150   &  129  & 52  & 56$^{+2}_{-2}$ & $30^{+25}_{-22}$ \\
\hline
 \\
\end{tabular}
 \begin{tablenotes}
      \small
      \item $^{\rm a}$ The maximal rotation velocity of the galaxy derived from fitting by``arctan"{} model.
      \item $^{\rm b,c}$ The elevation angle derived by the standard method and using the model of gas distribution around the galaxy, respectively.
    \end{tablenotes}
\end{center}
\end{table*} 

\subsubsection{Elevation angle and the position of absorbers}
\label{sec:azimodel}
The analysis of velocity maps gives us the orientation of the galactic disk relative to the quasar/or galaxy sight line. Here we determine the orientation of absorption system along the quasar sight line with respect to disk plane. To be consistent with \cite{Peroux2020} we adopt ($\phi$) to be the elevation angle ($90^{\circ}-{\rm polar\,angle}$) or latitude with respect to the disk plane and ($\theta$) to be deprojected angle in the disk plane with respect to the major axis\footnote{\cite{Peroux2020} the angle $\phi$ is referred to as the ``azimuthal angle{"}, but we reserve that term for the disk in-plane angle with respect to the major axis.
% between the absorber and the projected major axis of the galaxy, see the illustration in Fig.\,1 of their pape.
}.

We estimated the elevation angle ($\phi$) of absorption systems in two ways: \\
(a) using the standard approach as the angle between the galaxy's major axis and the line joining the galaxy center to the quasar on the sky plane \citep{Bouche2012}, and \\
(b) by integrating the elevation angle along the quasar sight line using a model for the gas distribution around the galaxy.
In this approach, we assume that the probability of detecting gas absorption along the sight line can be described as following:
\begin{equation}
    f(r,\phi) = C\times f_{\Omega} f_{r} f_{\phi},
    \label{eq1}
\end{equation}
where $(r,\phi)$ are the radial coordinate and elevation angle of the point along the sight line, $C$ is a normalization coefficient, $f_{\Omega} = 1/r^2$ characterizes the decrease in the gas cross section with increasing distance from the galaxy center (lower solid angles are probed at larger distances), $f_{r}$ and $f_{\phi}$ are model distributions of the gas density around the galaxy. For $f(r)$, we adopt the Navarro–Frenk–White halo density profile \citep{Navarro1997} 
\begin{equation}
f_{r}=r_s/r(r+r_s)^2, 
\end{equation}
with the parameter $r_s=6R_e$.
For the elevation angle distribution $f_{\phi}$, we adopt 
\begin{equation}
f_{\phi}=\mathcal{N}(0,\pi/6) + \mathcal{N}(\pi/2,\pi/6), 
\end{equation}
where $\mathcal{N}(\mu,\sigma)$ is a Gaussian distribution with a mean $\mu$ and a width $\sigma$. Thus, $f_{\phi}$ is a bimodal distribution with two peaks, one near the galaxy plane and the other near polar axis, with opening angles of about 30 degrees, consistent with the range of outﬂow opening angles from $\theta_{\rm max}=30$ to 45 degrees estimated from galaxy spectra with outﬂow detections \citep[e.g.,][]{Martin2012}. Using the probability function\,\eqref{eq1} we calculate the mean value of the elevation angle as:
\begin{equation}
    \overline{\phi} = \int_{-\infty}^{\infty} \phi(x) f(r(x),\phi(x))\,dx,
\end{equation}
where $x$ is the coordinate along the quasar sight line.
%XXXXPlease add the equation for azimthual angle using this f(r,phi).XXXXX

\begin{figure}
\begin{center}
        \includegraphics[width=0.45\textwidth]{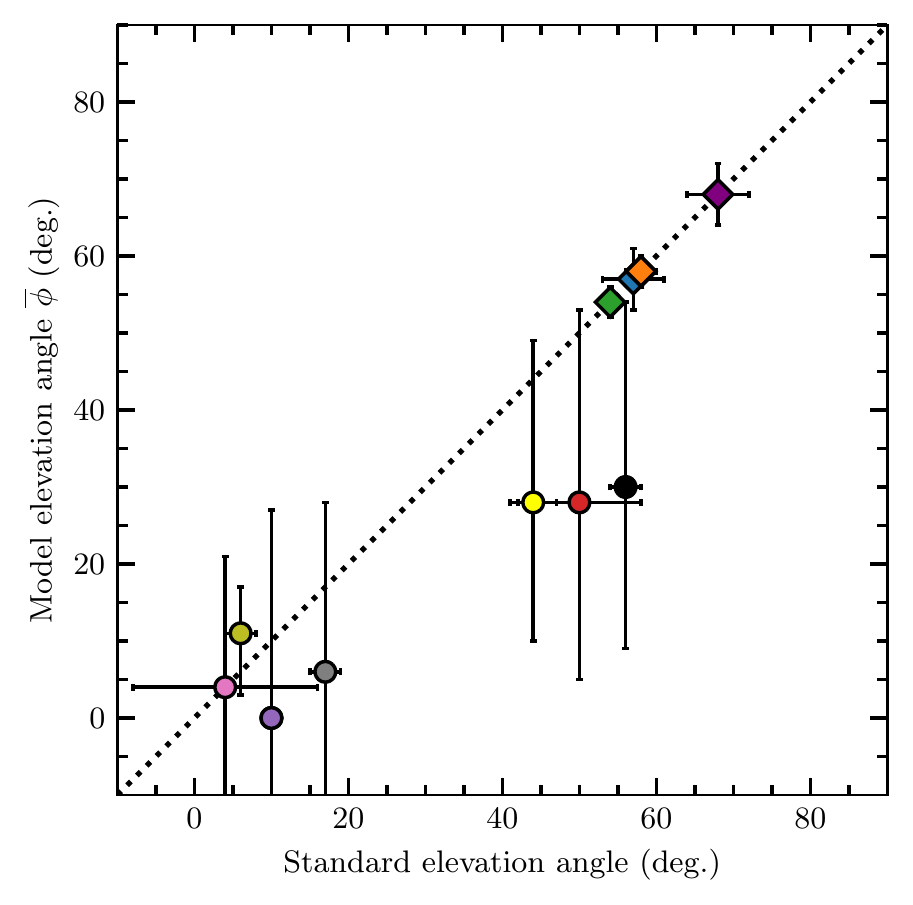}
        \caption{The comparison of elevation angles derived by the two methods:  from the modeling of gas distribution around the galaxy (vertical axis) and by the standard method (horizontal axis). See detail in text. The colors and shapes of symbols are the same as in previous figures.}
        \label{fig:Aziestimate}
\end{center}
\end{figure}

Fig.\,\ref{fig:Aziestimate} compares these two estimates of the elevation angle. The values agree mostly within the uncertainties. Approach (b) usually predicts lower elevation angles, because it takes into account a higher probability of detection for directions along the galactic plane, while the standard method corresponds to the direction with the smallest impact parameter. For galaxy sight lines (i.e., those with zero impact parameters), we estimated the elevation angle of absorption systems as ($\pi/2)-i$, where $i$ is the inclination angle of the galaxy. Using approach (b), we also calculated the deprojected radial coordinate of the absorption system in the disk plane ($d$) and height of absorption system above the disk plane ($h$) as follow: 
\begin{equation}
    d = r \cos \overline{\phi} \mbox{, } h = r \sin{\overline{\phi}},
\end{equation}
where $r$ is the radial coordinate of the point along the sight line with the highest probability $f(r,\phi)$.  
%XXXXPlease add equations for these.XXXX

\subsection{Gas kinematics}

\subsubsection{Quasar sight lines}
\begin{figure*}
\begin{center}
        \includegraphics[width=\textwidth]{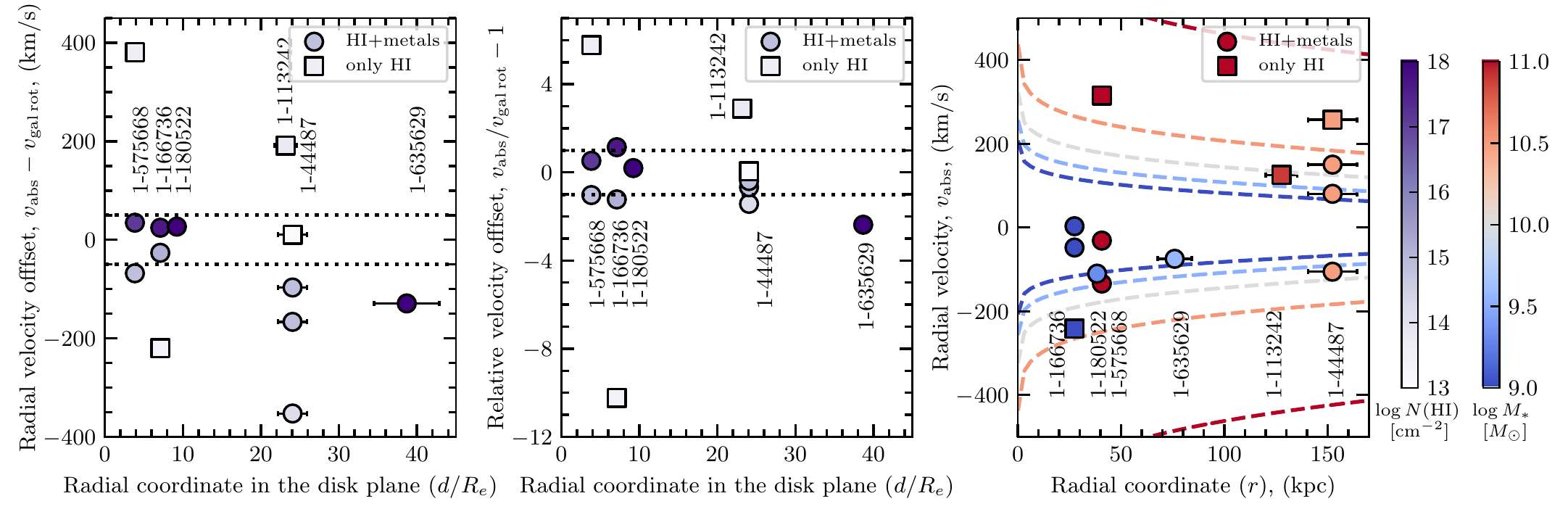}
        \caption{\rm 
        Left panel shows the difference between radial velocities of \HI\ absorption components { in quasar sight lines} and predicted velocity of galaxy rotational models are shown as a function of radial coordinate.  { Circles and squares represent components with both \HI\ and metal lines and only \HI, respectively. The color encodes the \HI\ column density of the components. The horizontal dotted lines show the range of velocities from $-50$\,km s$^{-1}$ to $50$\,km s$^{-1}$. The middle panel is the same as the left panel, but shows the velocity offset in units of the predicted velocity of galaxy rotational models. Dotted lines represent the $\pm1$ offsets relative to the predicted velocity (${\rm offset= 0}$). Right panel shows the radial velocity of absorption calculated in the galaxy rest frame as a function of the  radial coordinate. The symbols are the same as in other panels. Color encodes the stellar mass of nearby galaxies. The curves show the escape velocity as a function of the distance for stellar masses of $\log M_{*}/M_{\odot} = 9, 9.5, 10, 10.5, 11$. See the text for more details.}
        }
        \label{fig:kinematics}
\end{center}
\end{figure*}

We now compare the absorber velocities in the 5 quasar sight lines that show absorption detections with the corresponding best fit models of galactic disk rotation { in 6 MaNGA galaxies}. Using the fits to MaNGA emission-line velocities maps we calculate the radial velocity of the galactic disk along the direction towards the quasar sight line. The comparison is shown in Fig.\,\ref{fig:kinematics}. We show both the components seen in \HI\ alone, and the components seen in \HI\ as well as metals. Additionally, we show each of these quasar-galaxy pairs in Figs.\,\ref{Final-1-166736}, \ref{Final-1-180522},\ref{Final-1-44487}, \ref{Final-1-575668}, and \ref{Final-1-113242} in Appendix \,\ref{app:B}.

For three of the { six galaxies} (1-166736, 1-180522, 1-575668) 
%\st{five absorption systems (associated with galaxies 1-166736, 1-180522, 1-575668)} 
there is good agreement between the velocities of the strongest \HI\ absorption components and the predicted radial velocities of the galactic disks within $\pm50$\,km s$^{-1}$. We note that these { quasar} sight lines are located within 10 effective radii from the corresponding galaxies. { For two other galaxies (1-635629 and 1-113242), the absorption velocity is in opposite direction to that expected from the galactic disk rotation. 
%PLEASE CHECK--113242 DOESN'T SEEM TO HAVE ABSORPTION IN OPPOSITE DIRECTION. ALSO IT HAS NO Si II 1260, BUT THE FIGURE SHOWS A PROFILE. SIMILARLY FOR Si II 1260 FOR 1-385099, SEVERAL LINES IN FIG. 25, 28, 29, 30  --PLEASE ADD IN CAPTION OR TEXT THAT IN SUCH CASES PROFILES CORRESPOND TO 3 SIGMA UPPER LIMITS? (reply: 1-113242 has negative  rotational velocity in the direction to quasar sight line. The \HI\ absorption is seen at +130 km/s. The synthetic profile for SiII shows 1$\sigma$ upper limit, which is consistent with uncertainties (at S/N$\sim5$).3$\sigma$ line is much stronger.)
And for one galaxy (1-44487), the  absorption components are spread over a wide range $\sim350$\,km s$^{-1}$. However, this range is comparable with the velocity of galaxy rotation in the quasar sight line direction ($\sim250$\,km s$^{-1}$).
%, which is larger than for galaxies with a small impact parameter ($<10 R_e$). 

The middle panel of Fig.\,\ref{fig:kinematics} shows the velocity offset normalized by the predicted galactic disk rotation radial velocity at the appropriate distance. It is clear that this normalized 
velocity offset is within $\sim \pm$ 1 in most cases. In other words, the absorbing gas velocity is generally consistent with co-rotation with the galactic disk within $\sim$25 effective radii. }

%then the velocity offset for galaxies with smaller impact parameters ($<10 R_e$) will be similar to 1-44487 ($\sim25R_e$) and be within $\pm1$ galactic radial velocity. Then a good agreement is observed for the sight lines within 30 effective radii.}

{ We comment now on the difference between the kinematics of the absorbing gas  components seen in both \HI\ and metal lines, and those seen in only \HI\ lines. In all cases, only \HI\ absorption is observed in components with a low \HI\ column density ($\sim10^{13}\,{\rm cm^{-2}}$), whereas absorption in both \HI\ and metals is  observed in components with $N(\rm HI)>10^{15}\,{\rm cm^{-2}}$. Thus the absence of metal absorption in only \HI\ components is likley to be due to a limit in the sensitivity for detecting weak lines at the S/N reached. Second, 2 out of the 3 cases of large normalized velocity offsets are for H\,{\sc i}-only absorbers, possibly suggesting that the H\,{\sc i}-only absorption may be related to the galaxy halo or the IGM and thus not participate in the galaxy disk's rotation. The metal-bearing \HI\ absorbers are, however, likely to relate to the disk gas and hence cororate with it. }

%In all cases only \HI\ absorptions are observed in components with a low \HI\ column density ($\sim10^{13}\,{\rm cm^{-2}}$), whereas \HI\ and metal absorptions are observed in components with $N(\rm HI)>10^{15}\,{\rm cm^{-2}}$. Therefore the absence of metal absorptions in only \HI\ components may be due to a limit to the observed column density: the metal lines are to weak to be detected. The second,  
%Second, only \HI\ absorptions have systematically high velocity offset ($\sim200-400$ km s$^{-1}$, it is clearly seen for sightlines located within $10$ effective radii), while metal-bearing \HI\ absorptions have small velocity offsets. NOT TRUE FOR 1-635629. It can mean that only \HI\ absorptions relate to the galaxy halo or the IGM, while metal-bearing \HI\ absorptions are likely relate to gas in  the galactic disk. }

{ It is of interest to understand whether or not the \HI\ -only absorption is bound to the galaxies. To examine this, we compare  the velocity offsets of the absorbers relative to the systemic redshifts of the galaxies with the expected escape velocities at the impact parameters of the quasar sight lines. To estimate the escape velocity at a distance $r$ from the center of galaxy with stellar mass $M_{*}$ we used the methodology described in \cite{Kulkarni2022}. The right panel of Fig.\,\ref{fig:kinematics} shows the velocity offsets with respect to galaxy  redshifts\footnote{The galaxy redshift was corrected for the systematic velocity offset $V_{\rm sys}$ derived by fitting to MaNGA maps, see Sect.\,\ref{sec:mangavel}} 
for the quasar sight lines  in our sample. The curves show the escape velocity as a function of the distance for stellar masses of $\log M_{*}=9$, $9.5$, $10$, $10.5$, $11$. There are two galaxies (1-166736 and 1-44487), for which the velocity offset of only \HI\ components exceeds the escape velocity at the corresponding distance: these components may be associated with unbound outflow or be formed from the IGM. At the same time, the other two cases of only \HI\ absorption  correspond to more massive galaxies (1-575668 and 1-113242, $M_*\sim10^{11} M_{\odot}$), where the gas is likely bound to the galaxies. 
The metal-bearing \HI\ absorbers appear to be bound  to the galaxies (only one such absorber, associated with the galaxy 1-180522, has a radial velocity very close to the escape velocity).}

%\st{Two galaxies, 1-44487, and 1-113242 show velocity offsets, with the velocity of the strongest \HI\ components about $-100$ and $+200$\,km s$^{-1}$, respectively. In both of these cases, the quasar sight lines are located at  $\sim23$ effective radii, that is about twice the distance of the first group, { where we detect a good agreement}. In the case of 1-44487 we find at least 4 absorption components with  velocities spanning a wide range of $\sim300$\,km s$^{-1}$. Since the galaxy is relatively far from the quasar and the absorption has a complex structure, relatively high $N({\rm HI})\sim10^{15}\,{\rm cm^{-2}}$ and super-solar metallicity (${\rm [X /H]= +0.8}$ ), we checked the area around the quasar for other galaxies with similar redshift, but found none. However, we note that this galaxy is merging with another smaller galaxy and it is likely that the observed high metallicity is due to outflows in a region of enhanced star formation caused  by the merger.} 

The most interesting case is that of the quasar-galaxy pair, J2130$-$0025 and 1$-$180522. The galaxy is observed with a high inclination angle of about 70$^{\circ}$ (nearly ``edge-on''{}), and the quasar sight line is located very close to the galactic plane (elevation angle is $\sim3\pm7$$^{\circ}$) at $\sim8.5$ effective radii. In this case we find very good consistency between the absorption velocity and the galaxy disk rotation velocity.
The sight line of J2130$-$0025 is also close to another galaxy, 1-635629, which has a similar redshift as 1-180522, but is located at a distance of about 39 effective radii. In fact, the velocity of the absorption system is opposite to the expected disk velocity for 1-635629. We therefore infer that the absorption system corresponds to only one galaxy, 1-180522, and that 
there is no detection for 1-635629. 

In two cases, 1-166736 and 1-575668, we also detected high velocity components with velocity offset $v\simeq-200$\,km s$^{-1}$ and +400\,km s$^{-1}$, respectively. Since these components have low \HI\ column densities $\sim10^{13.5}$ cm$^{-2}$, they are likely to be highly ionized clouds. In the case of 1-166736, the direction of the cloud velocity is consistent with the direction of the gas outflow, which can be detected with the { targeted quasar} sight line (see Fig.\,\ref{Final-1-166736}). Assuming that the galaxy has two cone-shaped outflows from the center in both directions along the polar axis, the quasar sight line can probe the outflow only in the direction to the observer (with negative radial velocity) and cannot probe the outflow in the opposite direction (with positive radial velocity), because this part of sight line is located far from the galaxy center, see the ``$Z-Y${"} projection in Fig.\,\ref{Final-1-166736}. 
In the second case, the galaxy 1-575668 is observed nearly ``face-on" with a small inclination $\sim8$$^{\circ}$ (see Fig.\,\ref{Final-1-575668}). 
%IS THE FAIRLY STRONG VELOCITY VARIATION (+150 to -150 KM/S) SEEN IN THE STELLAR RADIAL VELOCITY MAP CONSISTENT WITH THE GALAXY HAVING ONLY AN 8 DEG INCLINATION? { (reply :yes. The galactic disk rotation velocity of the best fit model is $\sim800$ km/s, 8 degree corresponds to $\sim110$  km/s.)}
The quasar sight line can probe both outflows, however the distance between the sight line and galaxy polar axis is lower from the side of positive velocity outflow. Therefore, the probability of detecting an absorption system with a positive radial velocity is higher, which is in line with observations.

{ Two other galaxies, 1-44487, and 1-113242 show velocity offsets, with the velocity of the strongest \HI\ components about $-100$ and $+200$\,km s$^{-1}$, respectively. In both of these cases, the quasar sight lines are located at  $\sim23$ effective radii, that is about twice the distance of the first group, { where we detect a good agreement}. In the case of 1-44487 we find at least 4 absorption components with  velocities spanning a wide range of $\sim300$\,km s$^{-1}$. Since the galaxy is relatively far from the quasar and the absorption has a complex structure, relatively high $N({\rm HI})\sim10^{15}\,{\rm cm^{-2}}$ and super-solar metallicity (${\rm [X /H]= +0.8}$ ), we checked the area around the quasar for other galaxies with similar redshift, but found none. However, we note that this galaxy is merging with another smaller galaxy and it is likely that the observed high metallicity is due to outflows in a region of enhanced star formation caused  by the merger.}

Summing up, we find consistency with gas co-rotation along with the galactic disk within at least 10 effective radii in most cases. The sign of the velocity of higher-velocity absorption, when detected, is consistent with the direction of the central galactic outflows, which have a higher probability of detection in these sight lines. For quasar sight lines at larger impact parameters, the situation is less  clear.

\subsubsection{%Galaxy sight lines
AGN sight lines}

We now discuss the gas kinematics for { three of the four} %\st{four} %\st{galaxy}
AGN sight lines (those with zero impact parameters) in our sample that show detections of absorption.  
We observed the AGNs of these galaxies %galactic centers 
at elevation angles of about 60$^{\circ}$. 
%\st{three of the four cases and $\sim20^{\circ}$ in the fourth case.}
{ These directions are within the outﬂow opening angles ($\theta_{\rm max}=30^{\circ}$ to 45$^{\circ}$) reported by \cite{Martin2012}, and therefore these sight lines can probe gas in the central outflows. Of course, the AGN sight lines can also probe gas in the CGM/IGM at a large distance, and these scenarios can be difficult to distinguish.

For two sight lines (1-71974 and 1-385099),  the \HI\ absorption lines are weak ($N({\rm HI})\sim10^{13}-10^{14}\,{\rm cm^{-2}}$) and blue-shifted by $-50$ and $-250\,{\rm km \, s^{-1}}$, and by $-50$ and $-750\,{\rm km \, s^{-1}}$, respectively, with respect to the galaxy redshift. The velocity of the low-velocity components is comparable to the galactic disk rotation velocity measured by MaNGA ($\sim100\,{\rm km \, s^{-1}}$). The high-velocity components in these sight lines can not be described by the galactic disk rotation model. The absorption in these components is characterized by high ionization and high metallicity, about two orders higher than measured for the absorption in the quasar sight lines. The high ionization could potentially arise in either outflowing gas ionized by the AGN radiation, or in low-density IGM gas. However, the high metallicity 
suggests that the outflow scenario is more likely, since the IGM is not expected to be metal-rich.}    

%Since we do not know the distance to galaxy sight line absorptions, we can not examine whether or not they is bound to the galaxies. The physical conditions of absorption systems differ a lot between three sight lines. In one case (12-192116) we probe neutral gas with high $N({\rm HI})\sim10^{19}\,{\rm cm^{-2}}$, low ionization ($q=-3.1$) and low metallicity (${\rm [X/H]\simeq0}$), In two other cases the absorptions have low $N({\rm HI})\sim10^{13}\,{\rm cm^{-2}}$, and they are likely probe high ionization $q\simeq-1$) and high metallicity (${\rm [X/H]\simeq1-2}$) gas.} 

%{ There are several scenarios that can describe the observations. High ionized and high metallicity gas may be related to (i) circum-nuclear outflow at low impact parameter, in this case high ionization is supported by the AGN radiation, (ii) central outflow at the galactic scale, (iii) gas is originated from the IGM. The first two scenarios are possible. Unfortunately, we can not estimate the distance to the absorption system by comparing the values of $q=n_{\gamma}/n_{\rm H}$ parameter in the absorption system and in the MaNGA galaxy, because we do not have any constraints on the gas density ($n_H$). The third scenario is unlikely, because absorption has a high metallicity. 

{ The third AGN sight line, 12-192116, probes gas with high  neutral hydrogen content ($N({\rm HI})\sim10^{20.2}\,{\rm cm^{-2}}$), low ionization and low metallicity ($[{\rm X/H}]\simeq-1$). This absorption can not be related to the galactic disk due to the high elevation angle, but may arise in a  ``satellite''{} galaxy, similar to what may be observed by extragalactic observers as absorption from the Magellanic Clouds toward  the center of the Milky Way. Alternatively, this absorption could arise in gas clouds tidally interacting with the galaxy, similar to high velocity clouds (HVCs) with $N({\rm HI})> 10^{20}\,{\rm cm^{-2}}$ \citep[e.g.,][]{Putman2002, Hsu2011}. }

%\st{For the first three cases we detect blue-shifted absorption relative to the galaxy redshifts, which is in line with the model of central gas outflow. These outflow directions are consistent with the outﬂow opening angles 30 to 45 degree reported by Martin2012.}%  In the fourth case, 594755, we did not find any absorption, which may be due to the sight line being far from the outflow opening angle.

%\st{For galaxies 1-71974 and 1-385099, we find that the absorption systems are highly ionized, with a low \HI\ column density and super-solar metallicity. However, the properties of the absorption system towards the galaxy 12-192116 are completely different. The absorption system has the highest \HI\ column density among our sample ($10^{19.3}\,{\rm cm^{-2}}$), a low metallicity (${\rm [X/H]\simeq0}$) and the lowest ionization parameter ($\log q=-3.3$). Therefore, this absorption system corresponds to a neutral gas cloud, perhaps in a satellite of the galaxy. We speculate that the observed absorption may arise in the  ``satellite''{} galaxy, similar to what may be observed by extragalactic observers as the absorption of the Magellanic Clouds towards the center of the Milky Way.}

We also note that in the spectrum of the AGN of 1-385099, we find additional absorption of \HI\ at a very high velocity $v\simeq-800$\,km s$^{-1}$. Since this galaxy is part of a group along with at least 2 other galaxies, 1-585207 and SDSS\,J083804.94$+$245327, with similar redshifts and projected distances of $\sim50$\,kpc from the observed sight line, the high-velocity cloud we detect may correspond to the intra-group gas perturbed due to the interaction of these galaxies. The SDSS image of this region shows the presence of long tidal tails for all galaxies (see, e.g., Fig.\,\ref{fig:quasar-gal-pairs2}). As an analog from the local universe, we note that absorption at such high velocities (much higher than the velocities associated with the Milky Way's halo gas or the Magellanic Stream) is observed in the Sculptor group galaxies \citep[e.g.,][]{Putman2003}. 
% Putman, M. E., et al. 2003, ApJ, 586, 170

\subsection{Metallicity gradient}
\label{sec:metgradient}
Combining the cool-gas metallicity along with the warm-gas metallicity is essential to build a complete census of metals in and around galaxies. While such comparisons of cool-gas metallicity and warm-gas metallicity have been
performed in integral field spectroscopic studies of quasar absorbers at higher redshifts \citep[e.g.,][]{Peroux2012,Peroux2014}, such comparisons have not been performed for the $z\sim0$ galaxies that have much more detailed information. 
Our study of the CGM of MaNGA galaxies offers an opportunity to study differences in metallicity in the inner vs. outer parts of galaxies in some of the closest venues available, and can thus provide fresh insights into 
processes affecting galaxy evolution. 

\begin{figure*}
\begin{center}
        \includegraphics[width=1\textwidth]{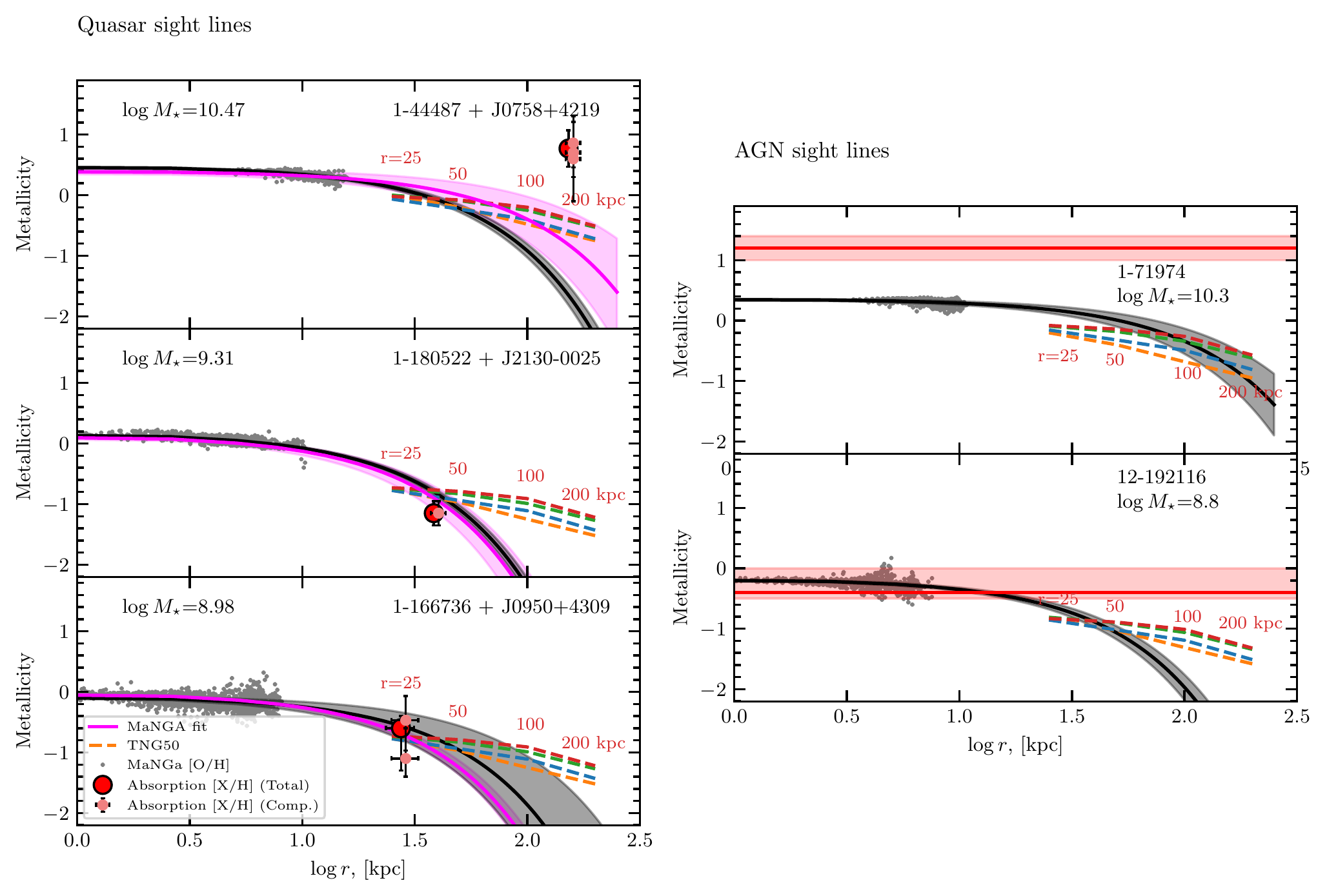}
        \caption{\rm The comparison of the gradient of IZI metallicity derived from fitting to MaNGA emission-line maps and metallicity measured in absorption systems. Left panels show the comparison for absorption systems in quasar sight lines with non-zero impact parameter. Right panels show the comparison for absorption systems along galaxy/AGN sight lines with zero impact parameter. Red and pink circles represent the average metallicity in absorption systems, { and the metallicity in individual components, respectively.} The black solid line and black shaded area in each panel show the linear gradient of IZI metallicity averaged over the elevation angle. The pink solid line and pink shaded area in left panels show the linear gradient of IZI metallicity in the direction to the quasar sight line. { The orange, blue, green and red dashed curves show the model distribution of metallicity in the CGM from the TNG50  simulation (see Fig.\,5 in \citealt{Peroux2020}), derived at different azimuthal angles 0, 30, 60, 90 deg, respectively, and four values of impact parameter (25, 50, 100, 200 kpc). The model is adopted to $z=0$ and the galaxy stellar mass.} 
        For two galaxies (1-180522 and 1-166736) quasar sight lines more likely probe gas near the galactic plane. The galaxy 1-44487 is an interacting galaxy, and this activity may result a high metallicity measured in the absorption system at high impact parameter. AGN sightlines likely probe a high-metallicity central outflow (1-71974) and gas in ”satellite” galaxy (12-192116), see text.}
        \label{fig:metgradient}
\end{center}
\end{figure*}

With this in mind, we study how the gradient of IZI metallicity derived from the fit to MaNGA emission-line maps within a few effective radii corresponds to the metallicity measured in the absorption systems along the targeted sight lines. 
{ Simulations predict a change in the metallicity gradient from an almost linear relation in the galactic disk \citep[e.g.][]{Mingozzi2020} to a  flatter behavior in the CGM \citep{Peroux2020}. Our sightlines probe the transition region between these two limits.  
}
Fig.\,\ref{fig:metgradient} shows the comparison for five galaxies, where we simultaneously measured  metallicity in the galaxy and the absorption system. For the absorption systems we %\st{use} 
show the average metallicity { and local metallicity in individual components} derived from fits with  
%\st{fitting to the total metal column densities by}
the {\sc cloudy} photoionization models (see section\,\ref{sec:met}). { For the studied systems, the average and local values are in good agreement with each other.} 
For absorption in quasar sight lines we use the deprojected radial coordinate of the absorption system in the disk plane ($d$) calculated in Section\,\ref{sec:azimodel}. 
For the galaxies, we calculate the gradient of the IZI metallicity { in the galactic disk} in two ways: (i) by averaging over all directions and (ii) by averaging over only the spaxels within $\pm15^{\circ}$ opening angle around the direction to the quasar sight line. The second way is possible only for quasar-galaxy pairs, when we have a preferred direction. The gradients are shown by grey and pink lines, respectively. 
In Figs.\,\ref{Final-1-166736}-\ref{Final-12-192116} we also present the IZI metallicity maps and fits to their radial profiles. The deprojected radial coordinate of each spaxel in the MaNGA maps was derived from the best fit model of the galaxy radial velocity map (see Section\,\ref{sec:mangavel}).
{ The model for the metallicity radial gradient in the CGM has been taken from  post-processing of the TNG50 galaxy simulation presented by \cite{Peroux2020} in their Fig.\,5. It represents measurements of the CGM metallicity in the TNG50 simulation at four values of impact parameter: $b = 25$, $50$, $100$ and $200$ kpc, and four values of azimuthal angle: 0$^{\circ}$, 30$^{\circ}$, 60$^{\circ}$ and 90$^{\circ}$. Other parameters were set to the appropriate values, which are: redshift $z=0$, stellar mass $M_{*}$ equal to stellar mass of the MaNGA\ galaxies. The metallicity is decreased by $0.4-0.7$\,dex  between 25 and 200 kpc in outflow (at 90$^{\circ}$) and galactic disk (at 0$^{\circ}$)  directions, respectively. It is greater than the metallicity gradient due to the elevation angle change, $0.1-0.4$\,dex at 25 and 200 kpc, respectively. \cite{Peroux2020} suggested that this is because fountains do not yet promote metal mixing over the full volume (i.e. range of elevation angles) at the distances $b\sim100$ kpc, as occurs closer to the galaxy.
%PLEASE ELABORATE MORE ON THE MODEL GRADIENTS PLOTTED IN FIG. 12. PEROUX20 SHOWS ONLY GRADIENT WITH ANGLE, NOT RADIAL GRADIENT. 
}
For the absorption systems in { the AGN} sight lines we { show the level of metallicity in the absorption system with the horizontal red line, since the distance of the absorbing region from the galaxy center is not known.}
%\st{set the radial coordinate equal to $1\pm1$ galaxy effective radius}.

%or the absorption systems, we set the radial coordinate equal to $1\pm1$ effective radius for absorption in galaxy sight lines, and calculate it from the azimuthal angle estimate for the quasar sight lines. 

Comparing the five panels of Fig.\,\ref{fig:metgradient}, we note that we confirm the increase of the galaxy central metallicity with the stellar mass, previously reported by \cite{Mingozzi2020} for star-forming galaxies from the MaNGA survey. 
Second, we detect a consistency of metallicity in the absorption system in the quasar sight lines with the prediction from the metallicity gradient for two galaxies (1-180522 and 1-166736). These quasar sight lines (J2130$-$0025 and J0950$+$4309, respectively) more likely probe gas near the galactic plane at $\sim7$ and $\sim8$ effective radii. In these cases,  the metallicity gradient measured from the fit to MaNGA\ maps (at $2-3$ effective radii) can persist over a longer distance.
{ The metallicities in these absorption systems are also consistent with the prediction for the CGM metallicity showing that the simulations may reproduce the CGM properties well.}
For the third galaxy, 1-44487, we measure about two orders of magnitude higher metallicity, than the predictions from the ``galaxy"{} gradient. { Since the impact parameter is high ($137$\,kpc), this quasar sight line should probe metallicity in the CGM, which is expected to be higher than that  predicted from the ``galaxy"{} gradient. For stellar mass $\sim10^{10.5} M_{\odot}$, the CGM metallicity 
is expected to be about $[{\rm X/H}]=-0.5$, which is still $\sim$1.5 orders of magnitude lower than the measured value.}  
This discrepancy may result from the activity of the merged galaxy. % in the first case, and a high-metallicity central outflow in the second case. and 12-192116). 
For the absorption systems located { toward}
%\st{near} 
the galactic centers of 1-71974 and 12-192116 galaxies, we found an excess of metallicity in the first case, which can be caused by  a high-metallicity central outflow, and  { one order lower metallicity in the second, which may be} 
%\st{the consistency with metallicity gradient in the second. For  12-192116, we probably measure}  
 the metallicity of the ``satellite"{} galaxy or an 
 HVC (see the discussion of this case in previous section). 
%Second, we detect a consistency of the absorption metallicity with the prediction from the metallicity gradient for three galaxies (1-180522, 1-166736 and 12-192116). For two of these galaxies (1-180522 and 1-166736), the quasar sight lines more likely probe gas near the galactic plane at $\sim7$ and $\sim8$ effective radii. In these cases, we find that the metallicity gradient measured from the fit to MaNGA maps can persist over a  longer distance. For the galaxy  12-192116, we probably measure the metallicity of the "sattelite"{} galaxy (see the discussion of this case in previous paragraph). 
%Third, for the remainingtwo cases, 1-44487 and 1-71974, we measure about two orders of magnitude higher metallicity, than the predictions from the "galaxy"{} gradients. These discrepancies may result from the activity of the merged galaxy in the first case, and a high-metallicity central outflow in the second case.

\subsection{Dependence of metallicity on the elevation angle}
\label{sec:metvsaziangle}

Hydrodynamic simulations of galaxy evolution predict an increase of CGM  metallicity with the elevation angle with respect to the disk plane \citep[see e.g.][]{Peroux2020}. Metal-free gas is expected to fall into the galaxy along directions close to the galactic disk plane, while metal-rich gas is expected to flows out of the galactic disk in directions near to the perpendicular to the plane by stellar winds and supernova explosions. Therefore, we can expect an increase of gas metallicity with the elevation angle (e.g. see Fig.\,5 in \citep{Peroux2020}). The angular gradient of metallicity is predicted to be around ${\rm +0.4\,dex/90^{\circ}}$ for galaxies at $z\simeq0$.

\begin{figure*}
\begin{center}
        \includegraphics[width=\textwidth]{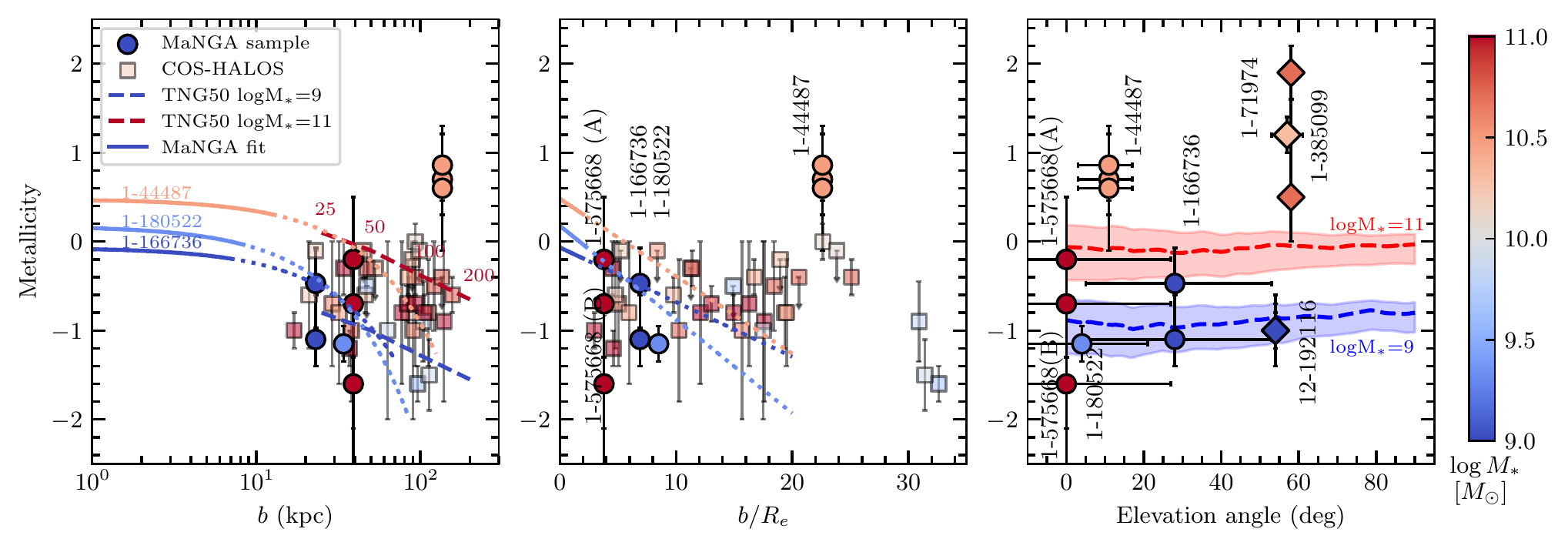}
        \caption{\rm The comparison of metallicity of absorption systems against the { impact parameter (left panel), impact parameter in units of effective radii (middle panel) and } elevation angle (right panel). Circles and diamonds represent { the metallicity in individual velocity components for the quasar and AGN absorbers, respectively, from our sample. Transparent squares show data from the COS-Halos survey \citep{Tumlinson2013}.} 
        %\st{elevation angles derived from the models of gas distribution (see text). Crosses indicate the elevation angles derived by the standard method.} 
        %{ The size of the symbols increases with \HI\ column density.}  
        The color of symbols denotes the  galaxy stellar mass. 
        { The solid and dotted curves in the left and middle panel show the metallicity gradient in MaNGA\ observations and its extrapolation to 20 effective radii (as in Fig.\,\ref{fig:metgradient}). The dashed lines in the left and right panels show the gradient of metallicity with distance (left) and with elevation angle (right) from  galaxy formation simulations  \citep{Peroux2020}.
        The radial gradient was derived at $z=0$, for an  elevation angle of 0 degrees, and for galaxy stellar masses of  $10^9\,M_{\odot}$ and $10^{11}\,M_{\odot}$.  The gradient with the elevation angle was derived at $z=0$, for $b = 25-100$\, kpc,  and for galaxy stellar masses of $10^9\,M_{\odot}$ and $10^{11}\,M_{\odot}$.
        }} %The cases 1-575668(A) and 1-575668(B) represent different fit results to \HI\ Ly$\alpha$ line profile in the quasar spectrum J\,1237$+$4447.}
        \label{fig:metazigradient}
\end{center}
\end{figure*}

Fig.\,\ref{fig:metazigradient} presents the dependence of metallicity in absorption systems on the elevation angle in our sample. For each system, we show { the metallicity measured in the velocity components to test the variation of physical conditions.} { We also show} the two estimates of the elevation angle discussed in Section\,\ref{sec:azimodel} ((a) from the standard method, and (b) from the model of gas distribution). 
{ We find that, if we consider the metallicity only in quasar absorption systems (apart from the case of the merged galaxy, 1-44487), we have a good agreement between the observed absorption metallicities and the simulations. The main difference in the measured metallicities is due to the difference in galaxy stellar masses, whereas the gradient with the elevation angle is small.
The metallicity measurements in the AGN sight lines should not follow the trend in the simulations, because they probably do not probe the CGM.}
%\st{We find that, if we analyze our quasar sight lines and galaxy sight lines (i.e. those with non-zero and zero impact parameters) together, the observed metallicity gradient is much stronger than that predicted by simulations. However, if we consider the metallicity only in quasar absorption systems (apart from the case of the merged galaxy, 1-44487), then we have a good agreement.}
The galaxy sight lines can probe gas at very low distances and in AGN central ejections, that could describe high metallicity and high ionization of these systems. However these processes were not considered in the simulations by \cite{Peroux2020, VandeVoort2021,Wendt2021}, meaning no AGN at low $M_{\star}$ in simulations. %, which would explain the difference. 
%\st{The case of galaxy 12-192116 is also unusual, because while it has similar elevation angle as other galaxy sight lines, its metallicity is found to be two orders of magnitude lower and consistent with the galaxy metallicity gradient (see Fig. fig:metgradient).} 
{ At the same time, the galaxy 12-192116 shows  good consistency. 
}
In this case, we may be dealing with a galactic absorption system with an unusually large elevation angle, and we suggest that it may be caused by absorption from a ``satellite''{} galaxy, or from metal-poor inflowing gas.

{ We also compare our results with metallicity measurements in the COS-Halos  survey \citep{Werk2014}. The effective radii of the COS-Halos galaxies were derived from the relation between the effective radius and the stellar mass by \cite{Mowla2019}. We do not detect a significant correlation between $[{\rm X/H}]$ and $b/R_e$ for the joint sample, with a Spearman rank order correlation coefficient of 0.1 and a $p$-value of $0.46$. }

\subsection{Ionization parameter}

\begin{figure*}
\begin{center}
        \includegraphics[width=1\textwidth]{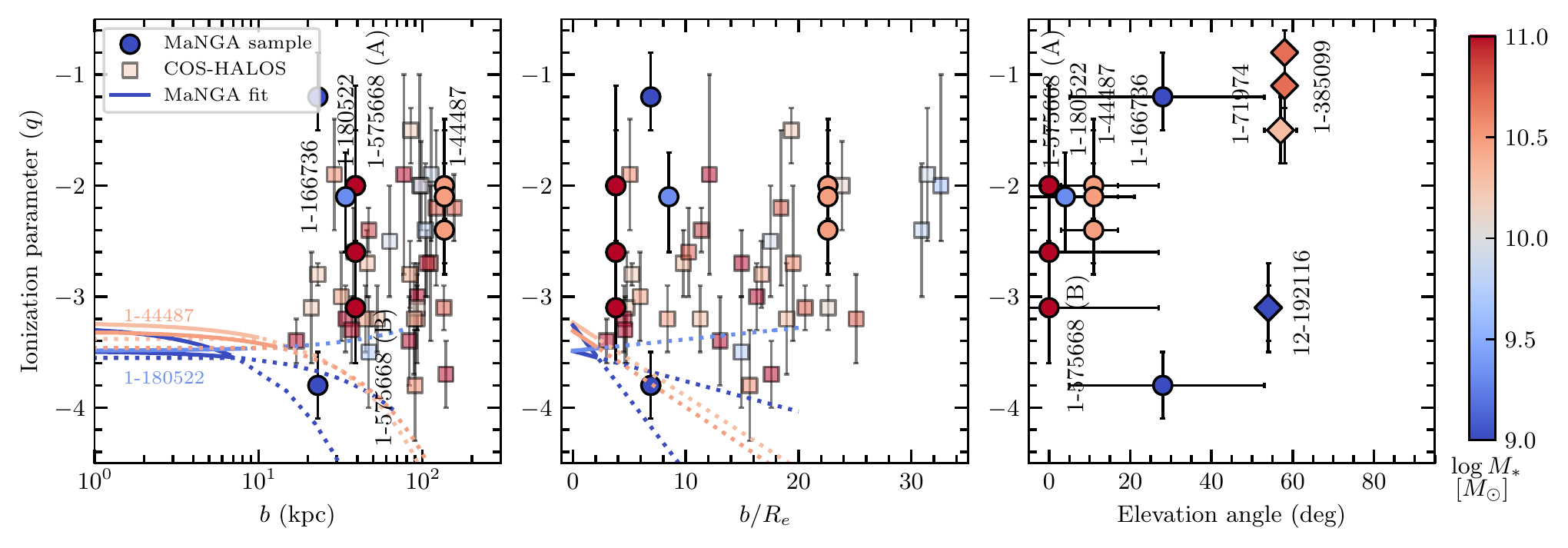}
        \caption{\rm The comparison of gas ionization parameter of absorption systems against the impact parameter (left panel), impact parameter to effective radius (middle panel) and the elevation angle (right panel). The symbols are the same as in Fig.\,\ref{fig:metazigradient}: circles and diamonds - our sample, transparent squares - from COS-Halos survey. The color of the symbols denotes the galaxy stellar mass. 
        { The solid and dotted curves in the left and middle panel show the ionization parameter gradient in MaNGA\ observations and its approximation to 20 effective radii (as in Fig.\,\ref{fig:metgradient}).}}
        \label{fig:q-azi-gradient}
\end{center}
\end{figure*}

Fig.\,\ref{fig:q-azi-gradient} shows the ionization parameter vs. impact parameter and elevation angle for our sample galaxies. We find the ionization parameter to be roughly constant for MaNGA\ galaxies with a mean of $\sim$$10^{-3.3}$ and the dispersion of $\sim$0.1\,dex (see Table\,\ref{tab:hst_results}). On the contrary,  the ionization parameter in the absorption systems spans over two orders of magnitude { above the average galactic ionization}. The difference maybe primarily due to the difference of gas number density between the ISM ($\sim10^2$\,cm$^{-3}$, \citealt[e.g.][]{Mingozzi2020}) and the CGM ($\sim10^{-1}-10^{-3}$\,cm$^{-3}$). The ionization parameter $q=n_{\gamma}/n_{\rm H}$ is proportional to the ratio of  $I_{\rm UV}/n_{\rm H}$. { Assuming the ionization of the CGM is due to the extragalactic background only (whose intensity is about $10^{-2}$ of the average UV galactic radiation), the difference in $q$ parameter is obtained from $(n_{\rm H}^{\rm CGM})^{-1}$, which gives a factor $10-10^{3}$ for the range of the CGM number densities. The factor will be lower if the galactic UV intensity is stronger than the average UV galactic radiation.}

We did not find much correlation of the ionization parameter of the absorption systems with galaxy properties such as stellar mass, SFR, or sSFR. However, it correlates with the %impact parameter and the 
elevation angle (see { the right panel of} Fig.\,\ref{fig:q-azi-gradient}). 
%THIS PARAGRAPH SEEMED  CONTRADICTORY WITH FOLLOWING PARAGRAPH. THE TEXT SOUNDED LIKE WE ARE SAYING THAT THERE IS A TREND WITH IMPACT PARAMETER IF WE CONSIDER ONLY OUR SAMPLE, BUT NOT AFTER COMBINING OUR SAMPLE WITH COS-Halos. I HAVE CHANGED IT TO SAY JUST ELEVATION ANGLE. 
%{ reply: I agree}

As we showed above quasar sight lines probe gas around galaxies at lower elevation angles and at higher impact parameters, than AGN sight lines (1-385099 and 1-71974), and for { the former,} %them 
we found lower ionization parameters. This is consistent with the picture that gas at small elevation angles corresponds to galactic disks (or inflowing gas) and has lower metallicity and lower ionization parameters than the gas observed in absorption at larger elevation angles (which may arise in outflows with higher ionization fractions and higher metallicity). { For the quasar sight line J0950$+$4309, which probes the gas around the galaxy 1-166736 at a moderate elevation angle, we found a large difference in the ionization parameter between the components seen in the \SiII\ and \SiIII\ absorption lines. The low-ionization gas traced by \SiII\ may correspond to the galactic disk, while the highly ionized gas observed only in \SiIII\ may correspond to the CGM. The higher ionization component in 1-166736 is consistent with the trend observed between the ionization parameter and the elevation angle.}

We also compare our data with measurements of the ionization parameter in the COS-Halos survey \citep{Werk2014}. Our results are consistent with %those reported by \citep{Werk2014} 
their results and cover the same range of ionization parameter. However, we do not confirm the  trend $\log q=-2.2\pm0.3 + (0.8\pm0.3)\times\log (R/R_{\rm vir})$ reported by \citep{Werk2014} (based on the points at very low and high impact parameters).
%The authors also found  the trend $\log q=-2.2\pm0.3 + (0.8\pm0.3)\times\log (R/R_{\rm vir})$, which is actually determined by the points at very low and high impact parameters. 
%For the joint sample of our data and the COS-HALOS survey, 
Combining our sample with the COS-Halos sample, 
we find that the correlation of ionization parameter with impact parameter is not statistically significant with a Spearman rank order correlation coefﬁcient of $0.2$ and $p$-value $0.16$. Indeed, high ionization parameters, on average, correspond to a higher impact parameter. However,  there is a large scatter of $q$ parameters which likely reflects the inhomogeneity of physical conditions in the CGM of different galaxies. 

%\st{Also for quasar sight lines we detect the increase of the ionization parameter with impact parameter. Assuming that the absorption system towards the galaxy 12-192116 probes a low-ionization gas in a nearby satellite galaxy at low impact parameter (see discussion in the previous section), we can consider this system together with quasar absorptions and find a strong correlation of ionization parameter with impact parameter with a Spearman rank order correlation  coefﬁcient of 1.0 and $p$-value $<10^{-5}$. Since the mean values of ionization parameter in galaxies are nearly the same, about $10^{-3.4}$, this correlation shows how  the ionization parameter decreases with the distance of the galaxy center.}

\section{Conclusions}
\label{sec:conclusions}
We have measured the CGM properties out to 25 effective radii using HST/COS spectroscopy of quasars and AGNs about a sample of low-redshift galaxies with exquisite data from the MaNGA survey. We detected the associated absorption for 11 of 14 galaxies in our sample.
{ In three cases, the absorption was detected in the sight lines toward the bright source near the galactic center, in other cases the absorption was detected in background quasar sight lines at impact parameter from 23 to 137 kpc. For the AGN sight lines we detected a strong \HI\ absorption ($N({\rm HI})\simeq10^{20.2}\,{\rm cm^{-2}}$) only in one case, in two other cases we found weak \HI\ absorption  ($N({\rm HI})\simeq10^{13}\,{\rm cm^{-2}}$) which may be related to high-metallicity and high-ionization gas in the central outflow. Our quasar sight lines show \HI\ absorption with a wide range of $N({\rm HI})\simeq10^{13}-10^{19}\,{\rm cm^{-2}}$.}
%\st{In two cases the absorptions are associated with central outflows, in one galaxy we probably detect the absorption of another satellite galaxy, in others the absorptions correspond to gas located close to galactic disks at high impact parameters.}

To summarize, our main results are as follows:

\begin{enumerate}
    \item
    { 
     The \HI\ column density vs. impact parameter measurements for quasar sight lines  correspond generally well with the radial \HI\ column density profile predicted from galaxy simulations \citep{VandeVoort2019, Nelson2020}. 
     \item 
     Our data also agree well with other spectroscopic studies of halos of galaxies at low redshift $z<0.3$ (COS-Halos by \citealt{Tumlinson2013}, COS-Weak by \citealt{Muzahid2018}) and of the gas in ``galaxies on top of quasars''  \citep[in the close vicinity of low-$z$ galaxies within impact parameters $\sim1-7$ kpc,][]{Kulkarni2022}. 
 %    Unlike previous studies  our sample is not biased by the galaxy stellar mass (as \citealt{Tumlinson2013}) or by high \HI\ column density of the associated absorptions (as \cite{Kulkarni20w2}). It results in more uniform $N({\rm HI})$ distribution than in previous studies. }
     
    \item{ We confirm the anticorrelation between the \HI\ column density and the galaxy stellar mass 
%    (high \HI\ column densities are more likely related to low stellar mass galaxies) that was 
    that was previously reported by \cite{Kulkarni2022}.    
    } 
    \item{We report a strong dependence of $N({\rm HI})$ decreasing with increasing $D_n{(4000)}$ index of the host galaxies for quasar sight lines, that may be a result of past star-formation activity having consumed or blown out cool gas from the CGM.}
    %\item The derived \HI\ column densities and ionization-corrected metallicities cover a large range of values from $N({\rm HI})=10^{13}$ to $10^{19}$ and from $[{X/H}]=-1.5$ to $\simeq +2$. We compared our data with other spectroscopic studies of halo of galaxies at low redshift $z<0.3$ and found good consistency for the dependence of $N({\rm HI})$ on the impact parameter and stellar mass. 
    }
    \item A comparison of absorption velocities with radial velocity maps of ionized gas line emission in galaxies shows { consistency with corotation of the strong \HI-absorption component with the disk out to $\sim$10 effective radii (within $\pm50\,{\rm km\,s^{-1}}$) and $\sim25$ effective radii (within $\pm1$ galactic disk rotational velocity).
    The components with only \HI\ absorption (without associated metal lines) are likely to have a high velocity shift and in some cases may be unbound to the galaxy.}  
    %\item
    %The sign of the velocity of the weaker, high-velocity \HI\ components is consistent with the direction of the central galactic outflow, which can be probed by the sight lines. 
    \item Comparing the observed CGM properties with the galaxy properties from MaNGA maps, we estimate the gradients in metallicity and ionization parameters. 
    { The measurements of absorption metallicities in individual quasar sight lines correspond well with the gradient of metallicity in the galactic disk derived from MaNGA\ observations. 
    %For these absorption systems we also detect a good agreement between the velocity of the absorption system and and galactic disk rotation.
    Overall, from our sample and previous studies we find a lower metallicity in the quasar sight lines with respect to the AGN sight lines. The difference is consistent with the predictions of the CGM metallicity from TNG50 simulations \citep{Peroux2020}.  
    
    The ionization parameter in absorption systems is on average  one order of magnitude higher than the galactic value ($q\sim10^{-3.5}$). The measurements in our sample and previous studies do not show a statistically significant gradient of the ionization parameter with distance from the galaxy. This indicates a strong inhomogeneity of the physical conditions in the CGM (number density and intensity of \HI-ionized radiation). However, the data are consistent with an increasing ionization parameter with increasing elevation angle.
    }
\end{enumerate}

Our data offer the first detailed comparisons of CGM properties with extrapolations of detailed galaxy maps. While our data offer a number of interesting insights into the exchange of gas and metals between galaxies and their CGM, our current sample is still small. Observations of the CGM of many more galaxies mapped with integral field spectroscopy are essential to more fully understand how galaxies interact with their CGM. 

\section{Data availability}

Data directly related to this publication and its figures can be requested
from the authors. The MaNGA data used in this paper can be downloaded from the MaNGA public archives. The {\it HST} data used in this paper can be found in MAST: \dataset[10.17909/zpy3-w565]{http://dx.doi.org/10.17909/zpy3-w565}.

\section*{Acknowledgements}
This work is supported by a grant from the Space Telescope
Science Institute for GO program 16242 (PI V. Kulkarni). Additional partial support is also gratefully acknowledged from the US National Science Foundation grant AST/2007538 and NASA grant 80NSSC20K0887 (PI V. Kulkarni).

We would like to thank Francesco Belfiore and Kyle Westfall for helpful advice on analysing the the MaNGA emission line ratios. We thank Sergei Balashev for sharing a version of the {\sc Spectro} Voigt profile fitting code, before its official release.
{ We are grateful to an anonymous referee for careful reading and constructive  suggestions that have helped to improve this paper.}

Funding for the Sloan Digital Sky Survey IV has been provided by the Alfred P. Sloan Foundation, the U.S. Department of Energy Office of Science, and the Participating Institutions. 
SDSS-IV acknowledges support and resources from the Center for High Performance Computing  at the University of Utah. The SDSS website is www.sdss4.org. 

SDSS-IV is managed by the Astrophysical Research Consortium for the Participating Institutions 
of the SDSS Collaboration including the Brazilian Participation Group, the Carnegie Institution for Science, Carnegie Mellon University, Center for Astrophysics | Harvard \& Smithsonian, the Chilean Participation Group, the French Participation Group, Instituto de Astrof\'isica de Canarias, The Johns Hopkins University, Kavli Institute for the Physics and Mathematics of the Universe (IPMU) / University of Tokyo, the Korean Participation Group, Lawrence Berkeley National Laboratory, Leibniz Institut f\"ur Astrophysik Potsdam (AIP),  Max-Planck-Institut 
f\"ur Astronomie (MPIA Heidelberg), Max-Planck-Institut f\"ur Astrophysik (MPA Garching), 
Max-Planck-Institut f\"ur Extraterrestrische Physik (MPE), National Astronomical Observatories of China, New Mexico State University, New York University, University of 
Notre Dame, Observat\'ario Nacional / MCTI, The Ohio State 
University, Pennsylvania State 
University, Shanghai Astronomical Observatory, United Kingdom Participation Group, 
Universidad Nacional Aut\'onoma de M\'exico, University of Arizona, 
University of Colorado Boulder, University of Oxford, University of 
Portsmouth, University of Utah, University of Virginia, University 
of Washington, University of Wisconsin, Vanderbilt University, 
and Yale University, and the Collaboration Overview Start Guide Affiliate Institutions Key People in SDSS Collaboration Council Committee on Inclusiveness Architects SDSS-IV Survey Science Teams and Working Groups Code of Conduct Publication Policy How to Cite SDSS External Collaborator Policy.

\newpage
\appendix

\section{Flux uncertainty estimate}
\label{app:A}
The flux errors of the HST/COS spectra originates in three sources: the errors associated with flat-field response, the Poisson error in the counts from the object flux (galaxy/quasar), and the Poisson error in the counts from the background flux \citep[e.g., ][]{COS-2021-03-v1}. In our case the first and third contributions are much smaller then the second. However, for faint sources, the object counts are low. Therefore, the problem is to estimate the upper and lower flux errors for the Poisson distribution in the case of a small number of counts.  
The standard \calcos pipeline uses an asymmetric uncertainty based on the frequentist-confidence method \citep[see][]{Gehrels1986} and described by
\begin{equation}
    \sigma_{N;{\rm upper}}=1+ \sqrt{N+\frac{3}{4}}
\end{equation}
and
\begin{equation}
    \sigma_{N;{\rm lower}}=N - \left[N\left(1-\frac{1}{9N}-\frac{1}{3\sqrt{N}}\right)^{3}\right],
\end{equation}
where $N$ is the number of observed counts. We found that for low count numbers ($<10$), these uncertainties are overestimated by the standard  \calcos pipeline. Therefore we reevaluated the uncertainties by the maximum probability  estimate (MPE) method.

\begin{figure}[!tbp]
    \begin{center}
        \includegraphics[width=0.5\linewidth]{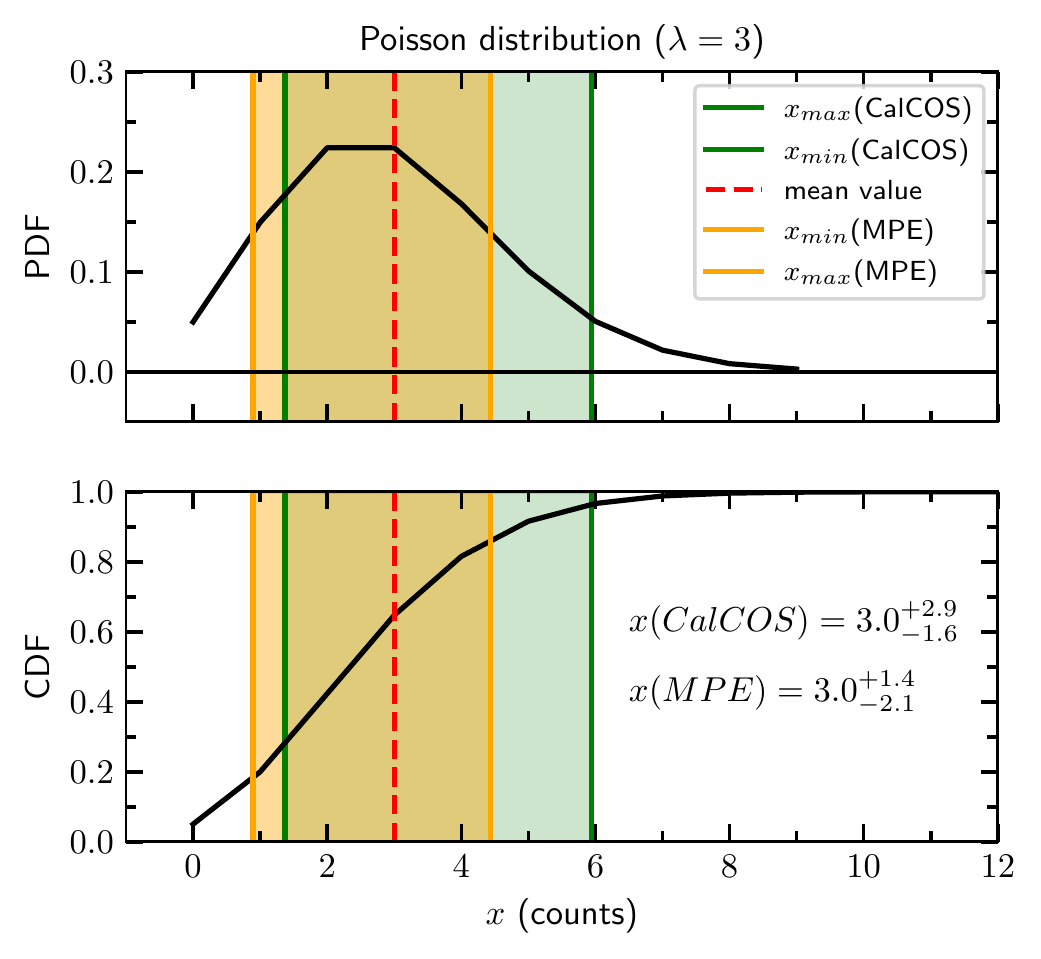}\hfill  
        \caption{\rm The comparison of the estimates of positive and negative uncertainties for the Poisson distribution in the case of a low number of counts %($\lambda=3$)
        ($N=3$). 
        The top and bottom panels show the probability distribution function (PDF) and the cumulative distribution function (CDF), respectively. the vertical red dashed lines show the mean value. The green lines and green dashed area represent the confidence interval derived by the \calcos pipeline. The yellow lines and yellow dashed area represent the confidence interval derived by the maximum probability estimate (MPE) method.}
        \label{fig:calcos-err1}
    \end{center}
\end{figure}

\begin{figure}[!tbp]
    \begin{center}
        \includegraphics[width=0.5\linewidth]{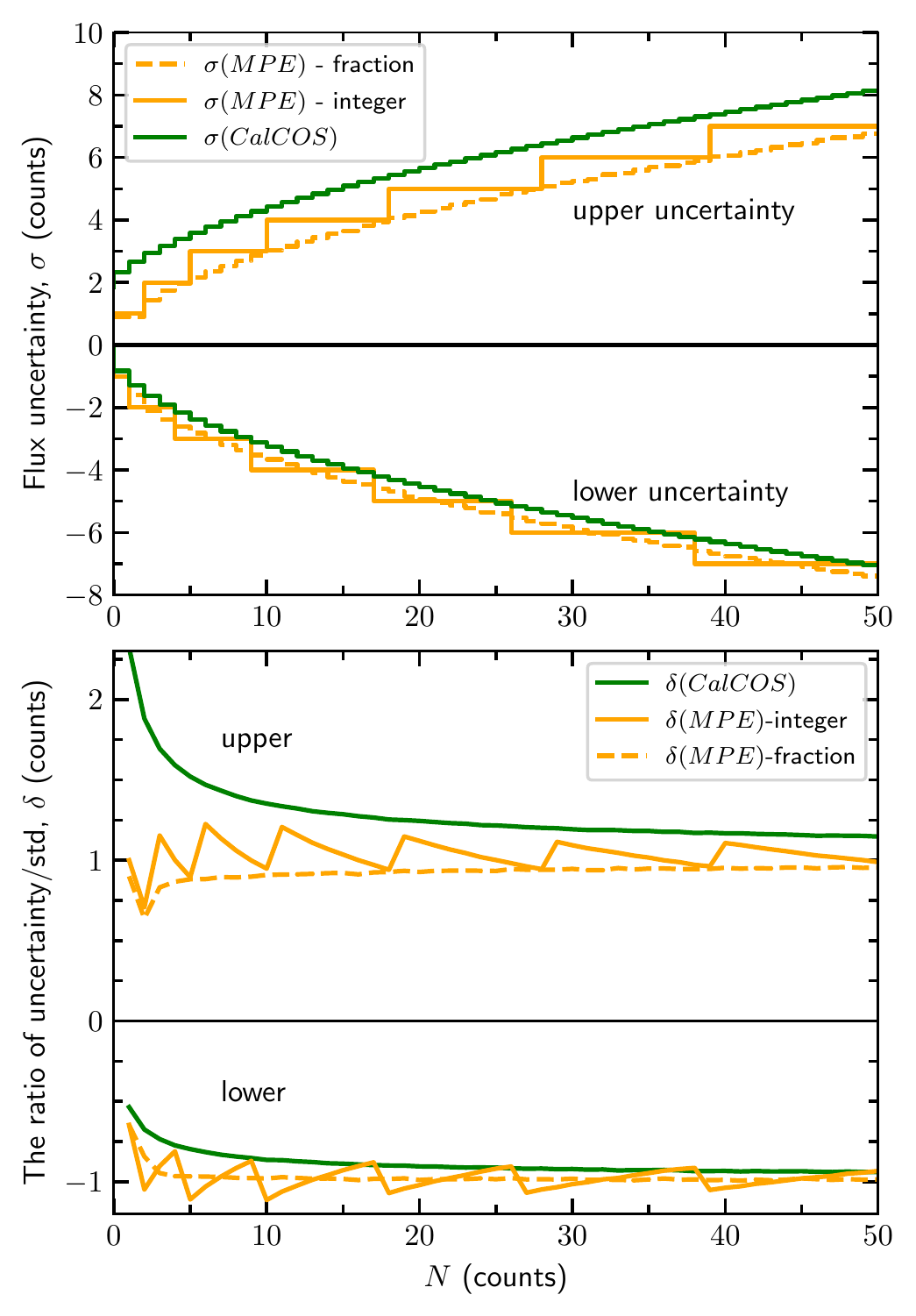}\hfill  
        \caption{\rm Top panel shows the comparison of  of positive and negative uncertainties for the Poisson distribution for different values of $N$. Dashed and solid yellow curves show our estimate, which is represented by a fraction in an integer number of samples.
        Green line represents the \calcos estimate. 
        Bottom panel shows the comparison of upper and lower estimates with the standard deviations for the Poisson distribution, calculated independently for the upper and lower outliers.}
        \label{fig:calcos-err2}
    \end{center}
\end{figure}

A comparison of the uncertainties estimated using the two methods is shown for the case $N=3$ in Fig.\,\ref{fig:calcos-err1}. The lower uncertainties are similar, while the upper uncertainties derived by the MPE method is about two times lower than ones derived by the frequentist-confidence method. We note that both estimates correspond to 68\% confidence interval (1$\sigma$), however the frequentist-confidence estimate is shifted to higher values.   
The relative difference between the uncertainty estimates decreases with an increase in the number of counts ($N$), and is small for $N>20$. The top panel of Fig.\,\ref{fig:calcos-err2} shows the upper and lower uncertainty estimates from the two methods. We adopt the uncertainty estimates in integral number of counts, since this seems more physical. The bottom panel shows the comparison of the ratio of uncertainties to the standard deviation for the Poisson distribution. The standard deviation was calculated independently for fluxes above and below the mean value ($N$). The estimates correspond well to the standard deviation over the entire range of $N$.

\section{ABSORPTION-LINE ANALYSIS DETAILS}
\label{app:B}

%\section{ABSORPTION-LINE ANALYSIS DETAILS}
%\label{app:C}

In this section we present the detailed results of analysis of each absorption system individually. Fig.\,\ref{fig:PDF} presents the posterior PDF of fitting parameters for systems shown in
Fig. 3. Figs.\,\ref{Final-1-166736}-\ref{Final-1-113242} show results for quasar-galaxy pairs,  Figs.\,\ref{fig:Final-1-71974}-\,\ref{Final-12-192116} show results for AGNs. 
In these figures we compare radial velocity, metallicity and ionization parameter derived from the fitting to MaNGA maps of the galaxies and ones measured from fitting to absorption lines in the absorption system in our HST/COS spectra. Below we describe the panels in these figures.

The panels in the top row show the HST COS data for the \HI, \SiIII, \SiII, \CII\ absorption lines and our best fits to these lines. The synthetic profile is shown in red and the contribution from each component is shown in green, blue, purple and orange.

The panels in the second row show the comparison of radial velocities. In each case, the left panel shows the MaNGA velocity field (determined from the H$_{\alpha}$-line emission or the stellar continuum). The black line represents the positional angle ($PA$), the pink line and pink shaded area represent the the direction to the quasar sight line within opening angle 15$^{\circ}$, the black cross represents the position of the center of the disk, and the orange cross represents the position of the AGN (only for AGN sight lines). The second panel shows the galaxy rotation velocity curve, reconstructed using the best fit to the radial velocity map. The circles show measurements from the MaNGA spaxels, and the pink line shows the model. The right panel compares the model of radial velocity in the direction to the quasar ($QA$) and the velocities of the absorption components. The dashed vertical line represents the value of the impact parameter. 

The panels in the third row show the orientation of the quasar sight line with respect to the disk plane. The left panel  shows a 3d plot: quasar sight line is shown by the black line (with the black star denoting the quasar), the observer is located at the top of the panel. The color of points in the galactic disk corresponds to the value of the radial velocity measured by the observer (same as in the MaNGA velocity map). The pink shaded area shows the range of the elevation angles corresponding to our probability estimate of the position of the absorption system along the quasar sight line (see Section\,\ref{sec:azimodel}). The dashed and solid pink lines represent the maximum probability value ($\overline{\phi}$) and its $1\,\sigma$ uncertainty. The middle and right panels show the $Y-Z$ and $X-Y$ projections of the 3-d plot, respectively. 

The panels in the fourth row show the comparison of the metallicity of the ionized gas measured from the IZI modelling of MaNGA maps of emission lines and the metallicity of the cool gas from the {\sc CLOUDY} fitting to metal column densities in the absorption system. The left panel shows the MaNGA maps, the lines and symbols are the same as in the first panel in the second row (radial velocity map). The middle panel shows the radial profile of the IZI metallicity (circles with errorbars) and the best fit to IZI metallicity gradients by a linear model (black line). The pink line corresponds to the IZI metallicity gradients in the direction to the quasar ($QA$) within a 15$^{\circ}$ opening angle. Shaded areas represent $1\sigma$ uncertainty. The right panel show the comparison of the IZI metallicity gradient with metallicity measured in the absorption system (small circles represent values for individual components, red circle shows the total value).

The panels in the fifth row are similar to the panels in the fourth row, but for the ionization parameter.

Figs.\,\ref{fig:fit:j0950-lines}-\ref{fig:fit:j1629-lines} show the best fits to absorption lines, and Fig.\,\ref{fig:_hist} shows the comparison of measured total column densities of metals with the values predicted by {\sc cloudy} photo-ionization models. 

\subsection{Comments to fit to absorption systems}

\subsubsection{J0755+3911}

We measured the \HI\ and metal species column densities of the absorption system at the redshift of the AGN 1-71974 ($z_{\rm gal}=0.0336$). The absorption system consists of at least four velocity components detected in \HI\ Ly$\alpha$ and \SiIII\ 1206\AA\ transitions. The \HI\ absorption consists of two weak components with column densities $\sim10^{12.8}$ and $10^{13.3}$ cm$^{-2}$, which are   blue-shifted by $-62$ and $-228$ km/s with respect to the galaxy redshift. The \SiIII\ 1206 absorption is detected in two components at $-130$ and $-200$ km/s, which are shifted with respect to \HI\ absorption lines. We fitted the absorption system with four velocity components with the redshifts tied to the redshift of the prominent \HI\ and \SiIII\ absorptions. The result of the fit is given in Table\,\ref{tab:detailedfit} and 
line profiles are shown in Fig.\,\ref{fig:fit:j0755-lines}.

\subsubsection{J0758+4219}
We detected the absorption system associated with the CGM of the galaxy 1-44487, which consists of four velocity components detected in \HI, \SiII, \SiIII, \SII\ absorption lines. The velocity components of metal absorption lines well correspond to the position of the \HI\ velocity components. Therefore we fitted this system with four components. One of the \HI\ components is blended with the Milky Way \SII\ 1253\AA\ absorption line, and we fitted the MW \SII\ absorptions consistently with the fit to the galaxy absorption lines. The result of the fit is given in Table\,\ref{tab:detailedfit} and 
line profiles are shown in Fig.\,\ref{fig:fit:j0758-lines}.

\subsubsection{J0838+2453}
In the spectrum of the AGN J0838+2453
we detected two \HI\ absorptions at  $-34$ and $-750$ km/s with respect to the redshift of the host galaxy. Because of low signal to noise ratio for this spectrum, the associated metal absorption lines are detected only in \SiIII\ 1206\,\AA\ transition, which is located close to the \HI\ Ly$\alpha$ quasar emission line. 
The \SiIII\ absorption has five velocity components. Two velocity components well correspond to the position of the \HI\ absorption components, whereas other three component are detected only in \SiIII. Therefore we fitted the absorption system with five velocity components. The result of the fit is given in Table\,\ref{tab:detailedfit} and 
line profiles are shown in Fig.\,\ref{fig:fit:j0838-lines}. 

\subsubsection{J0950+4309}
The absorption system at the redshift of the galaxy 1-166736 is well detected in \HI, \SiII, \SiIII, \CII\ transitions. The \HI\ absorption line is saturated, therefore we derived the velocity structure from the fit to metal absorption lines. We found that \SiII and \CII\ absorptions can be well fitted by one component which is red-shifted relative to the center of the strong \SiIII\ absorption line. The additional blue component seen in only the \SiIII\ absorption line also has a higher Doppler parameter, that indicates the difference in physical conditions between two components. 
The redshift of blue component has been chosen to well fit the blue wing of \HI\ Ly$\alpha$ absorption.
Additionally we detected the third weak component at $-247$ km/s with respect to the galaxy redshift, which is detected only in \HI\ profile.  The result of the fit is given in Table\,\ref{tab:detailedfit} and 
line profiles are shown in Fig.\,\ref{fig:fit:j0950-lines}.

\subsubsection{J1237+4447}
\begin{figure*}[ht]
\begin{center}
        \includegraphics[width=1\textwidth]
        {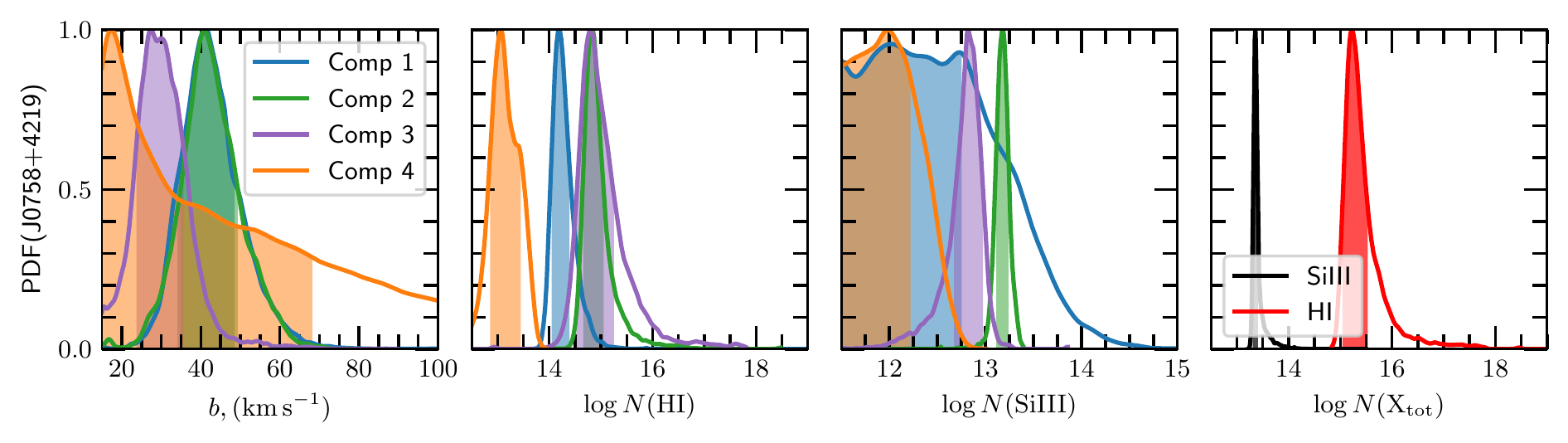}\\
        \vspace{-0.2cm}
        \includegraphics[width=1\textwidth]{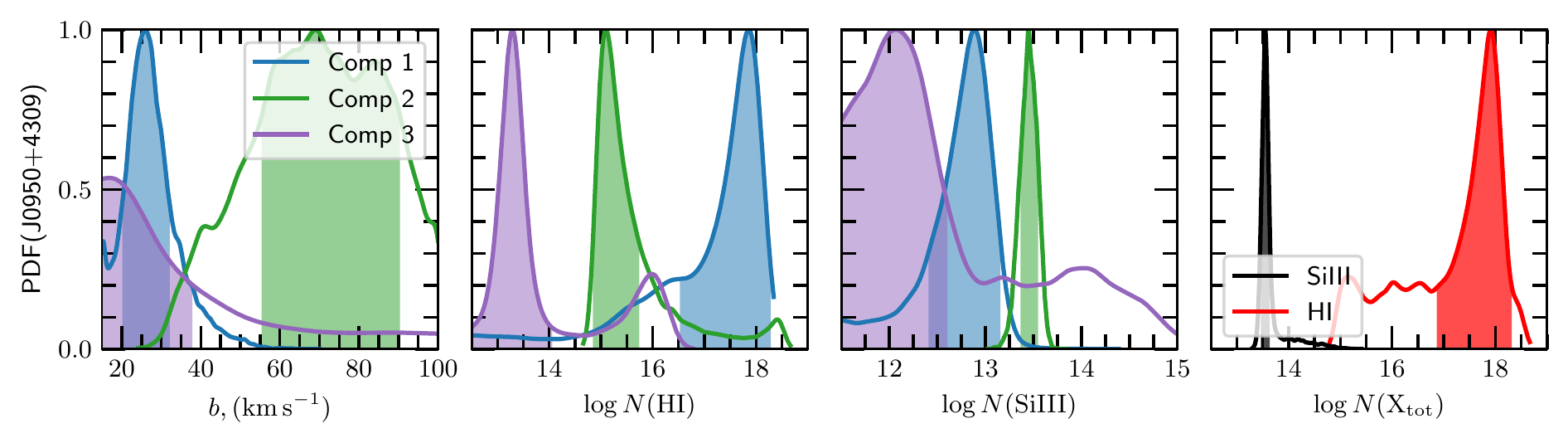}\\
        \vspace{-0.2cm}
        \includegraphics[width=1\textwidth]{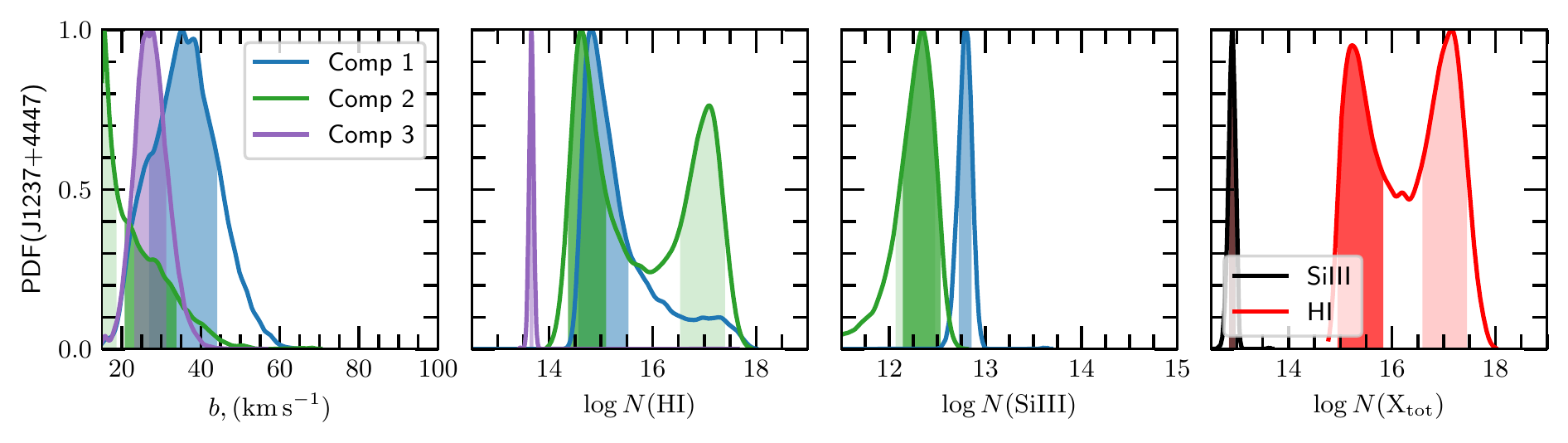}\\
        \vspace{-0.2cm}
        \includegraphics[width=1\textwidth]{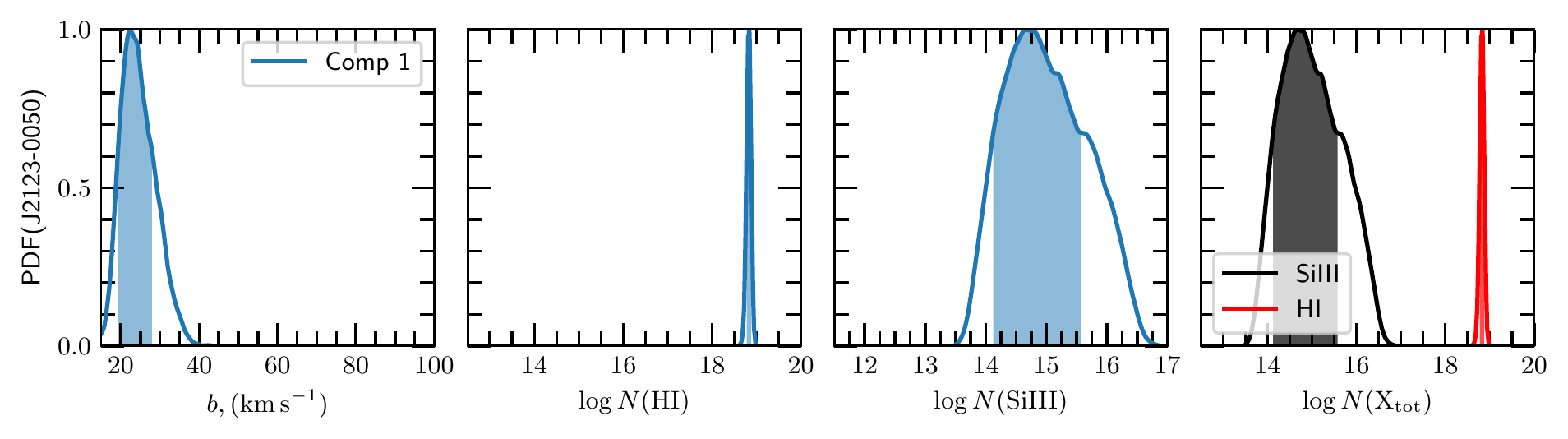}
        \caption{\rm  The posterior probability density function (PDF) of fitting parameters of \HI\ and \SiIII\ lines in four HST COS spectra shown in Fig.\,\ref{example:hst-data}. The first three columns from left to right show the PDF of $b$-parameter, $\log N({\rm HI})$,  $\log N({\rm SiIII})$. The color of curves correspond to the color of velocity components  in Fig.\,\ref{example:hst-data}. Dashed area represents the 68\% confidence interval around the value with the maximum probability. For the case of J1237$+$4447 the light and heavy dashed areas show two different solutions (see text). The rightmost column shows the PDF of the total \HI\ column density and the total \SiIII\ column density (summed over the components) by red and black curves, respectively.  
        }
        \label{fig:PDF}
\end{center}
\end{figure*}

\begin{figure*}
\begin{center}
        \includegraphics[width=1\textwidth]{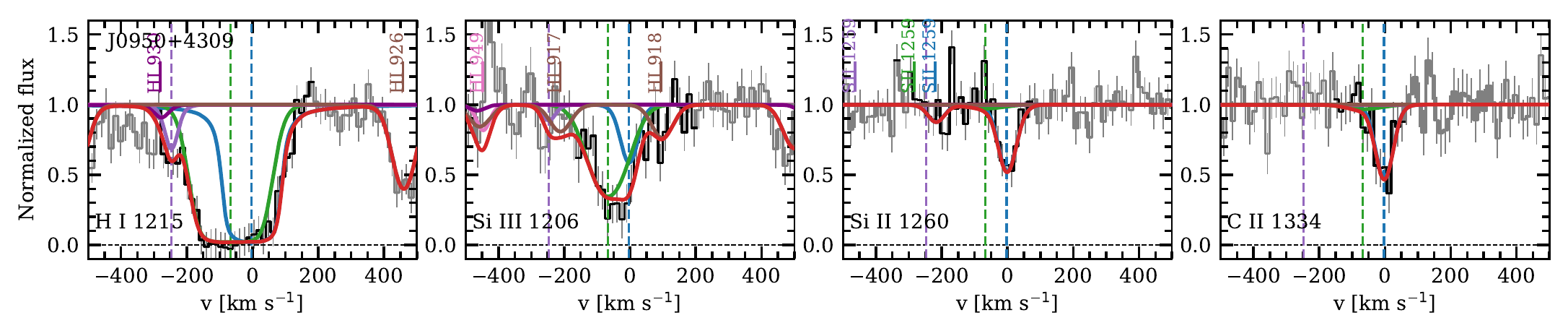}
        \includegraphics[width=0.97\textwidth]{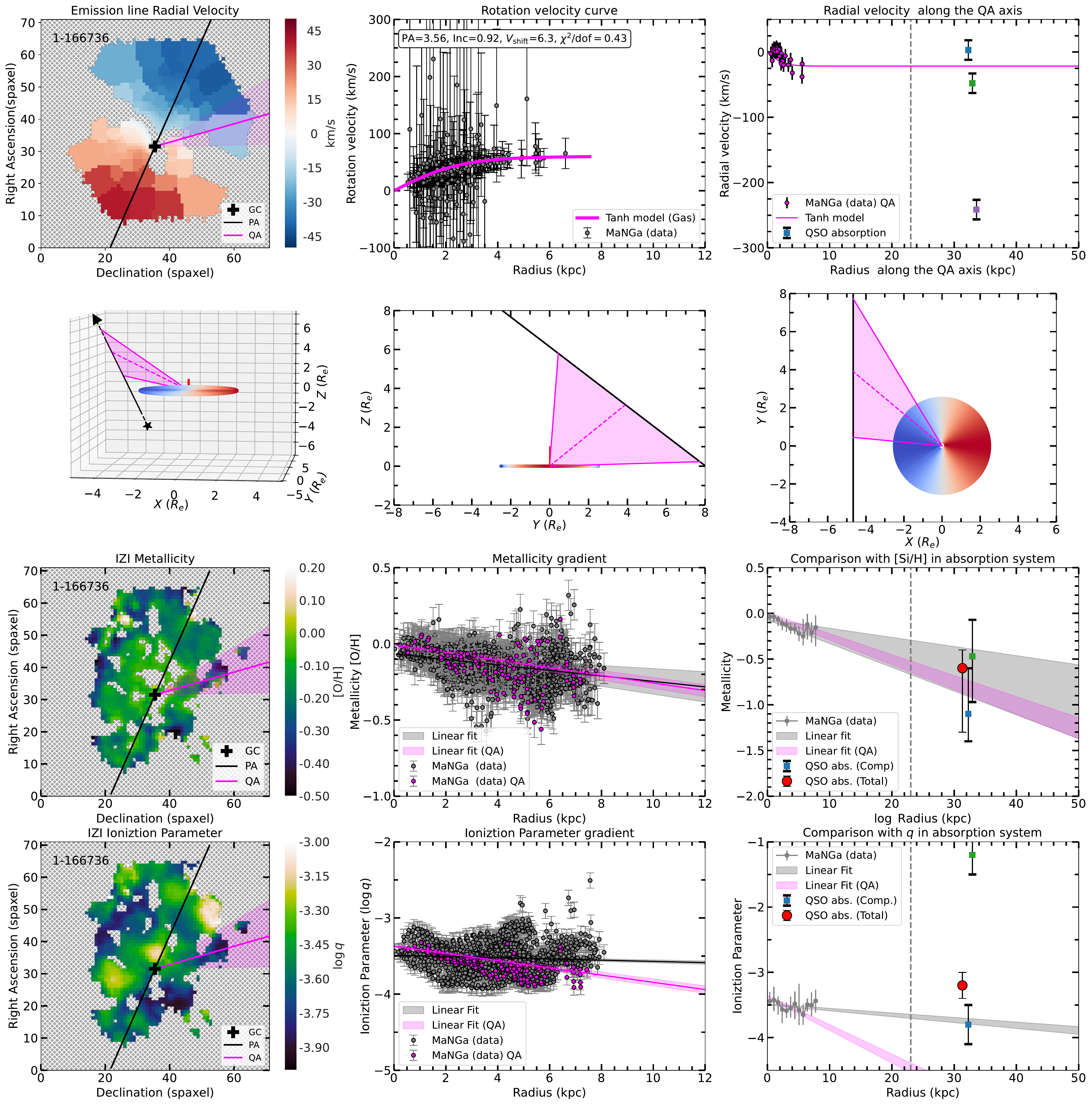}
        \caption{\rm The comparison of radial velocity, metallicity and ionization parameter derived from the fitting to MaNGA maps of the galaxy 1-166736 and ones measured in the absorption system towards the quasar J0950$+$4309. A detailed description of panels  is presented in the Appendix\,\ref{app:B}.}
        \label{Final-1-166736}
\end{center}
\end{figure*}

\begin{figure*}
\begin{center}
        \includegraphics[width=1\textwidth]{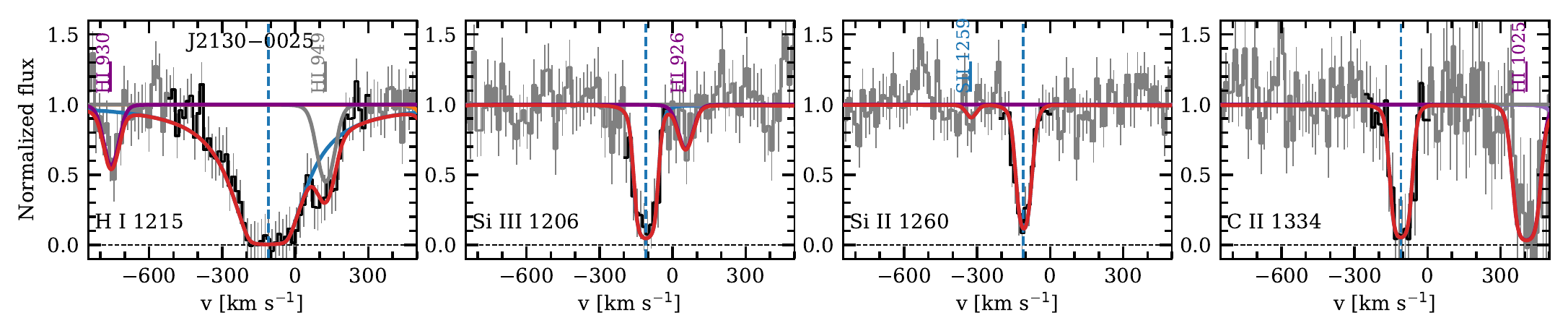}
        \includegraphics[width=0.97\textwidth]{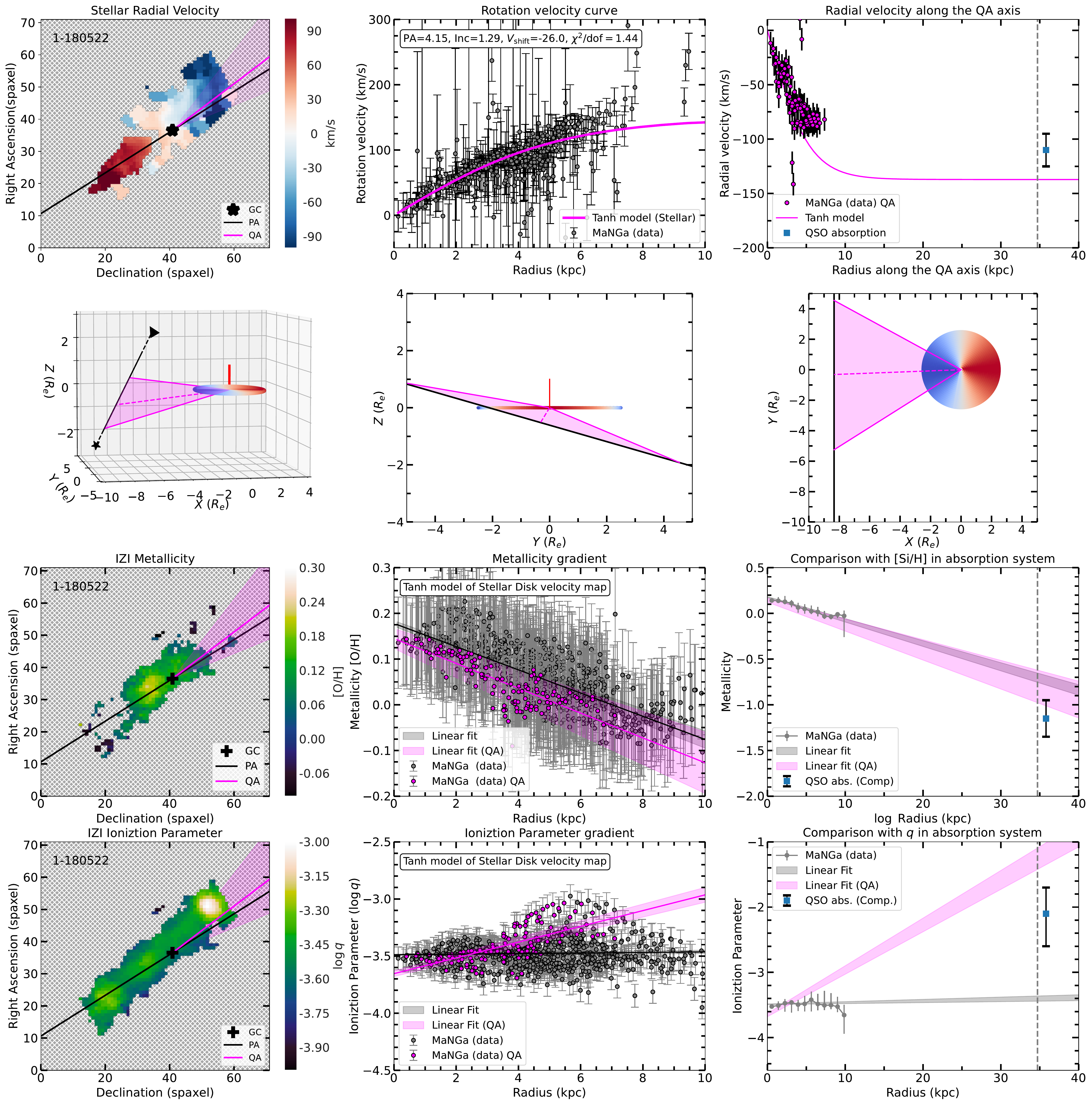}
        \caption{\rm 
        The comparison of radial velocity, metallicity and ionization parameter derived from the fitting to MaNGA maps of the galaxy 1-180522 and ones measured in the absorption system towards the quasar J2130$-$0025. The detailed description is presented in the Appendix\,\ref{app:B}.
        }
        \label{Final-1-180522}
\end{center}
\end{figure*}

\begin{figure*}
\begin{center}
        \includegraphics[width=1\textwidth]{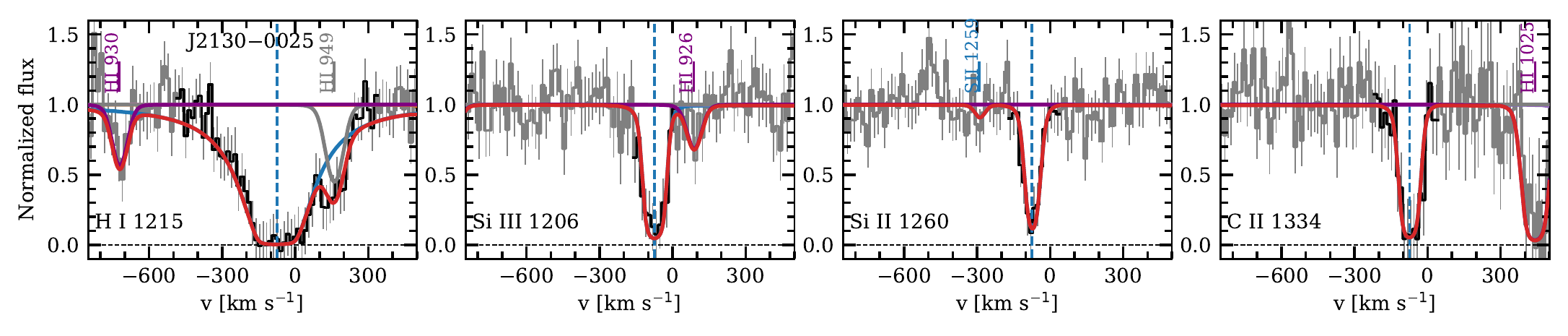}
        \includegraphics[width=0.97\textwidth]{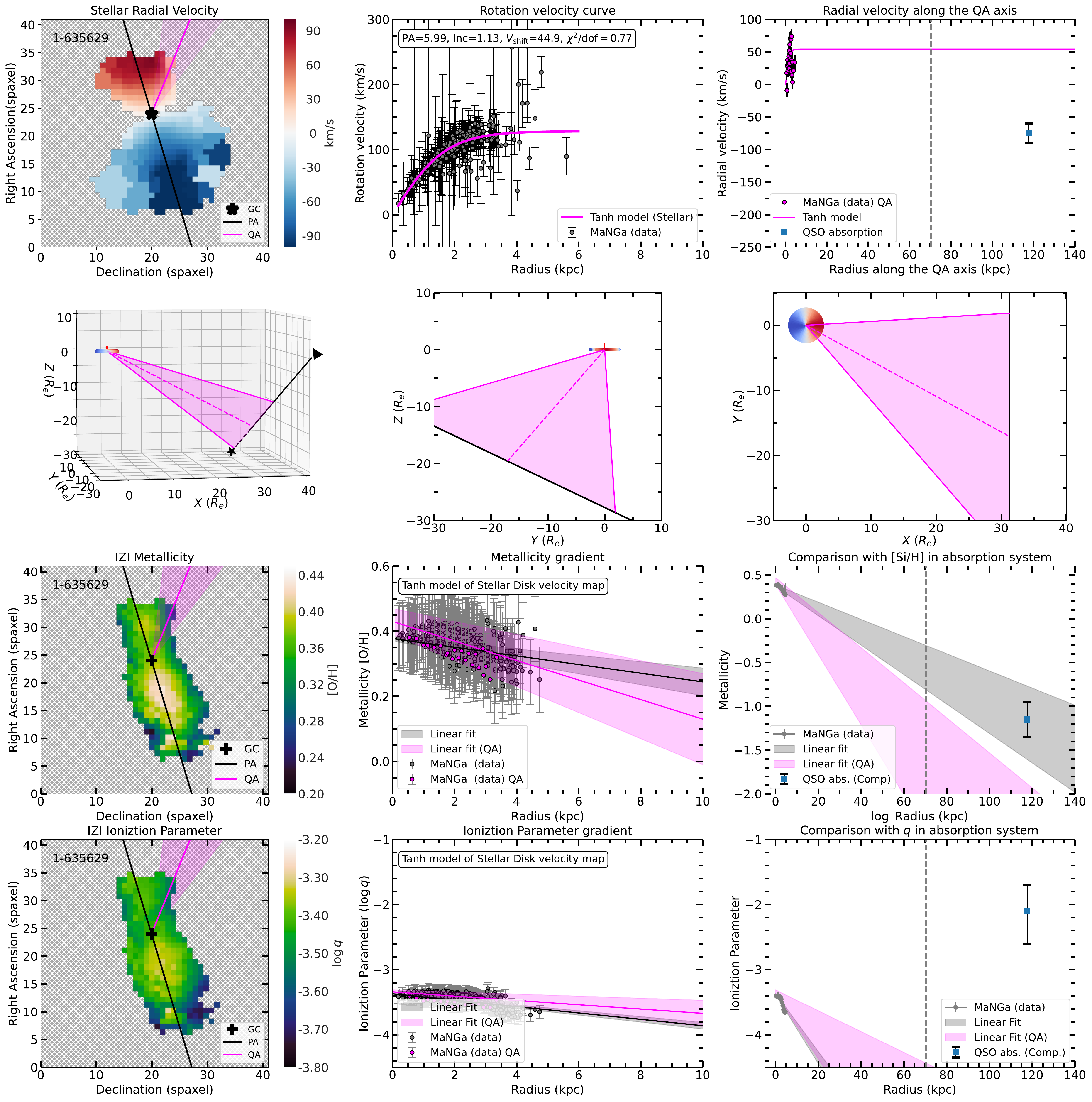}
        \caption{\rm 
        The comparison of radial velocity, metallicity and ionization parameter derived from the fitting to MaNGA maps of the galaxy 1-635629 and ones measured in the absorption system towards the quasar J2130$-$0025. The detailed description is presented in the Appendix\,\ref{app:B}.
        }
        \label{Final-1-635629}
\end{center}
\end{figure*}

\begin{figure*}
\begin{center}
        \includegraphics[width=1\textwidth]{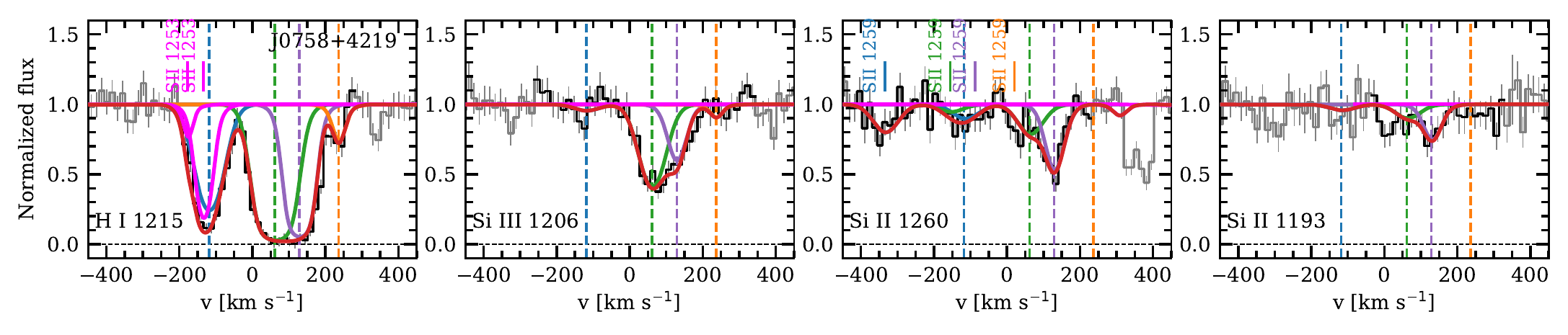}
        \includegraphics[width=0.97\textwidth]{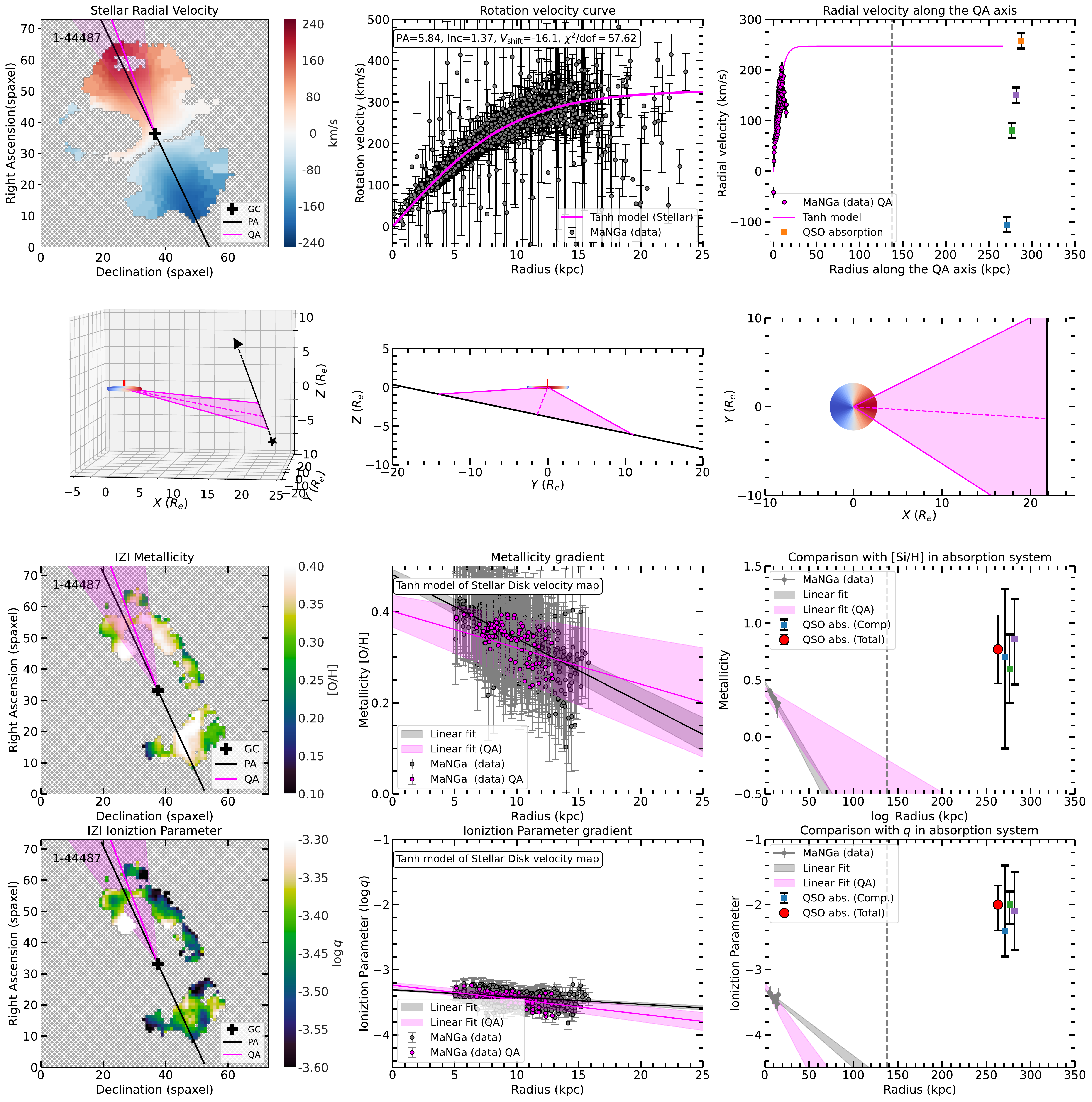}
        \caption{\rm         The comparison of radial velocity, metallicity and ionization parameter derived from the fitting to {MaNGA} maps of the galaxy 1-44487 and ones measured in the absorption system towards the quasar J0758$+$4219. The detailed description of panels  is presented in the Appendix\,\ref{app:B}.}
        \label{Final-1-44487}
\end{center}
\end{figure*}

\begin{figure*}
\begin{center}
        \includegraphics[width=1.0\textwidth]{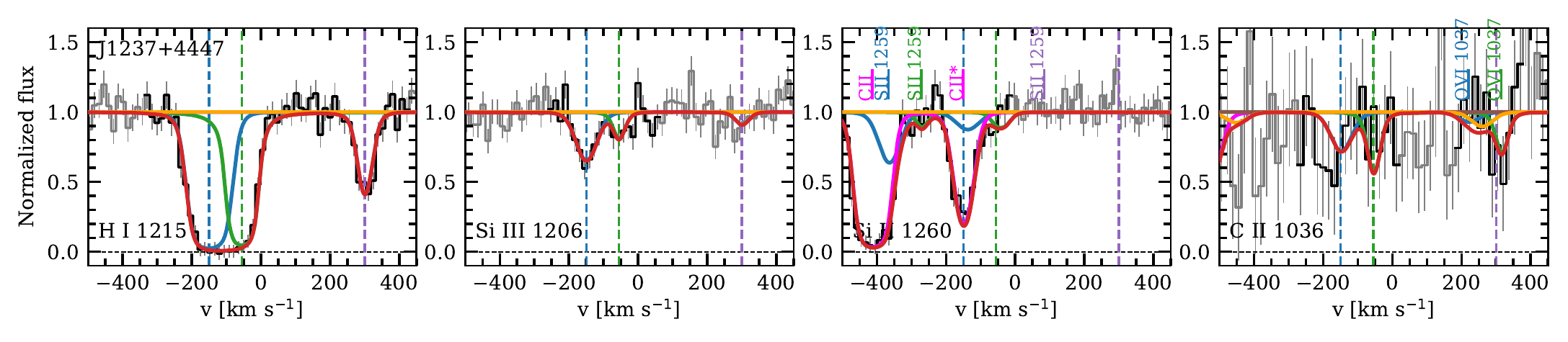}
        \includegraphics[width=1.0\textwidth]{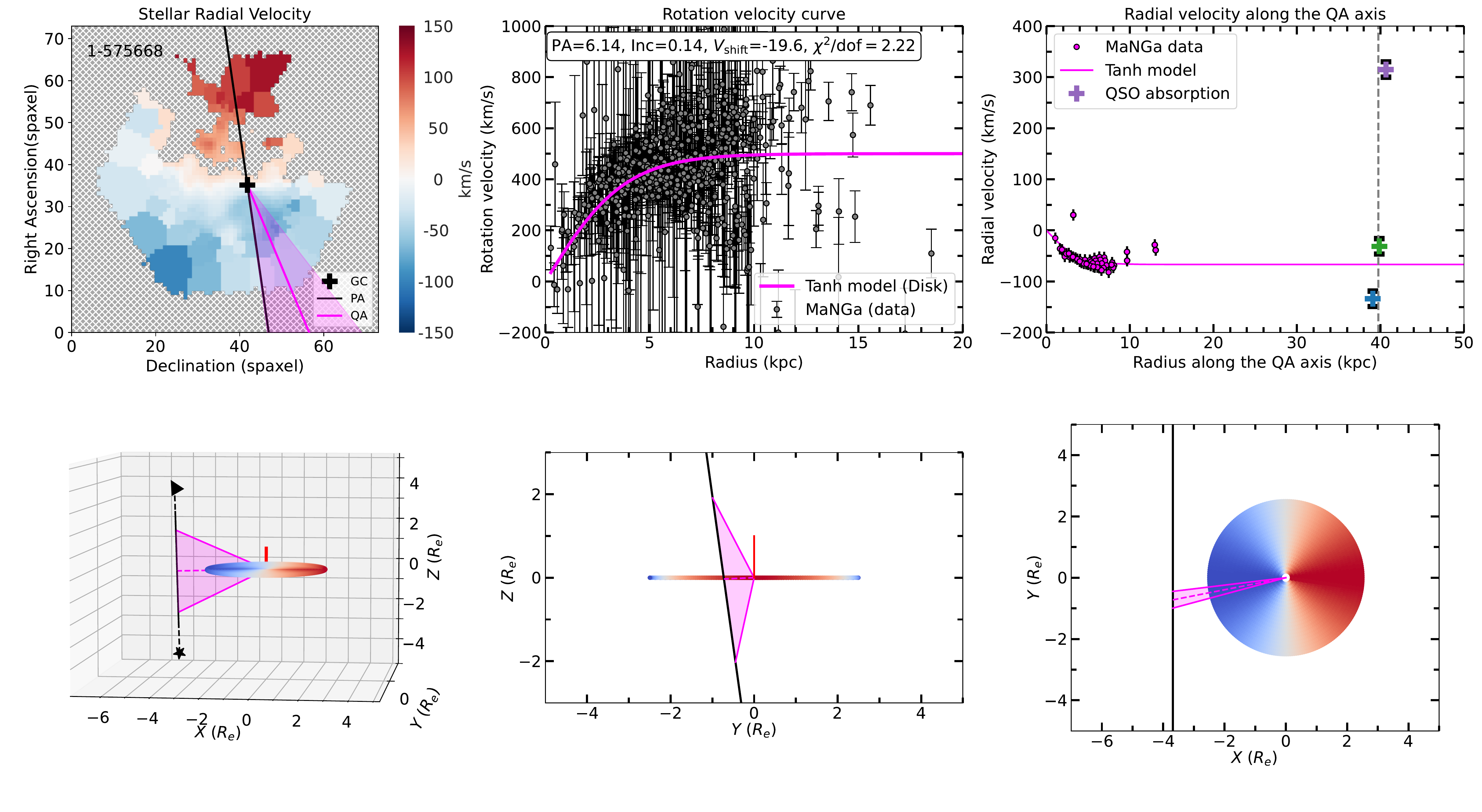}
        \caption{\rm 
        The comparison of radial velocity derived from the fitting to MaNGA maps of the galaxy 1-575668 and ones measured in the absorption system towards the quasar J1237$+$4447 The detailed description of panels is presented in the Appendix\,\ref{app:B}.}
        \label{Final-1-575668}
\end{center}
\end{figure*}

\begin{figure*}
\begin{center}
        \includegraphics[width=1.0\textwidth]{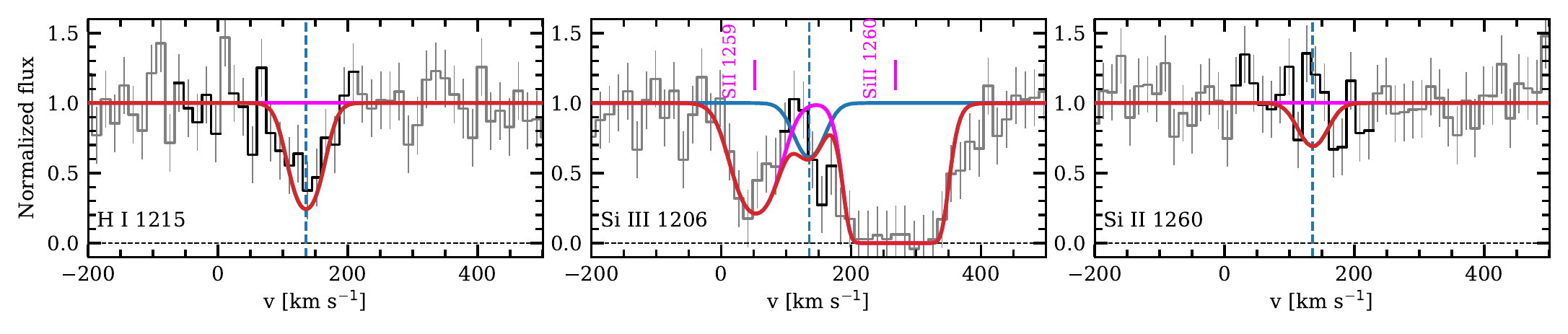}
        \includegraphics[width=1.0\textwidth]{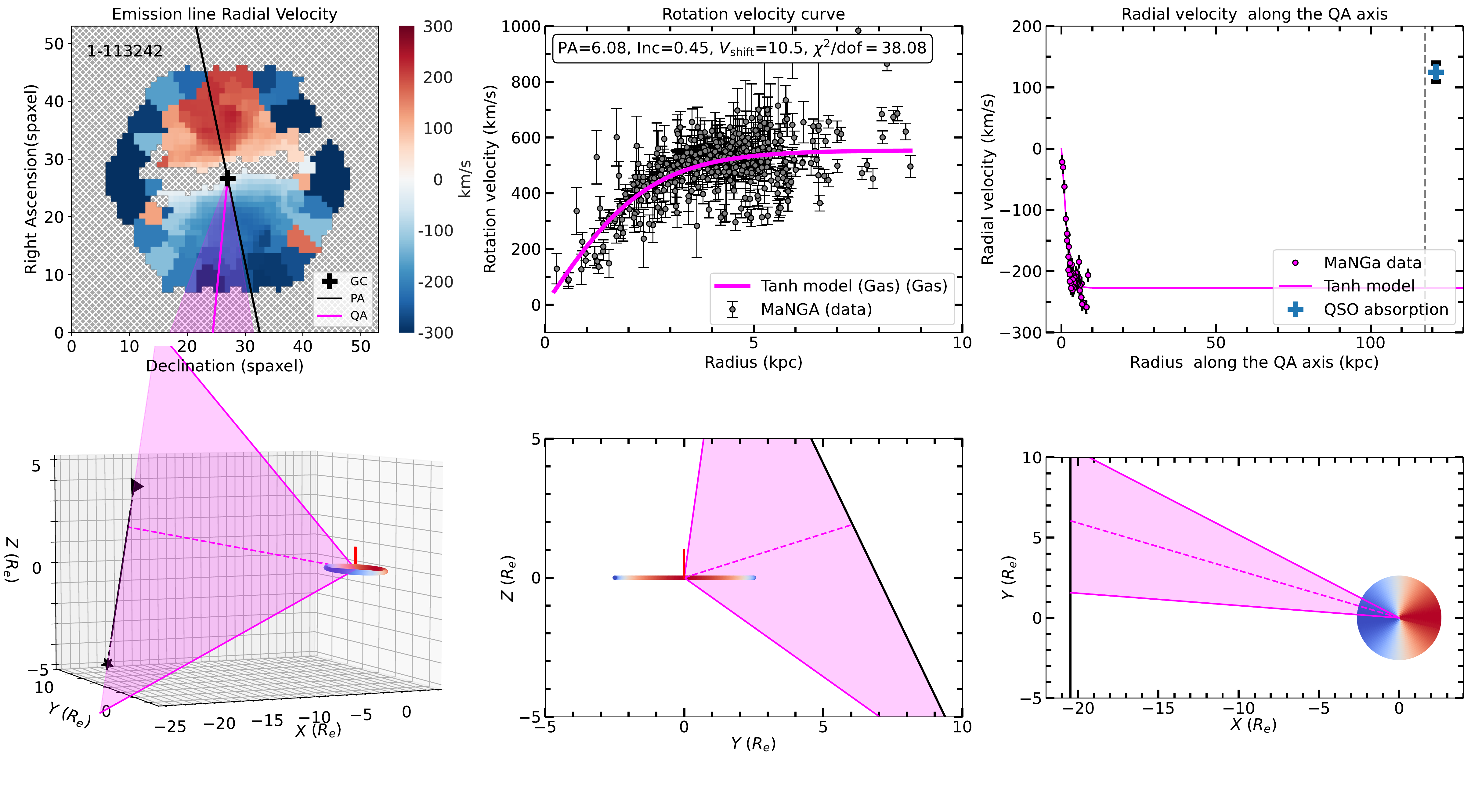}
        \caption{\rm 
        The comparison of radial velocity derived from the fitting to MaNGA maps of the galaxy 1-113242 and ones measured in the absorption system towards the quasar J2106$+$0909  The detailed description of panels is presented in the Appendix\,\ref{app:B}.
        }
        \label{Final-1-113242}
\end{center}
\end{figure*}

\begin{figure*}
\begin{center}
        \includegraphics[width=1.0\textwidth]{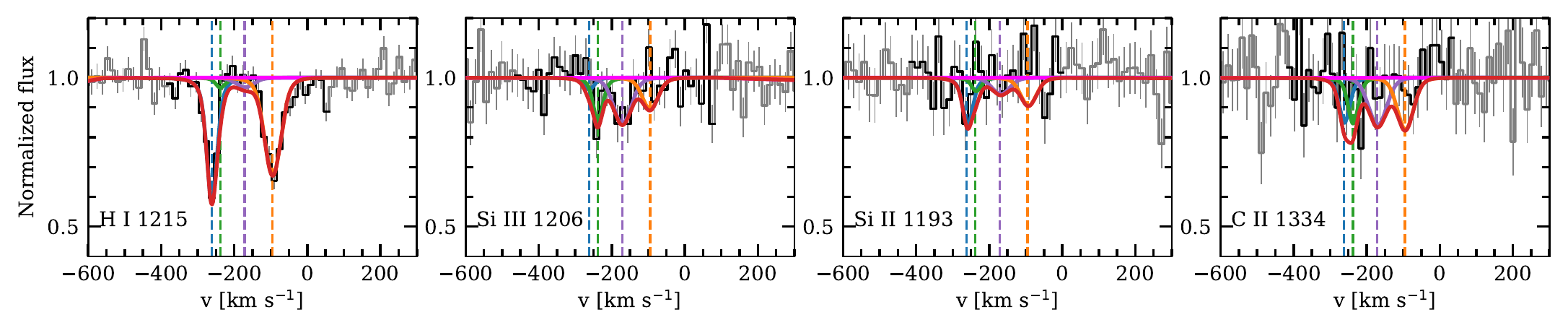}
        \includegraphics[width=1.0\textwidth]{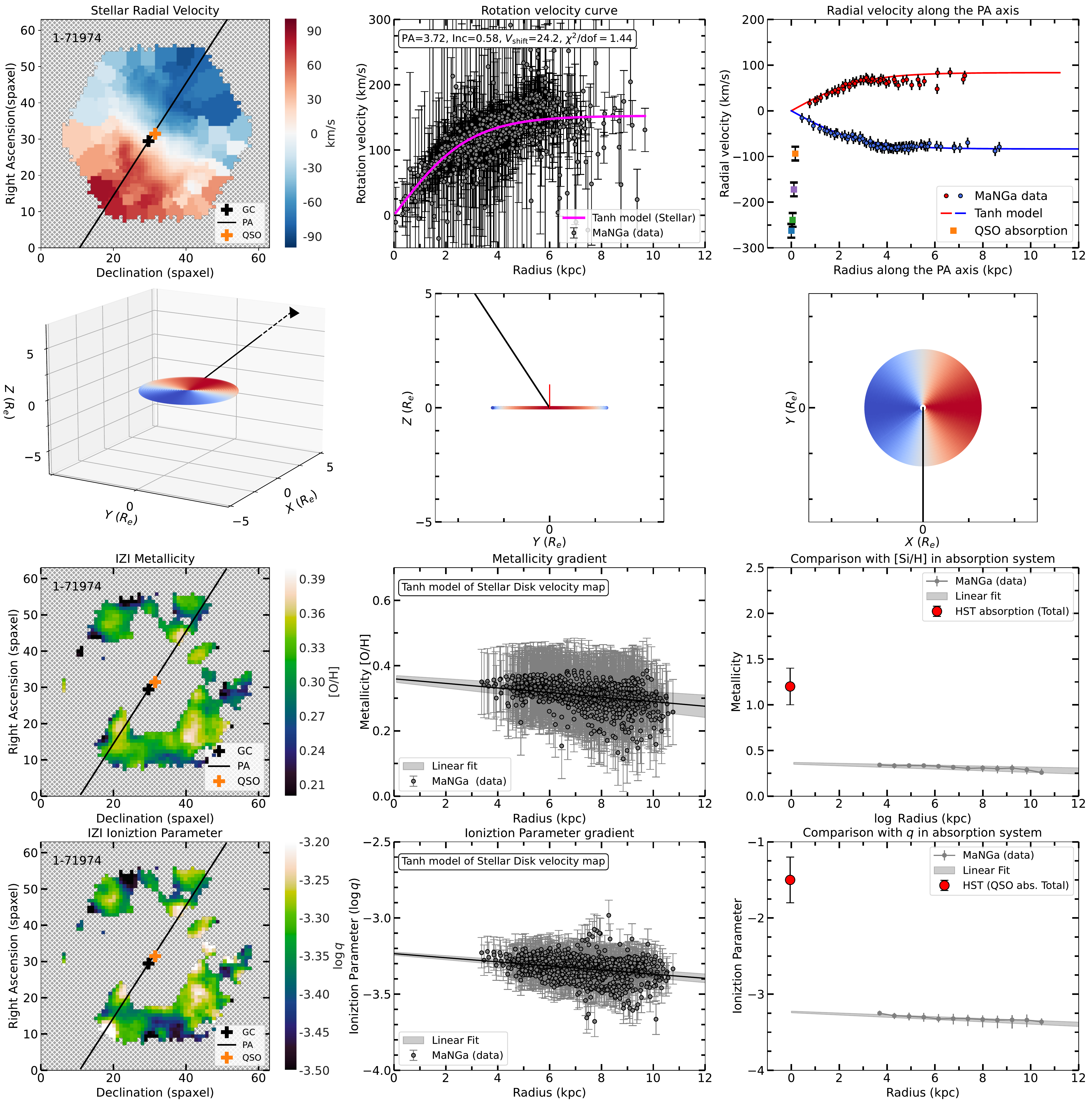}
        \caption{\rm The comparison of radial velocity, metallicity and ionization parameter derived from the fitting to MaNGA maps of the galaxy 1-71974 and ones measured in the absorption system towards the AGN J0755$+$3911. The detailed description of panels  is presented in the Appendix\,\ref{app:B}.}
        \label{fig:Final-1-71974}
\end{center}
\end{figure*}

The absorption system at the redshift of the galaxy is detected in \HI\ Ly$\alpha$ and Ly$\beta$ and \SiIII\ absorption lines. \SiII\ absorption line is blended by the saturated \CII\ and \CII$^*$ absorption lines of the MW. Since the \HI\  Ly$\alpha$ is saturated, we derived the velocity structure from the fit to \SiIII\ absorption line, which is well fitted by two velocity components. The fit to the red-shifted component in the \HI\ line profile is degenerate in the parameter space $N({\rm HI})-b$ and has two solutions with low and high $N({\rm HI})$. We consider both solutions named by them (case A) and (case B). The likelihood functions are shown in fig.\,\ref{fig:PDF}.
The result of the fit is given in Table\,\ref{tab:detailedfit} and 
line profiles are shown in Fig.\,\ref{fig:fit:j1237-lines}. 

\subsubsection{J1338+0311}
The absorption system associated with the host galaxy of the AGN 12-192116 has the strong damped \HI\ Ly$\alpha$ absorption ($N=10^{20.2}$ cm$^{-2}$). We described the fit to the \HI\ Ly$\alpha$ line in Section\,\ref{sec:spec_analysis}. The associated metal absorption lines are fitted by four velocity components. The result of the fit is given in Table\ \ref{tab:detailedfit} and 
line profiles are shown in Fig.\,\ref{fig:fit:j1338-lines}. 

\subsubsection{J2106+0909}
We detected only one weak absorption \HI\ Ly$\alpha$ line ($N=10^{13.7}$ cm$^{-2}$) at the redshift of the galaxy 1-113242. The associated \SiIII\ absorption line is blended with the MW \SII\ absorption lines, therefore we can set only upper limit to  \SiIII\ column density.  
The result of the fit is given in Table\ \ref{tab:detailedfit} and 
line profiles are shown in Fig.\,\ref{fig:fit:j2106-lines}. 

\subsubsection{J2130-0025}
We detected strong saturated absorption lines of \HI\ and metal species at the redshift of the galaxy 1-180522. The absorption lines are well fitted with one-component model. The result of the fit is given in Table\ \ref{tab:detailedfit} and 
line profiles are shown in Fig.\,\ref{fig:fit:j2130-lines}. 

\subsubsection{The case of nondetections}
In three cases (1653+3945, J1709+3421 and J1629+4007), we do not detect any absorption (in \HI\ or any of the metal ions) within the range of $\pm800$ km\,s$^{-1}$ relative to the galaxy redshifts (1-594755, 1-561034, 1-564490). Fig.\,\ref{fig:fit:j1629-lines} shows the parts of quasar spectra near the expected positions of \HI, \SiII, \SiIII, \NV\ lines. We set upper limits on $N({\rm HI})\sim10^{13}\,{\rm cm}^{-2}$.

%%%%%%%%%%%%%%%%%%%%%%%%%%%%%%%%%%%%%%%%%%%%%%%%%%%%%%%
%\afterpage{%
%\fontsize{9}{11}\selectfont
%\setlength{\arrayrulewidth}{0.5mm}
\setlength{\tabcolsep}{0.1pt}
\begin{table*}[h]
\caption{Fit results to absorption systems.}
\label{tab:detailedfit}
\resizebox{0.92\textwidth}{!}{
%\scalebox{0.9}{
\hskip-3.0cm\begin{tabular}{lcccccccccccccccc}
\hline
$z_{\rm abs}$ & v & $b$ & \multicolumn{14}{c}{$\log N({\rm X})$}  \\
&(km/s) & (km/s) & (\HI) & (\SiII) & (\SiIII) & (\CII) & (S\,{\sc ii}) & (S\,{\sc iv})  & (N\,{\sc i}) & (N\,{\sc ii}) & (N\,{\sc v}) & (O\,{\sc i})& (O\,{\sc vi}) & (Fe\,{\sc ii})& (Fe\,{\sc iii}) & (Ar\,{\sc i})\\
\hline
\multicolumn{17}{c}{J0755+3911, $z_{\rm em,corr}=0.03362$} \\
\hline
0.03284 & -228 &  $16^{+5}_{-1}$  &  $<12.8$ & $12.4^{+0.4}_{-1.0}$  & $12.7^{+0.5}_{-1.1}$ &  $<13.4$ & $<14.7$& N/C & $<13.5$ & N/C & N/C & $<13.8$ & N/C & $<13.7$& $<13.5$ & N/C \\
0.03292& -203  &  $17^{+20}_{-2}$ & $<13$  & $<12.7$  &  $12.0^{+0.3}_{-0.8}$             & $<13.4$  & $<14.0$ & N/C & $<13.3$ & N/C & N/C & $<13.1$ & N/C & $<13.7$ & $<13.5$ & N/C \\
0.03315& -136  &  $<20$             & $<12.2$             &  $<12.8$ & $12.3^{+0.3}_{-0.5}$ &  $<15$   & $<13.8$ & N/C & $<13.0$ & N/C& N/C  & $<13.8$ & N/C & $<13.7$ & $<13.2$ & N/C \\
0.03342& -62   &  $24^{+5}_{-5}$  & $13.3^{+0.1}_{-0.1}$ &  $<12.7$ & $<12.6$     &  $<13.6$ & $<14.5$ & N/C & $<13.7$ & N/C& N/C  & $<13.5$ & N/C & $<13.9$  & $<13.5$ & N/C\\
\multicolumn{3}{l}{Total:}          &$13.6^{+0.1}_{-0.1}$  & $12.7^{+0.3}_{-0.5}$ & $12.4^{+0.2}_{-0.5}$ & $13.0^{+0.2}_{-0.3}$ & $<15.0$ & N/C & $13.1^{+0.4}_{-0.9}$ & N/C & N/C &  $13.1^{+0.4}_{-0.6}$ & N/C & $13.2^{+0.3}_{-0.9}$ & $<13.5$ & N/C \\ 
\hline
\multicolumn{17}{c}{J0758$+$4219, $z_{\rm em,corr}=0.03181$} \\
\hline
0.03138 & -115  & $42^{+7}_{-7}$ & $14.2^{+0.3}_{-0.2}$ &  $12.3^{+0.2}_{-0.9}$ & $11.9^{+1.8}_{-0.6}$ & N/C & $<14.4$ & $<14.5$ & $<13.8$ & $<14.3$ & $<13.6$ & $<14.0$ & N/C & $<13.7$ & $<14.0$ & $<14.0$\\
{0.03202} & 52   & $41^{+8}_{-6}$  & $14.9^{+0.2}_{-0.2}$ &  $12.7^{+0.2}_{-0.2}$ & $13.2^{+0.1}_{-0.1}$ & N/C & $14.2^{+0.2}_{-0.5}$ & $<15.0$ & $<13.7$ & $<14.0$ & $<13.3$ & $<13.9$ & N/C & $<14.0$ & $<14.3$ & $<14.3$ \\
{0.03226} & 130 & $27^{+8}_{-4}$ & $14.8^{+0.5}_{-0.3}$   &  $13.0^{+0.1}_{-0.1}$ & $12.8^{+0.2}_{-0.2}$ & N/C & $13.8^{+0.5}_{-0.7}$ & $<15.0$ & $<13.6$ & $<14.1$ & $<13.3$ & $<13.9$ & N/C & $<13.8$ & $<14.4$ & $<14.3$ \\
{0.03263} & 230 & $17^{+50}_{-2}$ & $13.1^{+0.3}_{-0.2}$ &  N/A & $<12.3$ & N/C & N/A & N/A & N/A & N/A  & N/A & N/C & N/A & N/A & N/A & N/A\\
\multicolumn{3}{l}{Total:} & $15.2^{+0.3}_{-0.2}$ &  $13.2^{+0.1}_{-0.1}$ & $13.4^{+0.1}_{-0.1}$ & N/C & $14.7^{+0.1}_{-0.2}$ & $<15.0$ & $13.2^{+0.2}_{-2.0}$ & $13.5^{+0.4}_{-0.4}$ & $<13.5$ & $<14.0$ &N/C & $<13.9$ &$<14.4$& $<14.3$  \\ 
\hline
\multicolumn{17}{c}{J0838$+$2453(A), $z_{\rm em,corr}=0.02843$} \\
\hline
0.02834 & -34 &  $41^{+10}_{-10}$ & $13.2^{+0.1}_{-0.1}$      &  $<13.5$ & $12.5^{+0.4}_{-0.7}$ &  N/C & $<14.7$ & N/C & $<14.7$ & N/A & N/A & $<14.7$  &N/A &$<15.5$ &N/A & N/A \\
0.02812 & -93 &  $<20$            & N/A                       & $12.3^{+0.5}_{-1.0}$  &  $12.8^{+0.3}_{-0.3}$ &   N/C &  $<15$ & N/C & $<13.6$              & N/A& N/A  & $<14.5$ & $<14.1$    & N/A   &  N/A & N/A \\
0.02774 & -201 &  $20^{+10}_{-10}$           & N/A                       & $12.8^{+0.4}_{-0.3}$  &  $<13.5$    &  N/C & $1<13.5$ & N/A & $<13.7$ & N/A& N/A & $<14.3$ & N/A & N/A & N/A & N/A \\
{Total:} & $-100$ & &$13.2^{+0.1}_{-0.1}$ & $13.4^{+0.2}_{-0.2}$ & $13.8^{+0.6}_{-0.5}$ & N/C & $14.8^{+0.4}_{-1.0}$ & N/A & $<14.7$ &  N/A& N/A & $<15$ &N/A & $<15$ & N/A & N/A \\ 
\hline
\multicolumn{17}{c}{J0838$+$2453(B), $z_{\rm em,corr}=0.02843$} \\
\hline
0.02584 & -754 &  $55^{+15}_{-15}$ & $14.0^{+0.1}_{-0.1}$ & $12.8^{+0.5}_{-1.3}$ &  $13.2^{+0.8}_{-1.7}$ &  N/C & $14.2^{+0.8}_{-1.5}$ & N/A & $<13.7$  & N/A& N/A & $<13.0$ & N/A& $<16.0$& N/A & N/A\\
%0.02579 & -769 &  $<75$ & N/A & $<13.6$ &  $<13.1$ &   N/C & $13.8^{+1.1}_{-1.1}$ &N/A& $14.0^{+0.8}_{-1.2}$ & N/A& N/A& N/A & N/A& $<21$ &N/A&N/A\\
0.02556 & -853 &  $<80$ & N/A & $13.4^{+1.1}_{-0.7}$ &  $<13.3$ &   N/C & $13.4^{+0.5}_{-1.4}$ &N/A& $<14.4$ & N/A& N/A& $<13.0$ & N/A & N/A & N/A & N/A\\
{Total:} & $-800$ & &$14.0^{+0.1}_{-0.1}$ & $<13.6$ & $13.6^{+0.5}_{-1.1}$ & N/C & $14.5^{+0.5}_{-1.4}$ &N/A& $<14.4$ &  N/A& N/A & $<14.5$ & N/A & N/A &N/A & N/A\\ 
\hline
\multicolumn{17}{c}{{J0950+4309}, $z_{\rm em,corr}=0.01714$} \\
\hline
{0.01713} & -2 &  $28^{+6}_{-6}$ & $17.6^{+0.3}_{-0.8}$ &  $13.0^{+0.1}_{-0.1}$ & $13.0^{+0.2}_{-0.3}$ &  $14.0^{+0.1}_{-0.1}$ & $<14.7$& N/C & $<14.3$& N/A & $<13.8$ & $<14.0$ & N/C & $<14.3$ & $<14.0$ &  N/C\\
0.01696 & -52 & $70^{+10}_{-10}$ & $15.1^{+0.7}_{-0.2}$ & $<12.4$ & $13.4^{+0.1}_{-0.1}$ &  $<13.6$ & $<14.6$& N/C &$<14.4$& $<16.3$& $<14.1$ & $<14.0$ & N/C & $<14.2$ & $14.0^{+0.3}_{-1.2}$ & N/C \\
0.01630 & -247 & $<100$ & $13.4^{+0.2}_{-0.2}$ & $<12.7$ & $<12.7$ & $<14.1$ & $<14.6$ & N/C & $<14.0$& $<16.0$& $<13.5$& $<14.6$ & N/C & $<14.0$ & N/A& N/C \\
\multicolumn{3}{l}{{Total:}} &$17.6^{+0.3}_{-0.7}$ & $13.0^{+0.1}_{-0.1}$ & $13.6^{+0.1}_{-0.1}$ & $14.1^{+0.2}_{-0.2}$ & $14.2^{+0.2}_{-0.2}$ & N/C & $13.6^{+0.4}_{-0.8}$ & $<16.2$& $14.0^{+0.2}_{-0.4}$ & $13.6^{+0.3}_{-0.6}$ & N/C & $13.6^{+0.3}_{-0.6}$ & $14.0^{+0.3}_{-1.2}$ & N/C\\
%\multicolumn{3}{l}{Total(B):} &$15.2^{+0.3}_{-0.2}$ & $13.1^{+0.2}_{-0.2}$ & $13.6^{+0.2}_{-0.2}$ & $14.1^{+0.2}_{-0.2}$ & $14.2^{+0.3}_{-0.6}$ & $13.8^{+0.4}_{-0.8}$ & $<16.2$& $14.00^{+0.2}_{-0.4}$& $13.6^{+0.6}_{-0.6}$ & $13.6^{+0.3}_{-0.6}$ & $-2.0^{+0.3}_{-0.8}$& $-2.4^{+0.3}_{-0.3}$ &  $20.7^{+0.4}_{-0.4}$\\
\hline
\multicolumn{17}{c}{{J1237$+$4447(A)}, $z_{\rm em,corr}=0.06013$} \\
\hline
{0.05960}& -151 & {$38^{+7}_{-19}$} & {$14.8^{+0.6}_{-0.3}$} &  $<13.0$ & $12.8^{+0.1}_{-0.1}$ & $13.9^{+0.3}_{-3.0}$ & $14.3^{+0.5}_{-1.5}$ & $<14.3$ & $<13.3$ & $<13.5$ & $13.2^{+0.1}_{-2.5}$ & N/A  & $<14.3$ & $<13.8$ & $<13.8$ & $<13.4$\\
{0.05993} & -57 & {$16^{+4}_{-1}$} & {$17.1^{+0.3}_{-0.7}$} &  $<12.5$ & {$12.3^{+0.2}_{-0.3}$} & $<14.7$ & $<14.3$ & $<14.1$ & $<13.4$ & $<13.8$ & $<13.2$ & N/C & $<15.3$ & $13.5^{+0.4}_{-1.5}$ & $<14.0$ & $<13.3$ \\
%{ 0.05993 (B)} & -57 & {$28^{+6}_{-6}$} & {$14.8^{+0.8}_{-0.2}$} &  $<12.6$ & $12.4^{+0.2}_{-0.2}$ & $<14.8$ & $<14.2$ & $<14.1$ & $<13.3$ & $<13.6$ & $<13.2$ & N/C & $14.0^{+0.2}_{-1.5}$ & $13.7^{+0.3}_{-1.5}$ & $<14.2$ & $<13.2$\\
0.06119& 302 & {$20^{+9}_{-9}$} & {$13.6^{+0.1}_{-0.1}$} &  N/A & $<12$ & N/A & N/A& N/A & N/A & N/A & N/A & N/C& N/A & N/A & N/A& N/A\\
\multicolumn{3}{l}{{Total:}} & {$17.2^{+0.3}_{-0.4}$} &  $<13$ & $12.9^{+0.1}_{-0.1}$ & $<14.7$ & $14.3^{+0.5}_{-0.8}$ & $<14.2$ & $12.9^{+0.5}_{-1.3}$ & $<13.5$ & $13.2^{+0.3}_{-0.5}$ & N/C & $<14.3$ & $13.5^{+0.5}_{-0.6}$ & $<11.0$ & $<13.5$ \\
\hline
\multicolumn{17}{c}{{J1237$+$4447(B)}, $z_{\rm em,corr}=0.06013$} \\
\hline
{0.05960}& -151 & {$38^{+7}_{-19}$} & {$14.8^{+0.6}_{-0.3}$} &  $<13.0$ & $12.8^{+0.1}_{-0.1}$ & $13.9^{+0.3}_{-3.0}$ & $14.3^{+0.5}_{-1.5}$ & $<14.3$ & $<13.3$ & $<13.5$ & $13.2^{+0.1}_{-2.5}$ & N/A  & $<14.3$ & $<13.8$ & $<13.8$ & $<13.4$\\
%{ 0.05993 (A)} & -57 & {$16^{+4}_{-1}$} & {$17.1^{+0.3}_{-0.7}$} &  $<12.5$ & {$12.3^{+0.2}_{-0.3}$} & $<14.7$ & $<14.3$ & $<14.1$ & $<13.4$ & $<13.8$ & $<13.2$ & N/C & $<15.3$ & $13.5^{+0.4}_{-1.5}$ & $<14.0$ & $<13.3$ \\
{0.05993} & -57 & {$28^{+6}_{-6}$} & {$14.8^{+0.8}_{-0.2}$} &  $<12.6$ & $12.4^{+0.2}_{-0.2}$ & $<14.8$ & $<14.2$ & $<14.1$ & $<13.3$ & $<13.6$ & $<13.2$ & N/C & $14.0^{+0.2}_{-1.5}$ & $13.7^{+0.3}_{-1.5}$ & $<14.2$ & $<13.2$\\
0.06119& 302 & {$20^{+9}_{-9}$} & {$13.6^{+0.1}_{-0.1}$} &  N/A & $<12$ & N/A & N/A& N/A & N/A & N/A & N/A & N/C& N/A & N/A & N/A& N/A\\
\multicolumn{3}{l}{{Total:}} & {$15.2^{+0.5}_{-0.3}$} &  $12.4^{+0.4}_{-0.8}$ & $12.9^{+0.1}_{-0.1}$ & $14.0^{+0.5}_{-1.4}$ & $14.3^{+0.3}_{-1.2}$ & $<13.9$ & $12.9^{+0.3}_{-1.0}$ & $<13.5$ & $13.2^{+0.3}_{-0.8}$ & N/C & $14.1^{+0.5}_{-1.5}$ & $13.7^{+0.3}_{-0.9}$ & $<11.0$ & $<13.2$ \\
\hline
\multicolumn{17}{c}{ J1338$+$0311, $z_{\rm em,corr}=0.02604$} \\
\hline
0.02543 & -178 &  $15^{+13}_{-5}$ & N/A &  $12.9^{+0.6}_{-1.6}$ & $13.0^{+1.1}_{-1.7}$ &  N/C & $10.1^{+2.4}_{-0.1}$ & N/C & $14.1^{+0.4}_{-0.9}$ & $13.3^{+1.6}_{-2.4}$ & $13.2^{+0.5}_{-1.4}$ & $<15.0$ &N/C & $13.2^{+0.7}_{-0.7}$ & $14.6^{+0.6}_{-1.6}$ & N/A \\ %NOI*=14.6+0.8-0.9
0.02566 & -110 &  $80^{+30}_{-20}$ & N/A & $13.4^{+0.2}_{-0.2}$ & $13.6^{+0.3}_{-0.5}$ &  N/C & $10.1^{+2.5}_{-0.1}$ &N/C & $13.7^{+0.6}_{-1.6}$ & $10.1^{+4.6}_{-0.1}$ & $10.3^{+1.1}_{-0.3}$ & $15.0^{+0.2}_{-0.3}$ & N/C & $10.5^{+2.0}_{-0.5}$ & N/A &N/A \\ %NOI*=14.7+0.2-0.2
0.02603 & 2 &  $77^{+18}_{-13}$ & $20.2^{+0.1}_{-0.1}$ &  $13.6^{+0.2}_{-0.3}$ & $13.5^{+0.3}_{-0.3}$ &  N/C & $15.2^{+0.1}_{-0.1}$ & N/C &$14.2^{+0.2}_{-0.6}$ & $14.9^{+0.2}_{-0.2}$ & $11.5^{+0.8}_{-1.5}$ & $15.0^{+0.2}_{-0.2}$ &N/C & $14.6^{+0.4}_{-1.0}$ & $<14.6$ &N/A\\%NOI*=<14.3
0.02639 & 146 &  $50^{+10}_{-13}$ & N/A &  $12.7^{+0.4}_{-0.3}$ & $12.6^{+0.3}_{-2.6}$ &  N/C & $14.3^{+0.3}_{-2.5}$ & N/C & $14.2^{+0.2}_{-0.3}$ & $13.9^{+1.0}_{-2.8}$& $<12$ & $14.2^{+0.4}_{-0.8}$ & N/C & $<14.0$ & $<14$ & N/A\\ %NOI*<14.1
\multicolumn{3}{l}{Total:}  & $20.2^{+0.1}_{-0.1}$ & $13.9^{+0.2}_{-0.2}$ & $13.9^{+0.4}_{-0.2}$ & N/C & $15.2^{+0.1}_{-0.2}$ & N/C & $<14.3$ &  $<16$ & $<13.8$ & $15.5^{+0.1}_{-0.1}$ &N/C & $<14.6$ & $<14.3$ & N/A\\ 
\hline
\multicolumn{17}{c}{J2106$+$0909, $z_{\rm em,corr}=0.0437$} \\
\hline
0.04423& 135 & $30^{+10}_{-10}$ & $13.7^{+0.2}_{-0.2}$ &  $<13.0$ & $<13.2$ & N/C & $<14.8$ & N/A & N/A & N/A & N/A & N/A & N/A & N/A & N/A & N/A\\
\hline %15.0+0.2-0.2 NOI*=15.3+0.1-0.1
\multicolumn{17}{c}{{ J2130$-$0025}, $z_{\rm em,corr}=0.0200$} \\
\hline
{0.01967} & -146 & $26^{+6}_{-4}$ & $18.8^{+0.1}_{-0.1}$ &  $13.8^{+0.1}_{-0.2}$ & $14.6^{+0.9}_{-0.5}$ &  $15.3^{+0.9}_{-0.5}$ & $<14.6$ & N/C & $13.1^{+0.3}_{-1.1}$ & $<14.7$ & $<13.7$ & $14.6^{+0.9}_{-0.5}$ & N/C & $13.6^{+0.4}_{-1.9}$ & $<11$ &N/A \\
\hline
\end{tabular}
}
\end{table*}  

\begin{table*}
\begin{center}
\caption{Fit results to {\sc cloudy} photo-ionization model of absorption systems} 
\label{tab:cloudyfitparam}
\begin{tabular}{lcccccccc}
\hline
Quasar & $z_{\rm abs}$ & v & $\log N_{\rm HI}$ & $\log N_{\rm H_{tot}}$ & [X/H] & $\log q$ & $\log I_{\rm UV}/n_{\rm H}$ & $F_{*}$\\
& & km s$^{-1}$ & cm$^{-2}$ & cm$^{-2}$ &  &  & Drain/cm$^{-3}$ & \\
\hline
J0755+3911 & \multicolumn{2}{l}{Average:} & $13.5^{+0.3}_{-0.3}$ & $16.5^{+0.3}_{-0.3}$ & $1.2^{+0.2}_{-0.2}$ & $-1.5^{+0.3}_{-0.3}$ & $2.0^{+0.3}_{-0.2}$ & $0.0^{+0.4}_{-0.0}$\\
\hline
J0758$+$4219 & 0.03138 & -115  & $14.2^{+0.3}_{-0.2}$ &  $16.9^{+1.0}_{-0.7}$ & $0.7^{+0.6}_{-0.8}$ & $-2.4^{+1.0}_{-0.4}$ & $1.0^{+1.0}_{-0.5}$ & $<1$ \\
& 0.03202 & 52   & $14.2^{+0.3}_{-0.1}$ &  $17.8^{+0.2}_{-0.2}$ & $0.6^{+0.3}_{-0.3}$ & $-2.0^{+0.3}_{-0.3}$ & $1.4^{+0.3}_{-0.3}$ & $0.0^{+0.4}_{-0.0}$\\
&0.03226 & 130 & $15.0^{+0.4}_{-0.3}$ & $17.6^{+0.3}_{-0.3}$ & $0.9^{+0.4}_{-0.4}$ & $-2.1^{+0.6}_{-0.6}$ & $1.4^{+0.4}_{-0.6}$  & $0.2^{+0.4}_{-0.2}$\\
&0.03261 & 230 & N/A&  N/A & N/A & N/A& N/A & N/A\\
&\multicolumn{2}{l}{Average:} & $15.3^{+0.3}_{-0.2}$ &  $18.1^{+0.3}_{-0.4}$ & $0.8^{+0.3}_{-0.3}$ & $-2.0^{+0.3}_{-0.4}$ & $1.4^{+0.3}_{-0.4}$ & $0.1^{+0.5}_{-0.1}$ \\ 
\hline
J0838$+$2453 & \multicolumn{2}{l}{Average(A):} & $13.2^{+0.2}_{-0.3}$      &  $16.7^{+0.3}_{-0.3}$ & $1.9^{+0.3}_{-0.3}$ & $-0.8^{+0.2}_{-0.5}$ &  $2.7^{+0.3}_{-0.5}$& $0.0^{+0.1}_{-0.0}$\\ 
 & \multicolumn{2}{l}{Average(B):} & $13.8^{+0.2}_{-0.3}$ & $17.1^{+1.5}_{-0.3}$ &  $0.5^{+1.5}_{-0.5}$ & $-1.1^{+0.5}_{-0.7}$ & $2.4^{+0.5}_{-0.7}$ & $<1$\\ 
\hline
J0950+4309 & 0.01713 & -2 &  $17.6^{+0.3}_{-0.7}$ &  $18.9^{+0.3}_{-0.5}$ & $-1.1^{+0.5}_{-0.3}$ & $-3.8^{+0.3}_{-0.3}$ & $-0.5^{+0.3}_{-0.2}$  & $0.0^{+0.2}_{-0.0}$\\
& 0.01696 & -52 & $14.9^{+0.3}_{-0.3}$ & $19.4^{+0.4}_{-0.5}$ & $-0.5^{+0.4}_{-0.5}$ & $-1.2^{+0.4}_{-0.3}$ & $2.3^{+0.3}_{-0.4}$ & $0.0^{+0.3}_{-0.0}$\\
& 0.01630 & -247 & N/A & N/A & N/A & N/A & N/A & N/A\\
& \multicolumn{2}{l}{Average:} &$16.8^{+0.8}_{-0.3}$ & $19.0^{+0.6}_{-0.2}$ & $-0.6^{+0.2}_{-0.7}$& $-3.2^{+0.2}_{-0.2}$ & $-0.1^{+0.4}_{-0.2}$ & $0.0^{+0.2}_{-0.0}$\\
%\multicolumn{3}{l}{Total(B):} &$15.2^{+0.3}_{-0.2}$ & $13.1^{+0.2}_{-0.2}$ & $13.6^{+0.2}_{-0.2}$ & $14.1^{+0.2}_{-0.2}$ & $14.2^{+0.3}_{-0.6}$ & $13.8^{+0.4}_{-0.8}$ & $<16.2$& $14.00^{+0.2}_{-0.4}$& $13.6^{+0.6}_{-0.6}$ & $13.6^{+0.3}_{-0.6}$ & $-2.0^{+0.3}_{-0.8}$& $-2.4^{+0.3}_{-0.3}$ &  $20.7^{+0.4}_{-0.4}$\\
\hline
J1237$+$4447 & 0.05960 & -151 & $14.9^{+0.6}_{-0.3}$ &  $18.2^{+1.3}_{-0.6}$ & $-0.2^{+0.7}_{-1.1}$ & $-2.0^{+0.9}_{-0.5}$ & $1.5^{+0.6}_{-0.8}$ & $<1$\\
 & 0.05993(A) & -57 & $17.0^{+0.4}_{-0.3}$ &  $19.3^{+0.8}_{-0.9}$ & $-1.6^{+1.0}_{-1.0}$ & $-3.1^{+0.6}_{-0.5}$ & $0.0^{+0.9}_{-0.3}$& $0.0^{+0.2}_{-0.0}$\\
 & 0.05993(B) & -57 & $14.8^{+0.7}_{-0.3}$ &  $18.1^{+1.4}_{-0.7}$ & $-0.7^{+0.6}_{-1.4}$ & $-2.6^{+1.1}_{-0.5}$ & $0.9^{+1.2}_{-0.5}$ & $0.0^{+0.2}_{-0.0}$\\
%{ 0.05993 (B)} & -57 & { $28^{+6}_{-6}$} & { $14.8^{+0.8}_{-0.2}$} &  $<12.6$ & $12.4^{+0.2}_{-0.2}$ & $<14.8$ & $<14.2$ & $<14.1$ & $<13.3$ & $<13.6$ & $<13.2$ & N/C & $14.0^{+0.2}_{-1.5}$ & $13.7^{+0.3}_{-1.5}$ & $<14.2$ & $<13.2$ & $-0.5^{+0.3}_{-0.2}$ & $-3.3^{+0.3}_{-0.3}$ & $18.0^{+0.3}_{-0.3}$ & $0.3^{+0.3}_{-0.3}$\\
& 0.06119& 302 &  N/A& N/A & N/A & N/A & N/A & N/A\\
& \multicolumn{2}{l}{Average(A):} & $17.2^{+0.2}_{-0.7}$ &  $19.4^{+0.8}_{-0.5}$  & $-1.8^{+0.8}_{-0.8}$ & $-2.9^{+0.9}_{-0.6}$ & $0.3^{+0.9}_{-0.6}$ & $0.2^{+0.0}_{-0.2}$\\
& \multicolumn{2}{l}{Average(B):} & $15.2^{+0.5}_{-0.3}$ & $18.1^{+0.9}_{-0.5}$ & $-0.2^{+0.5}_{-1.0}$ & $-2.5^{+0.6}_{-0.4}$ & $0.9^{+0.6}_{-0.5}$  & $0.0^{+0.2}_{-0.0}$\\
\hline
J1338$+$0311 & Average: & & $20.2^{+0.1}_{-0.1}$ &  $20.5^{+0.2}_{-0.2}$ & $-0.4^{+0.4}_{-0.3}$ & $-2.8^{+0.3}_{-0.2}$ &  & $0.5^{+0.3}_{-0.3}$\\ 
\hline
J2106$+$0909 & 0.04423& 135 & N/A & N/A & N/A & N/A & N/A & N/A \\
\hline %15.0+0.2-0.2 NOI*=15.3+0.1-0.1
J2130$-$0025 & 0.01967 & -110 & $18.8^{+0.1}_{-0.1}$ &$21.1^{+0.4}_{-0.6}$ & $-1.1^{+0.2}_{-0.2}$ & $-2.1^{+0.4}_{-0.5}$ & $1.4^{+0.4}_{-0.6}$ & $0.1^{+0.3}_{-0.1}$\\
\hline
\end{tabular}
\end{center}
\end{table*}
%}

%%%%%%%%%%%%%%%%%%%%%%%%%%%%%%%%%%%%%%%%%%%%%%%%%%%%%%%

\begin{figure*}
\begin{center}
        \includegraphics[width=1.0\textwidth]{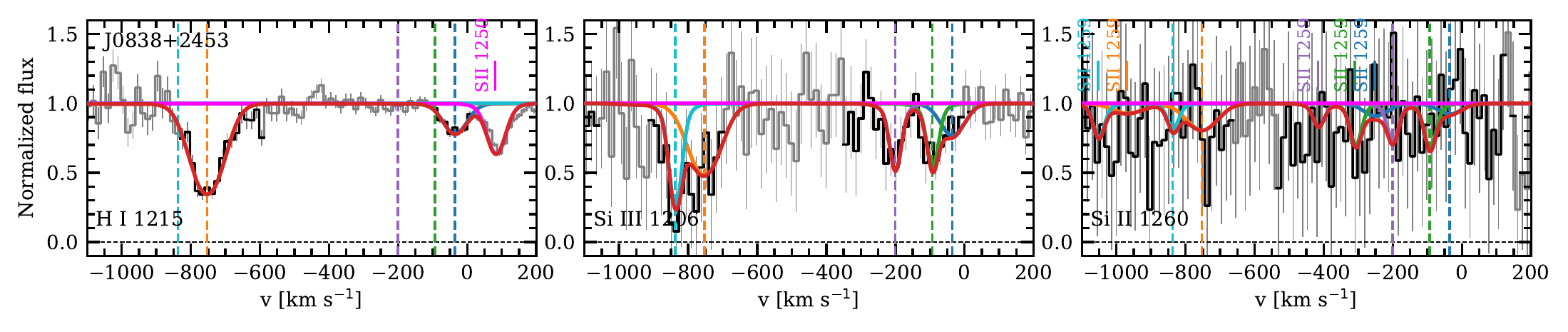}
        \includegraphics[width=1.0\textwidth]{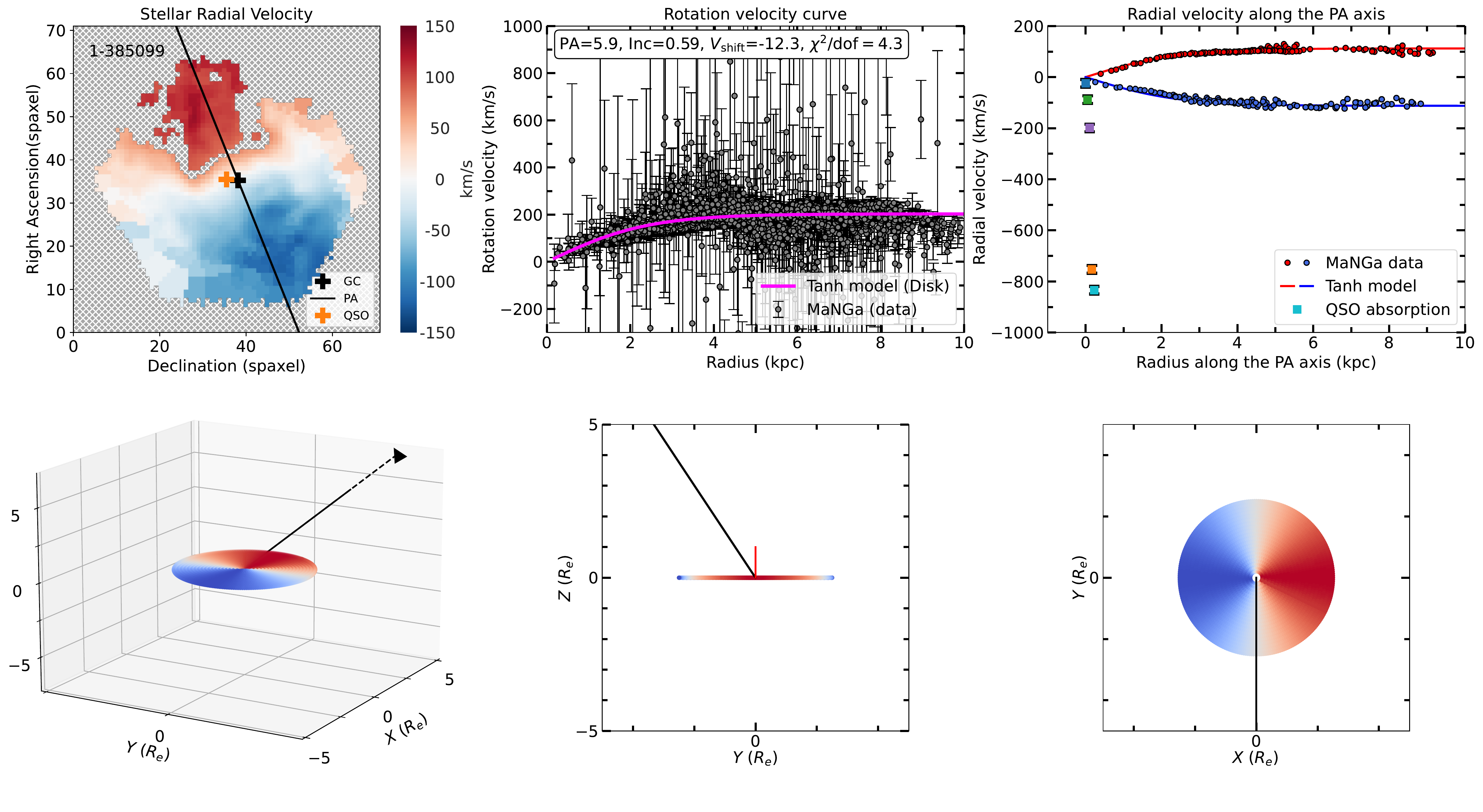}
        \caption{\rm 
        The comparison of radial velocity derived from the fitting to MaNGA maps of the galaxy 1-385099 and ones measured in the absorption system towards the quasar J0838$+$2453  The detailed description of panels is presented in the Appendix\,\ref{app:B}.
        }
        \label{Final-1-385099}
\end{center}
\end{figure*}

\begin{figure*}
\begin{center}
        \includegraphics[width=1.0\textwidth]{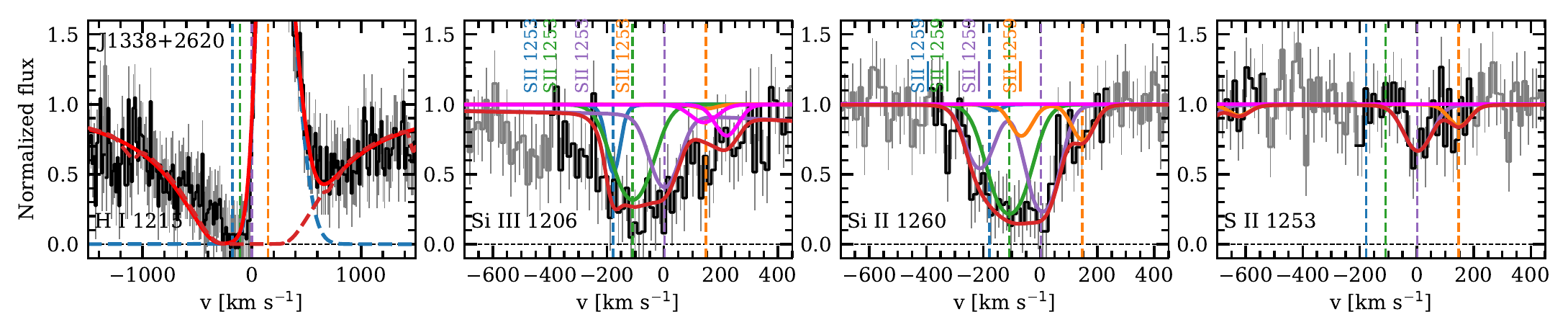}
        \includegraphics[width=1.0\textwidth]{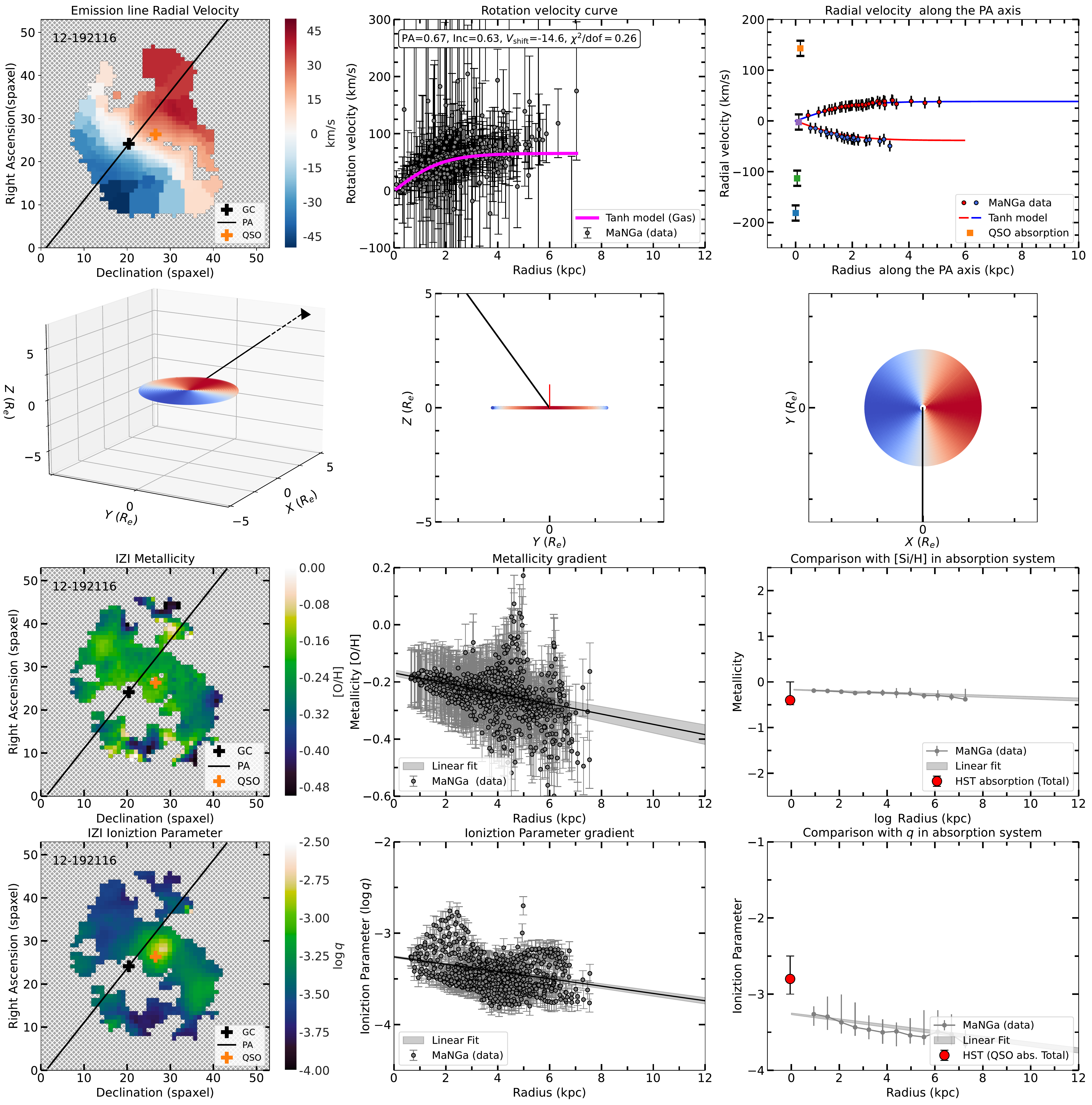}
        \caption{\rm 
        The comparison of radial velocity, metallicity and ionization parameter derived from the fitting to MaNGA maps of the galaxy 12-192116 and ones measured in the absorption system towards the AGN J1338$+$2620. The detailed description of panels  is presented in the Appendix\,\ref{app:B}.
        }
        \label{Final-12-192116}
\end{center}
\end{figure*}

\begin{figure*}
\begin{center}
        \includegraphics[width=1.0\textwidth]{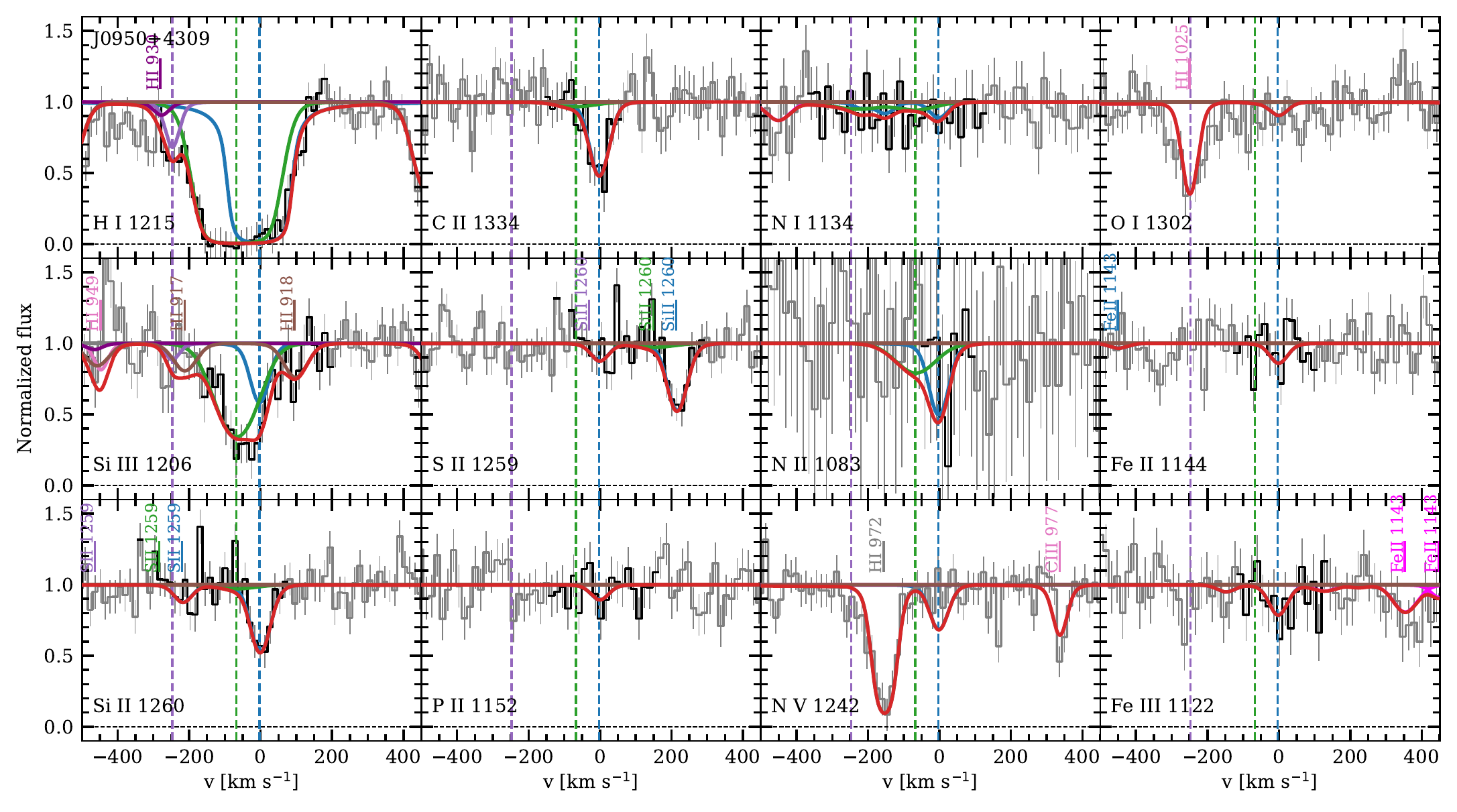}
        \caption{\rm Fit to absorption lines at $z_{\rm abs}=0.01714$  in the spectrum of J0950$+$4309. The synthetic profile is shown in red and the contribution from each component, associated with the studied galaxy, is shown in green, blue and purple (and orange). Dashed vertical lines represent the position of each component. Vertical sticks indicate the position of other absorption lines, associated with the MW (magenta sticks) and remote galaxies.}
        \label{fig:fit:j0950-lines}
\end{center}
\end{figure*}

\begin{figure*}
\begin{center}
        \includegraphics[width=1.0\textwidth]{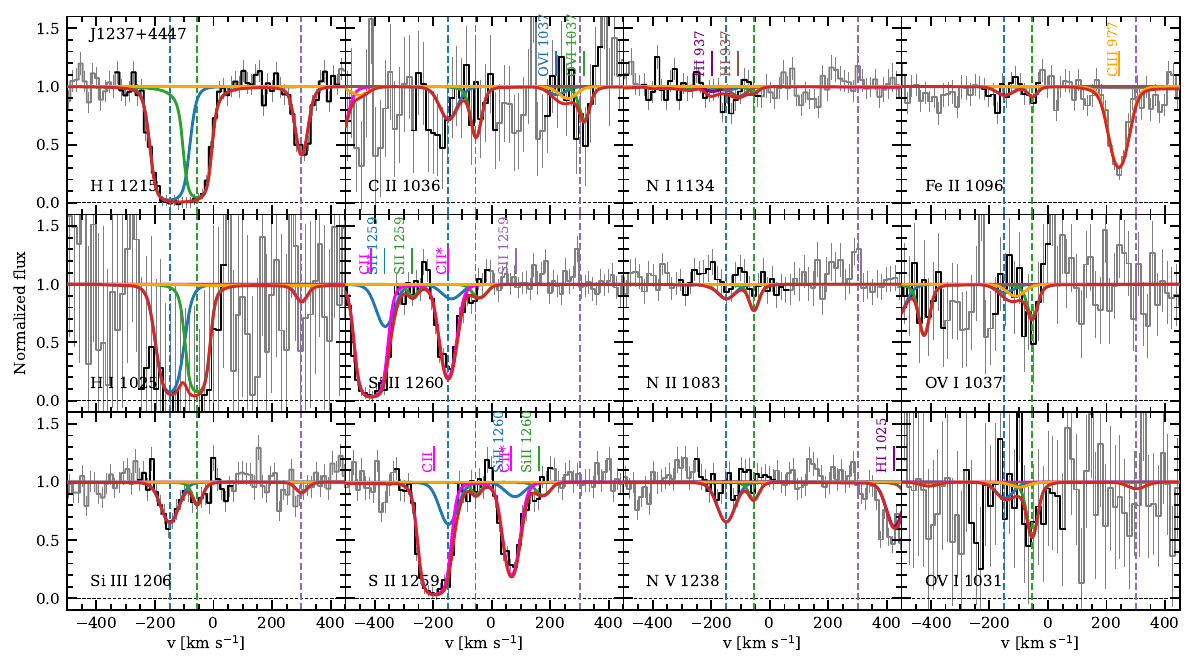}
        \caption{\rm Fit to absorption lines at $z_{\rm abs}=0.06013$  in the spectrum of J1237$+$4447. Lines are  as in Fig.\,\ref{fig:fit:j0950-lines}. }
        \label{fig:fit:j1237-lines}
\end{center}
\end{figure*}

\begin{figure*}
\begin{center}
        \includegraphics[width=1.0\textwidth]{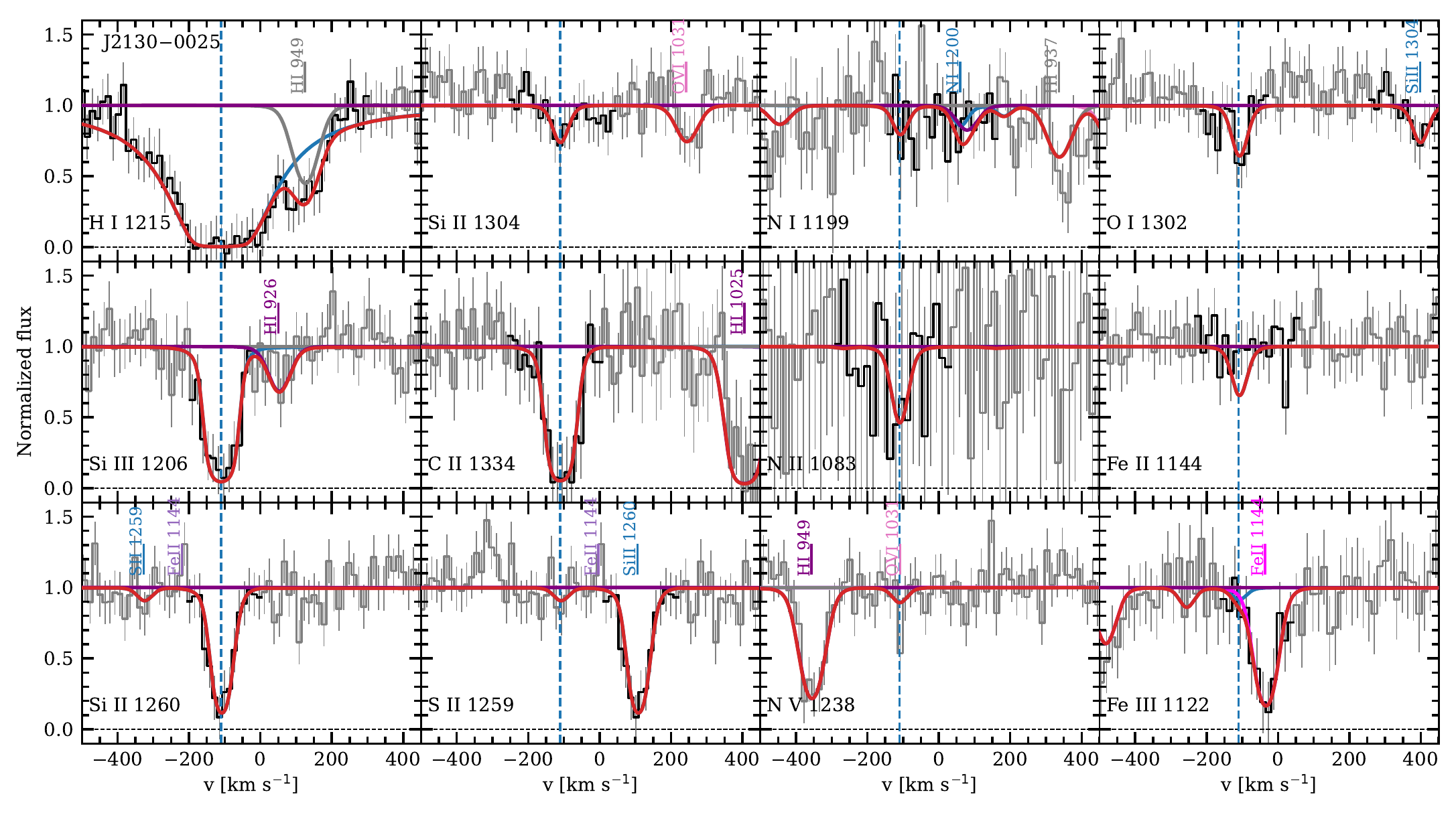}
        \caption{\rm Fit to absorption lines at $z_{\rm abs}=0.01966$  in the spectrum of J2130$-$0025. Lines are as in Fig.\,\ref{fig:fit:j0950-lines}.}
        \label{fig:fit:j2130-lines}
\end{center}
\end{figure*}

\begin{figure*}
\begin{center}
        \includegraphics[width=1.0\textwidth]{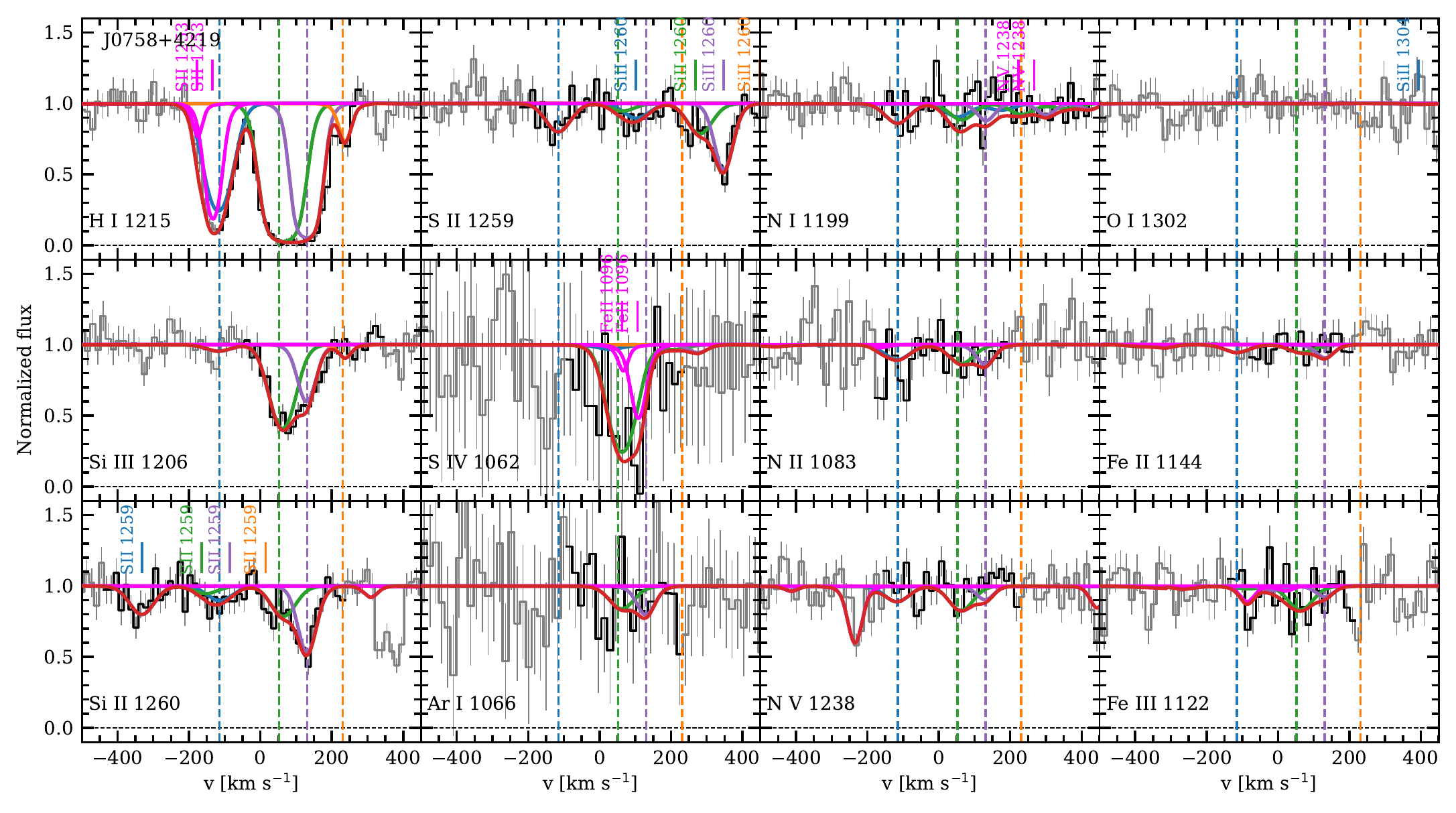}
        \caption{\rm Fit to absorption lines at $z_{\rm abs}=0.03181$  in the spectrum of J0758$+$4219. Lines are as in Fig.\,\ref{fig:fit:j0950-lines}.}
        \label{fig:fit:j0758-lines}
\end{center}
\end{figure*}

\begin{figure*}
\begin{center}
        \includegraphics[width=1.0\textwidth]{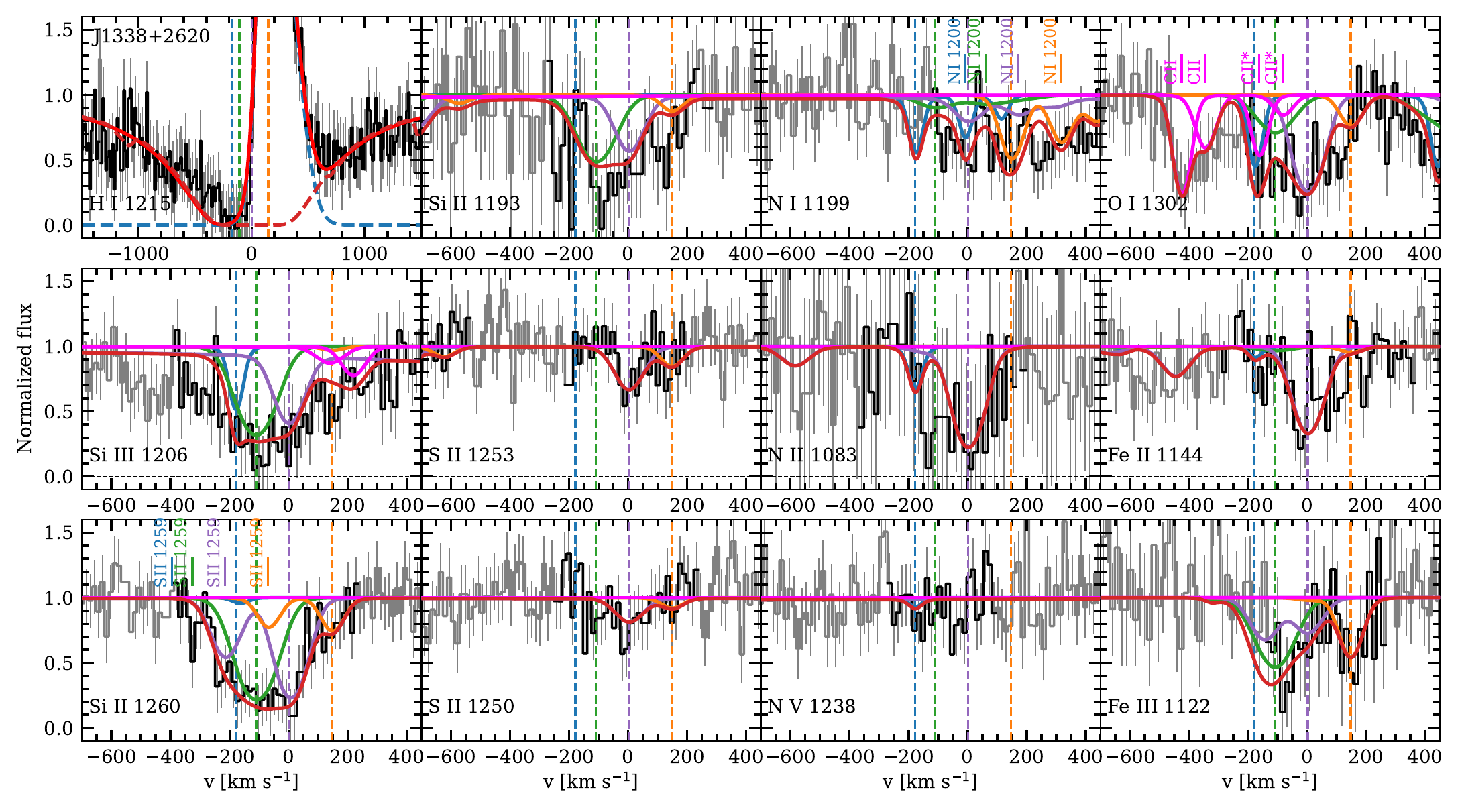}
        \caption{\rm Fit to absorption lines at $z_{\rm abs}=0.02604$  in the spectrum of J1338$+$2620. Lines are as in Fig.\,\ref{fig:fit:j0950-lines}. }
        \label{fig:fit:j1338-lines}
\end{center}
\end{figure*}

\begin{figure*}
\begin{center}
        \includegraphics[width=1.0\textwidth]{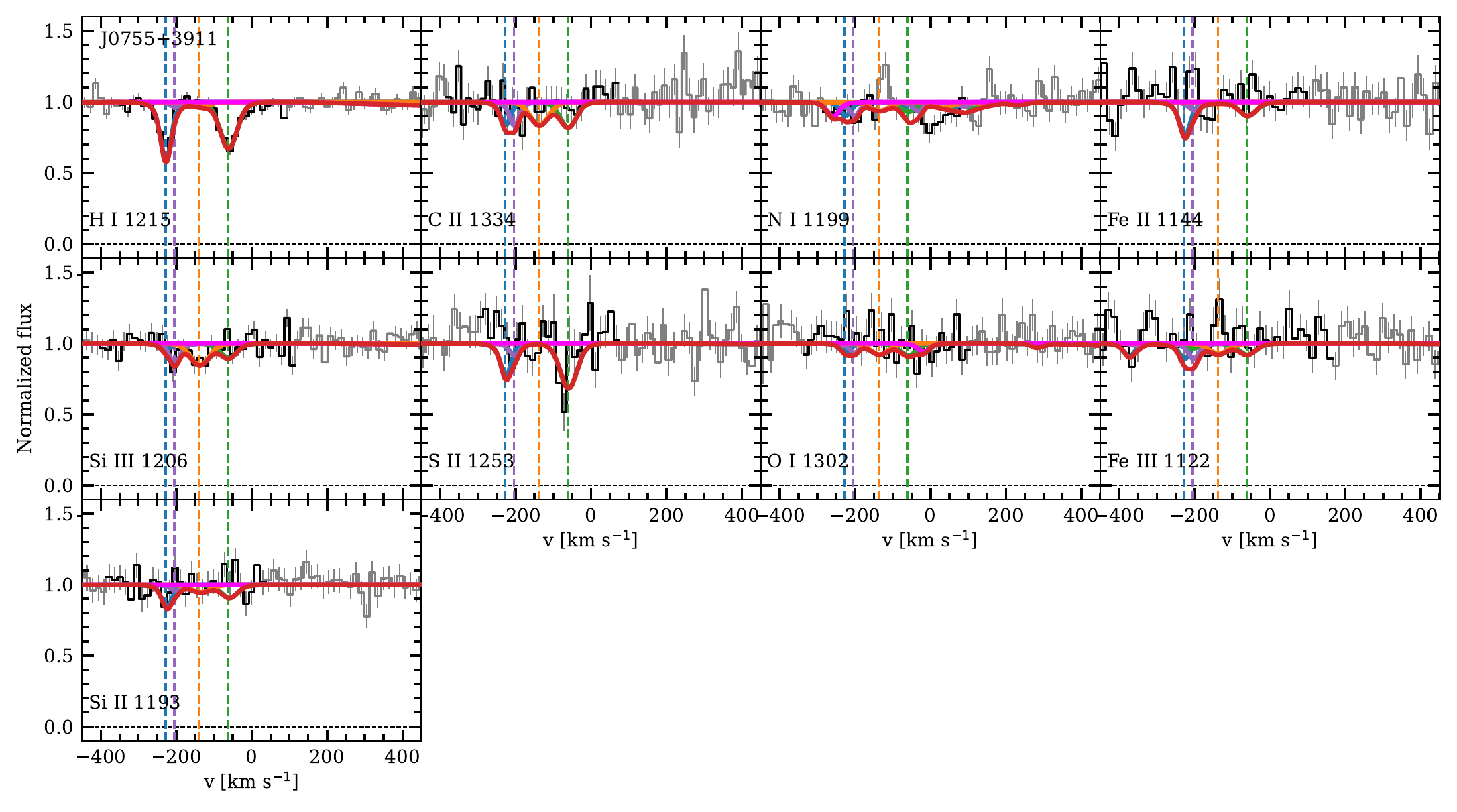}
        \caption{\rm Fit to absorption lines at $z_{\rm abs}=0.03362$  in the spectrum of J0755$+$3911. Lines are as in Fig.\,\ref{fig:fit:j0950-lines}. }
        \label{fig:fit:j0755-lines}
\end{center}
\end{figure*}

\begin{figure}
\begin{center}
        \includegraphics[width=1\textwidth]{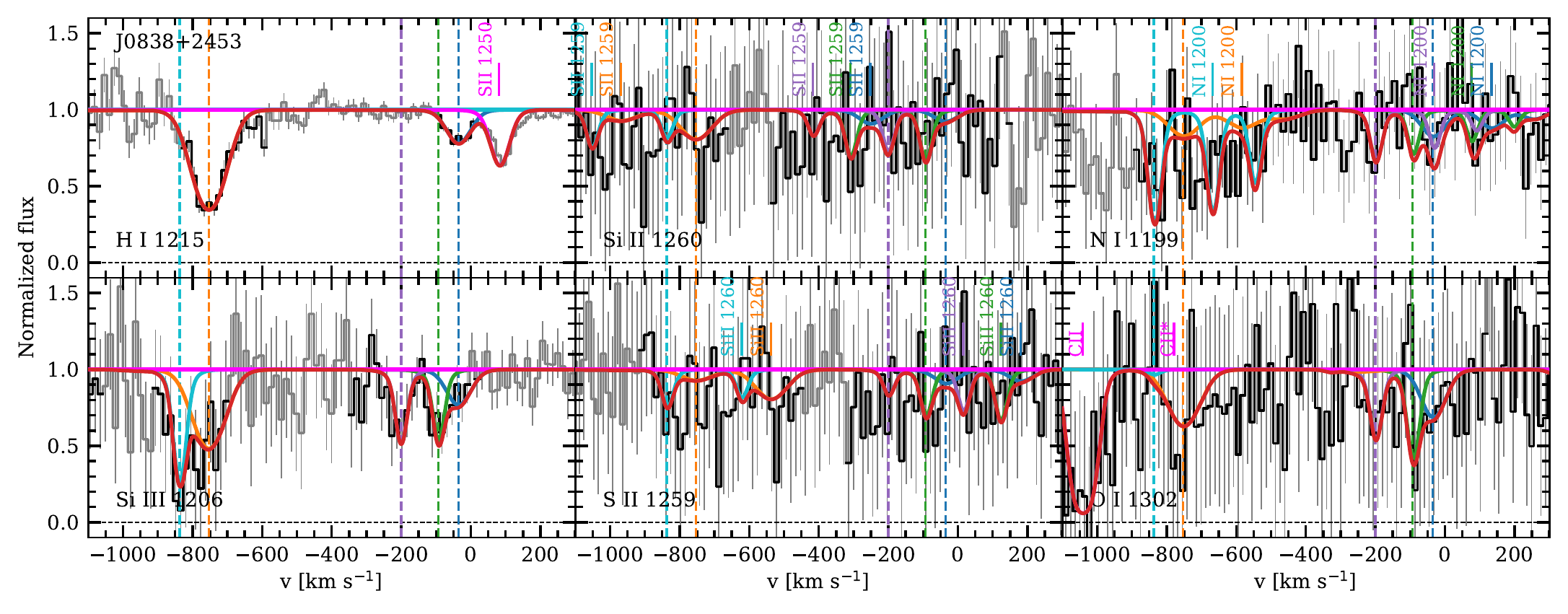}
        \caption{\rm Fit to absorption lines at $z_{\rm abs}=0.02843$  in the spectrum of J0838$+$2453. Lines are the same as in Fig.\,\ref{fig:fit:j0950-lines}. }
        \label{fig:fit:j0838-lines}
\end{center}
\end{figure}

\begin{figure}
\begin{center}
        \includegraphics[width=1\textwidth]{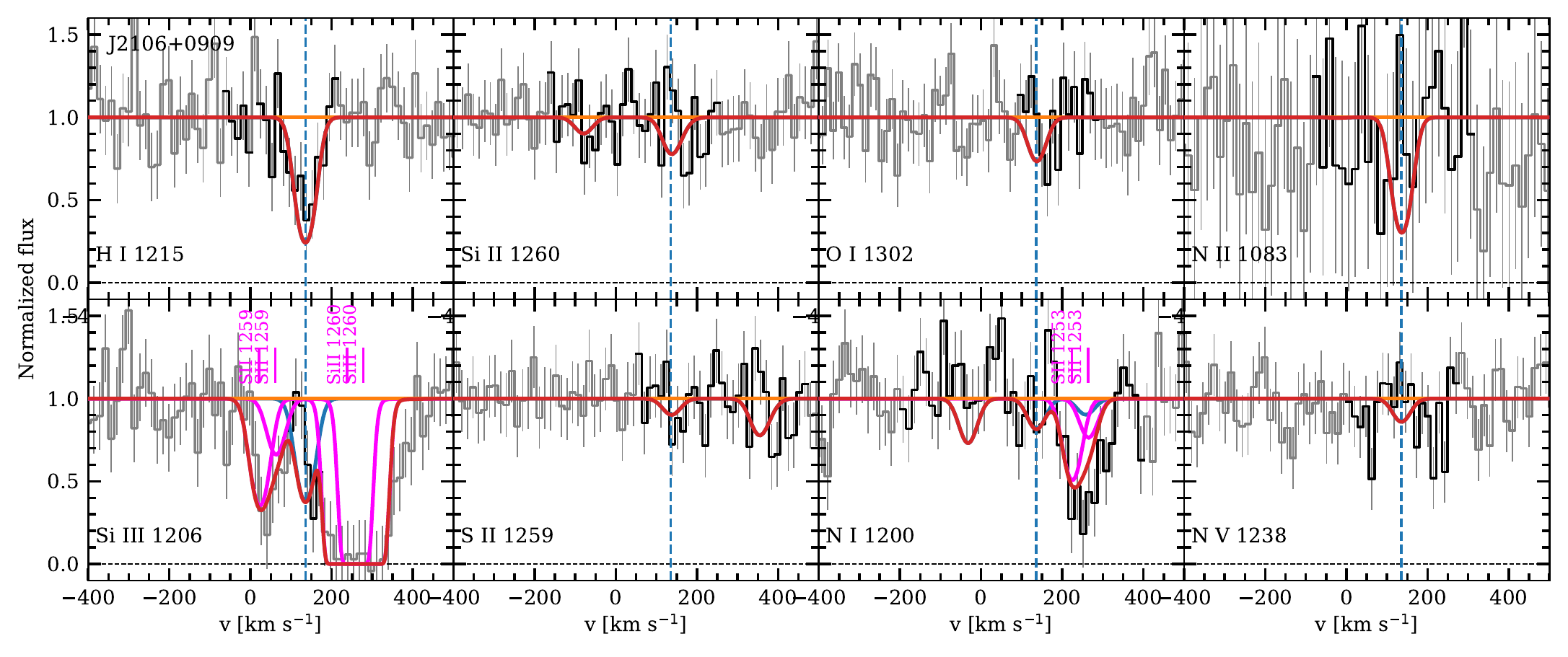}
        \caption{\rm Fit to absorption lines at $z_{\rm abs}=0.04375$  in the spectrum of J2106$+$0909. Lines are as in Fig.\,\ref{fig:fit:j0950-lines}. }
        \label{fig:fit:j2106-lines}
\end{center}
\end{figure}

\begin{figure}
\begin{center}
        \includegraphics[width=1\textwidth]{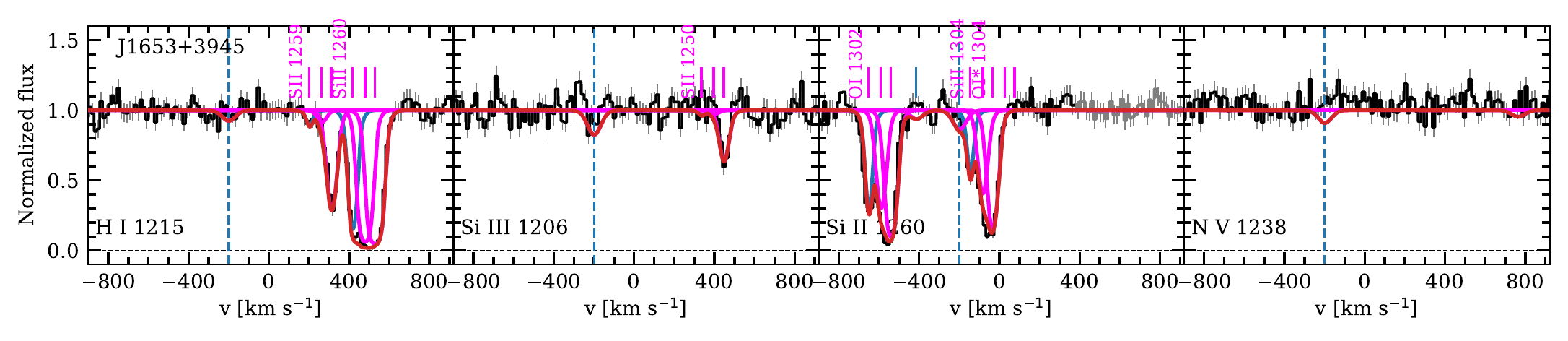}
        \includegraphics[width=1\textwidth]{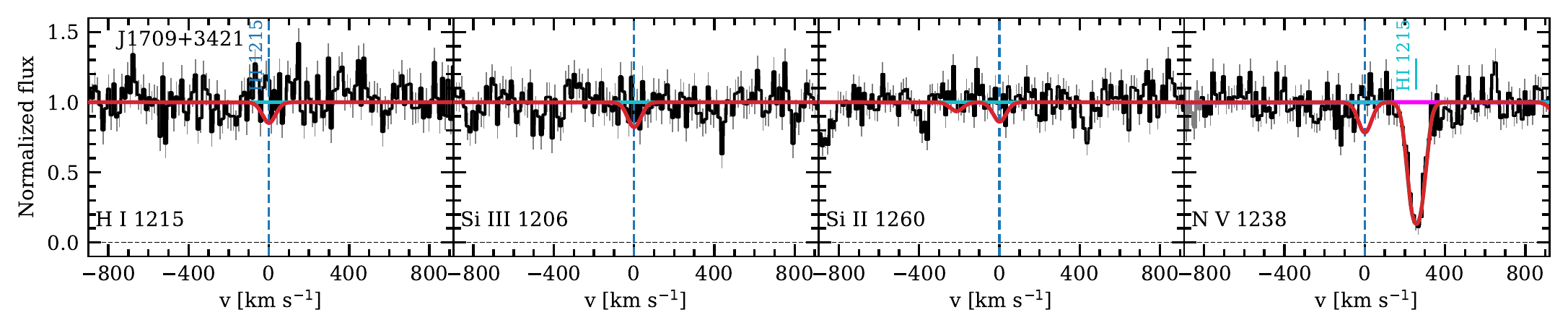}
        \includegraphics[width=1\textwidth]{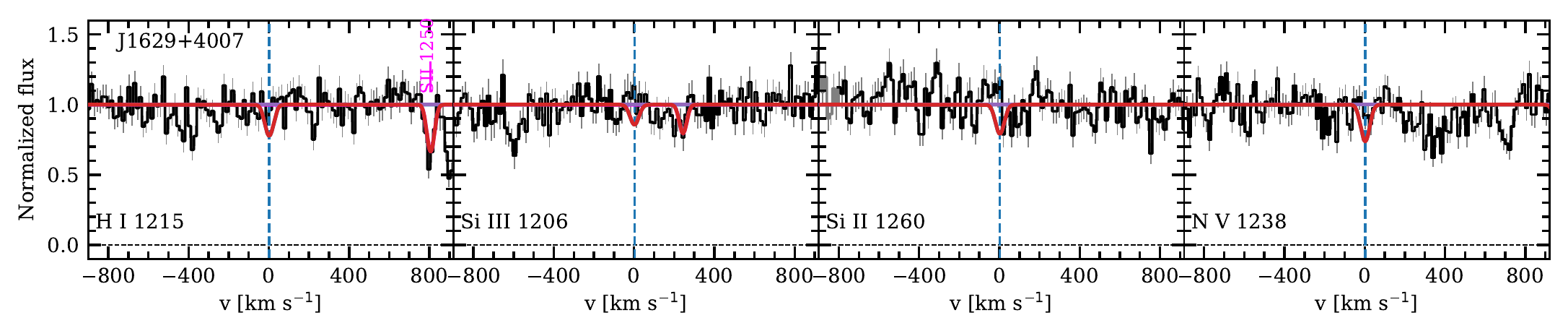}
        \caption{\rm The cases of non-detections. The absorption lines of \HI, \SiIII, \SiII, \NV\ in the HST COS spectra of  J1653$+$3945, J1709$+$3421 and J1629$+$4007  at the redshifts of the corresponding galaxies (1-594755, 1-561034, 1-564490, respectively). The synthetic profile is shown in red. The dashed vertical line represents the likely the position of absorption system. Vertical sticks indicate the position of absorption lines, associated with the Milky Way (MW, magenta sticks) and remote galaxies.}
        \label{fig:fit:j1629-lines}
\end{center}
\end{figure}

\begin{figure*}
    \begin{center}
        \includegraphics[width=0.97\linewidth]{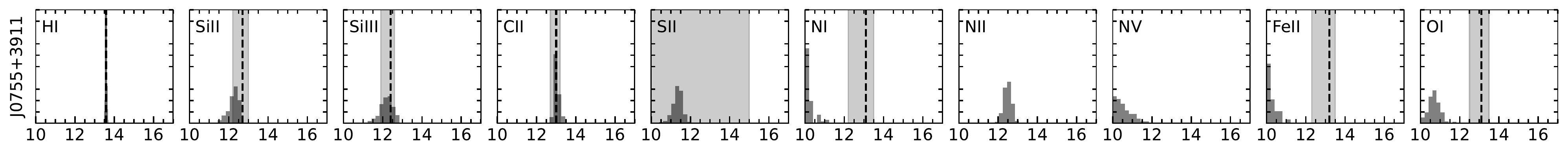}\hfill  
        \includegraphics[width=0.97\linewidth]{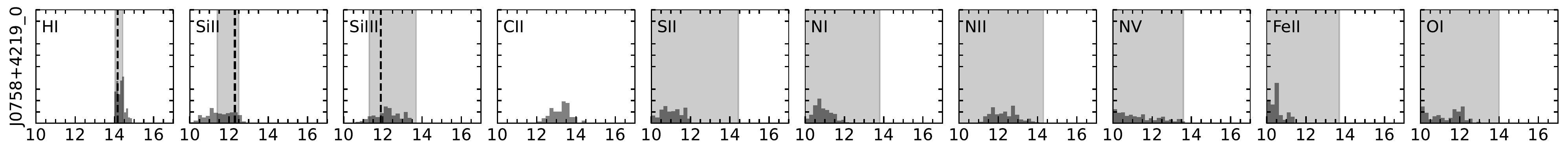}\hfill  
        \includegraphics[width=0.97\linewidth]{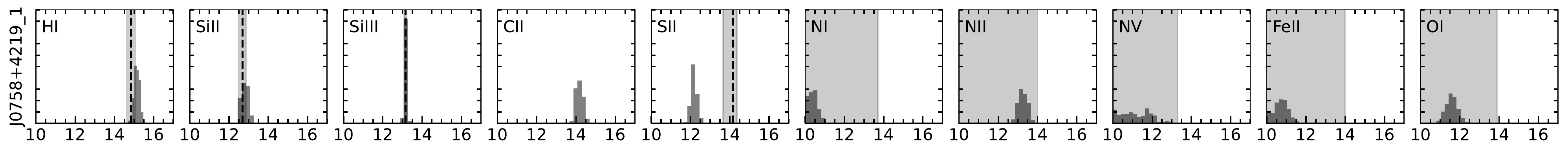}\hfill  
        \includegraphics[width=0.97\linewidth]{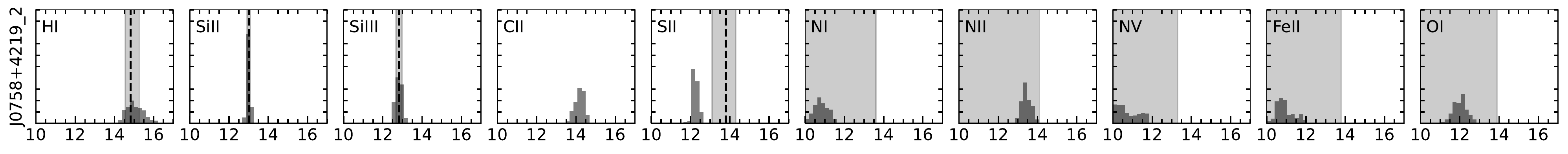}\hfill  
        \includegraphics[width=0.97\linewidth]{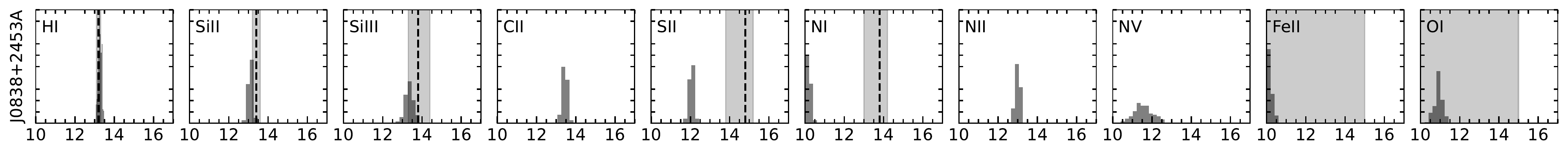}\hfill  
        \includegraphics[width=0.97\linewidth]{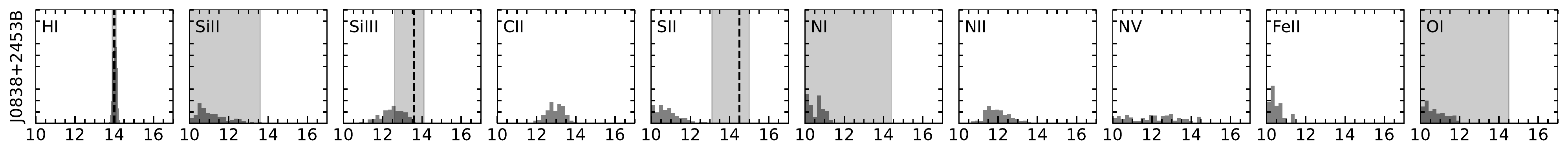}\hfill 
        \includegraphics[width=0.97\linewidth]{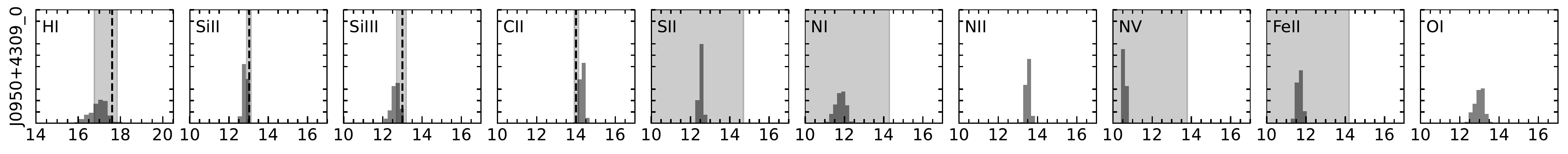}\hfill
        \includegraphics[width=0.97\linewidth]{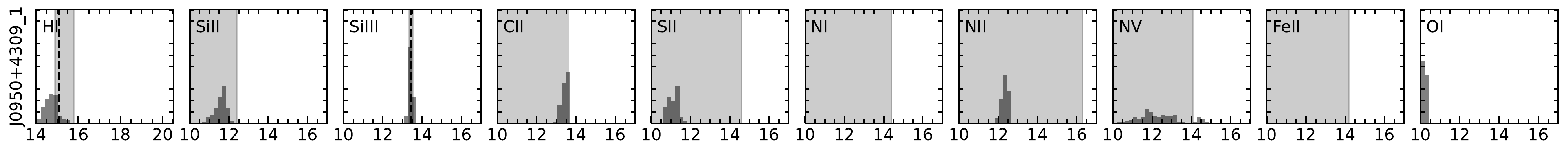}\hfill
        \includegraphics[width=0.97\linewidth]{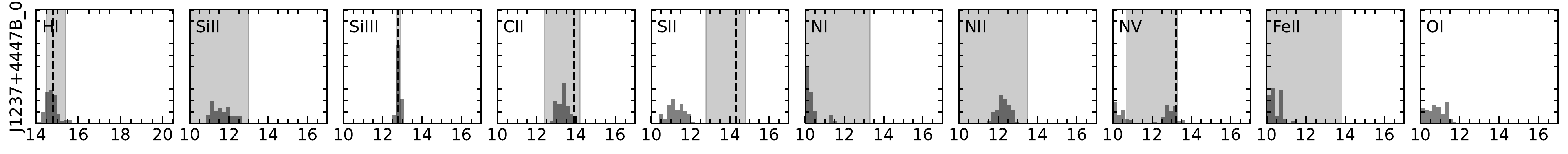}\hfill
        \includegraphics[width=0.97\linewidth]{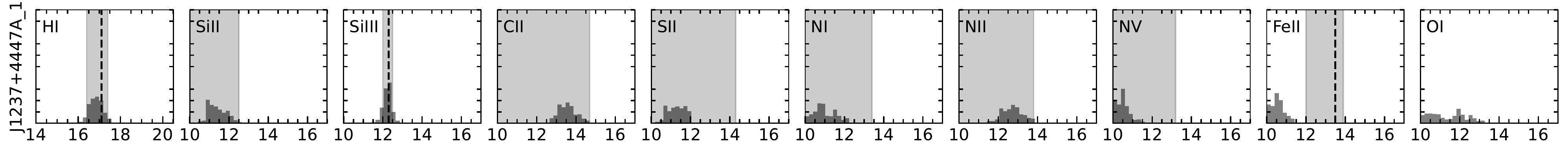}
        \hfill
        \includegraphics[width=0.97\linewidth]{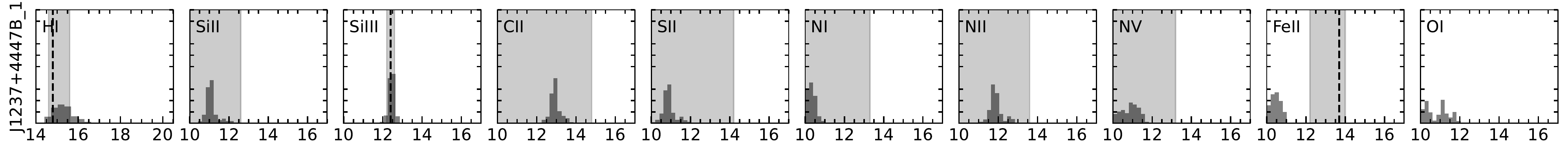}
        \hfill
        \includegraphics[width=0.97\linewidth]{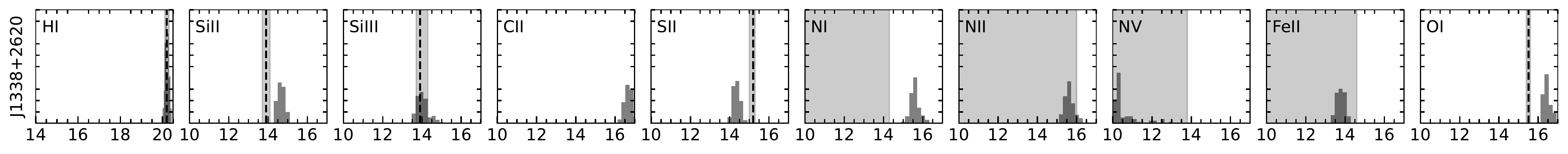}%
        \hfill
        \includegraphics[width=0.97\linewidth]{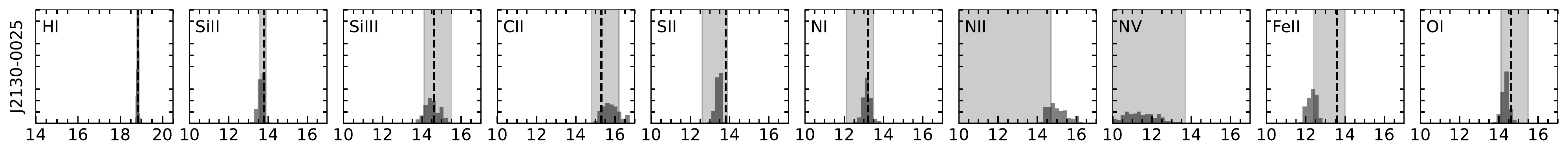}\hfill  
        \caption{\rm The fit results to measured species column densities in our sample with {\sc cloudy} simulation. Dashed vertical lines and light shaded areas represent the most probable value and its uncertainty. The light shaded area without the line represent upper limit. The histograms show distributions of species column densities in  {\sc cloudy} simulation.}
        \label{fig:_hist}
    \end{center}
\end{figure*}

\bibliography{Library}{}
\bibliographystyle{aasjournal}
%\bibliographystyle{aasjournal}
 % if your bibtex file is called example.bib
%\bsp	% typesetting comment 
%label{lastpage}

\end{document}